\documentclass[useAMS,usenatbib,usegraphicx]{mn2e}

\title[I: Dynamical Properties]{Southern GEMS Groups I: Dynamical Properties}
\author[Brough et al.]
	{Sarah Brough$^{1}$\thanks{E-mail: sbrough@astro.swin.edu.au},
	Duncan A.~Forbes$^{1}$, Virginia A.~Kilborn$^{1,2}$, Warrick
	Couch$^{3}$
\\$^{1}$Centre for Astrophysics and Supercomputing, Swinburne University of Technology, Hawthorn, VIC 3122, Australia
\\$^2$Australia Telescope National Facility, CSIRO, P.O. Box 76, Epping, NSW 1710, Australia
\\$^{3}$School of Physics, The University of New South Wales, Sydney, NSW 2052, Australia
}

\begin{document}

\date{Accepted... Received...; in original form 2005}

\pagerange{\pageref{firstpage}--\pageref{lastpage}} \pubyear{2005}

\maketitle

\label{firstpage}

\begin{abstract}
Here we present an investigation of the properties of 16 nearby galaxy
groups and their constituent galaxies. The groups are selected from
the Group Evolution Multi-wavelength Study (GEMS) and all have X-ray
as well as wide-field neutral hydrogen (HI) observations.  Group
membership is determined using a friends-of-friends algorithm on the
positions and velocities from the 6-degree Field Galaxy Survey (6dFGS)
and NASA/IPAC Extra-galactic Database (NED).  For each group we derive
their physical properties using this membership, including: velocity
dispersions ($\sigma_v$), virial masses ($M_V$), total K-band
luminosities ($L_K(Tot)$) and early-type fractions ($f_{early}$) and
present these data for the individual groups.  We find that the GEMS
X-ray luminosity is proportional to the group velocity dispersions and
virial masses: $L_X(r_{500})\propto\sigma_v^{3.11\pm0.59}$ and
$L_X(r_{500})\propto M_V^{1.13\pm0.27}$, consistent with the
predictions of self-similarity between group and clusters.  We also
 find that $M_V\propto L_K(Tot)^{2.0\pm0.9}$, i.e. mass grows faster
than light and that the fraction of early-type galaxies in the groups
is correlated with the group X-ray luminosities and velocity
dispersions.  We examine the brightest group galaxies (BGGs), finding
that, while the luminosity of the BGG correlates with its total group
luminosity, the fraction of group luminosity contained in the BGG
decreases with increasing total group luminosity. This suggests that
BGGs grow by mergers at early times in group evolution while the group
continues to grow by accreting infalling galaxies.  We form a
composite galaxy group in order to examine the properties of the
constituent galaxies and compare their properties with those of field
galaxies. There are clear radial trends, with group galaxies becoming
fainter, bluer and morphologically later types with increasing radius
from the group centre, reaching field levels at radii
$>r_{500}(>0.7r_{200})$.  We divide the composite group by group X-ray
luminosity and find that galaxies in high X-ray luminosity groups
(log$_{10}L_X(r_{500})\geq41.7$ erg s$^{-1}$) are redder with a higher
giant-to-dwarf ratio and are more likely to be early-type galaxies
than are those galaxies in low X-ray luminosity groups. We conclude
that harrassment and ram pressure stripping processes are unlikely to
cause these differences.  The differences are more likely to be due to
galaxy-galaxy mergers and possibly some further mechanism such as
strangulation.  If mergers are the dominant mechanism then the
properties of galaxies in the higher X-ray luminosity groups are a
result of mergers at earlier epochs in smaller mass groups that have
since merged to become the structures we observe today, while lower
X-ray luminosity groups are still undergoing mergers today.
\end{abstract}


\section{Introduction}

It has been known for many years that galaxies in denser environments
are in general redder and brighter
(e.g. \citealt{faber73,oemler74,visvanathan77,butcher84}) and are more
likely to be early-type galaxies showing little signs of recent star
formation than those in less dense environments
(e.g. \citealt{dressler80}).  Recent galaxy surveys have shown that
the density at which the observed properties of galaxies change from
those of the field to those of more dense environments is $\sim2-3$
virial radii in clusters (e.g. \citealt{lewis02}), and that the galaxy
density at these radii is equivalent to that of the density in poor
galaxy groups (e.g. \citealt{gomez03}).

The proposed mechanisms to transform the properties of galaxies are,
therefore, environmentally dependent.  The processes acting on
galaxies in the group environment are different to those acting in the
cluster environment: Ram pressure stripping
(e.g. \citealt{gunn72,quilis00}) and harassment
\citep{moore96} are more likely to be dominant in the dense, but rare,
environments of clusters.  However, mergers and strangulation are more
likely in the group environment, where the velocity dispersion of the
group is similar to that of its constituent galaxies
\citep{barnes85,zabludoff98,hashimoto00} and galaxies are predicted to 
be falling into a dense environment from the field for the first time.

Due to the small numbers of galaxies in an individual group and their
wide spatial distribution compared to a cluster, groups are a
relatively understudied environment.  However, galaxy groups are
significantly more abundant than galaxy clusters and most galaxies in
the local Universe are found in group, rather than cluster,
environments (e.g. \citealt{eke04}).  Considering the importance of
this environment it is vital that this is remedied.  It is therefore
important to study the properties of groups and their constituent
galaxies.  Multi-wavelength observations are vital to determine which
evolutionary processes are dominant in this environment.  While
optical spectra and near-infrared imaging provide dynamical
information on the groups and morphologies of individual galaxies,
X-ray observations provide information on the mass and dynamical state
of the group and neutral hydrogen (HI) observations show us regions of
potential star formation, and provide direct evidence of interactions
through tidal material.

\section{Group Sample}

The Group Evolution Multiwavelength Study (GEMS;
\citealt{osmond04,forbes06}) is an on-going study of 60 groups with a
range of optical and X-ray properties.  The group selection is
described in detail in \cite{osmond04}.  In addition to calculating
the X-ray luminosity of each group, which is proportional to the gas
density in the potential and hence mass, \cite{osmond04} also
determine whether the groups have intra-group X-ray emission, X-ray
emission solely from a galaxy halo or, if the X-ray flux was
$<3\sigma$ above the background level, defined the group as
undetected.  Wide-field neutral hydrogen (HI) observations of 16 GEMS
groups in the Southern hemisphere were made with the Parkes
radiotelescope (Kilborn et al. in preparation).


\cite{osmond04} defined the optical properties of the GEMS
groups within a radius corresponding to an overdensity of 500 times
the critical density -- $r_{500}$.  Galaxies were obtained from the
NASA/IPAC Extragalactic Database (NED) within $r_{500}$.
The optical data were used to calculate the mean velocity, $\bar{v}$,
and velocity dispersion, $\sigma_v$, of the groups and the updated
values were used to refine the search criteria.  The selection and
recalculation were repeated until the values of $\bar{v}$ and
$\sigma_v$ were stable.  The dynamical properties of the groups were
then calculated for the galaxies within $r_{500}$.  In some cases
there was only 1 galaxy member within $r_{500}$ preventing a measurement of
dynamical properties.

Here we re-analyse the optical properties of the 16 groups with HI
observations using a `friends-of-friends' percolation algorithm (FOF;
\citealt{huchra82}) to determine group membership.  We then compare 
the groups' optical properties with their X-ray properties.  The HI
properties of these groups will be analysed in Kilborn et al. (in
preparation).

Unlike \cite{osmond04} where galaxies are defined as group members if
they are within a specific radius and velocity range of each centroid,
FOF determines which galaxies are associated with one another in
position and velocity space and does not rely on any {\it a priori}
assumption about the geometrical shape of groups.  We use the 6-degree
Field Galaxy Survey (6dFGS; \citealt{jones04}) to obtain galaxy
velocities for this sample of 16 groups and emphasize that every
galaxy in our analysis has measured recession velocities, freeing our
statistical analyses from projection effects and providing a
significant advantage over some previous studies.  We examine the
Two-micron All Sky Survey (2MASS; \citealt{jarrett00}) near-infrared
magnitudes and colours of the constituent galaxies.  These magnitudes
provide an advantage over the $B$-band magnitudes used by
\cite{osmond04} as the light from the near-infrared is more closely
related to galaxy mass.  We also stack galaxies in the groups to
create a composite group with galaxy numbers comparable to those in a
cluster, in order to better understand the properties of the groups.

The dynamical properties of two of these groups, NGC 1332 and NGC 1407
were discussed in \cite{brough05}.  Those results are updated and the
remaining 14 groups presented and discussed in this paper.  The group
properties 
and X-ray attributes of the groups derived by \cite{osmond04} are 
given in Table~\ref{osmond_groups} -- G indicates groups with
intra-group X-ray emission (i.e. X-ray extent $>60$ kpc), H indicates
groups with only galaxy-halo emission (i.e. emission $\leq60$ kpc) and
U indicates that the X-ray flux from that region was $<3\sigma$ above
the background level and the group is therefore undetected in X-rays.  Eight of
the groups in this sample show X-ray emission, 6 show solely
galaxy-halo emission and 2 are undetected in X-rays.
The NGC 3783, NGC 7144 and NGC 7714 groups
had less than 4 members within $r_{500}$ in \cite{osmond04} and do not
appear in their final statistical analysis.

The outline of the paper is as follows: We introduce our data in
Section~\ref{data_sect} and in Section~\ref{members_sect} we discuss
our method of determining group membership using these data.  The
dynamical properties of the groups based on that membership are
presented in Section~\ref{dynamics_sect} and the properties of the
individual groups are discussed in Section~\ref{indiv_sect}.  The
relations between the X-ray and dynamical properties of the groups are
presented in Section~\ref{scaling_sect} and the properties of the
brightest group galaxies in Section~\ref{section_bggs}.  Our analysis
of the galaxies in each group stacked to create a composite group is
presented in Section~\ref{composite_sect} and we draw our conclusions
in Section~\ref{concl_sect}.  As in previous GEMS papers we assume
$H_0=70$ km s$^{-1}$ Mpc$^{-1}$.

\begin{table*}
\begin{center}
\caption{Group properties defined by Osmond \& Ponman (2004) using the 
distances calculated in that paper. Dashes indicate where information is
not available.}
\label{osmond_groups}
\begin{tabular}{lccccccc}
\hline
Group Name&N&$\bar{v}$&$\sigma_v$&$r_{500}$&log$_{10}L_X~(r_{500})$&$T_X$&X-ray emission\\ 
&&(km s$^{-1}$)&(km s$^{-1}$)&(Mpc)&(erg s$^{-1}$)&(keV)&\\
\hline
NGC 524  & 10&         2470$\pm 55$&175$\pm 42$&0.45&41.33$\pm0.05$&0.65$\pm0.07$&H\\
NGC 720  &  4&         1640$\pm136$&273$\pm122$&0.40&41.43$\pm0.02$&0.52$\pm0.03$&G\\
NGC 1052 &  5&         1366$\pm 41$& 91$\pm 35$&0.36&40.53$\pm0.15$&0.41$\pm0.15$&H\\
NGC 1332 & 10&         1489$\pm 59$&186$\pm 45$&0.42&40.93$\pm0.02$&0.56$\pm0.03$&H\\
NGC 1407 & 20&         1682$\pm 71$&319$\pm 52$&0.57&41.92$\pm0.02$&1.02$\pm0.04$&G\\
NGC 1566 &  9&         1402$\pm 61$&184$\pm 47$&0.47&40.85$\pm0.05$&0.70$\pm0.11$&H\\
NGC 1808 &  4&         1071$\pm 52$&104$\pm 47$&0.32&$<40.59$& -- &U\\
NGC 3557 & 14&         2858$\pm 80$&300$\pm 60$&0.27&42.11$\pm0.04$&0.24$\pm0.02$&G\\
NGC 3783 &  1&         2917       &        --  &0.25&40.94$\pm0.11$& -- &G\\
NGC 3923 &  8&         1764$\pm 85$&239$\pm 66$&0.40&41.07$\pm0.02$&0.52$\pm0.03$&H\\
NGC 4636 &  9&          936$\pm 95$&284$\pm 73$&0.51&41.71$\pm0.02$&0.84$\pm0.02$&G\\
NGC 5044 & 18&         2518$\pm100$&426$\pm 74$&0.62&43.09$\pm0.01$&1.21$\pm0.02$&G\\
NGC 7144 & 2&          1912 $\pm29$& 41$\pm 41$&0.38&40.71$\pm0.13$& -- & H\\
HCG 90   & 15&         2559$\pm 34$&131$\pm 25$&0.38&41.79$\pm0.05$&0.46$\pm0.07$&G\\
IC 1459  &  8 &        1835$\pm 79$&223$\pm 62$&0.35&41.46$\pm0.04$&0.39$\pm0.04$&G\\
NGC 7714 &  2&         2784$\pm 20$& 28$\pm 28$&0.22&$<40.48$& -- &U\\
\hline
\end{tabular}
\flushleft
The columns indicate (1) Group name, (2) Number of group members
within $r_{500}$, (3) Mean velocity of group and $1\sigma$ error, (4)
Velocity dispersion and $1\sigma$ error, (5) $r_{500}$ radius, (6)
$ROSAT$ PSPC X-ray luminosity extrapolated to the $r_{500}$ radius and
$1\sigma$ error, (7) $ROSAT$ PSPC X-ray temperature and $1\sigma$ error, (8)
indicates whether the X-ray emission is from intra-group gas (G), a
galaxy halo (H) or is $<3\sigma$ background (U).
\end{center}
\end{table*}

		

\section{Data}
\label{data_sect}
The 6dFGS is a wide-area (the entire southern sky with $|b|>10^{o}$),
primarily $K_s$-band selected galaxy redshift survey.  The catalogue
provides positions, recession velocities, and spectra for the
galaxies, along with total $K_s$-band magnitudes adapted from the
2MASS catalogue \citep{jones04}.

The second data release of the 6dFGS (DR2; \citealt{jones05}) contains
71,627 unique galaxies.  We obtained data for 13 of the 16 groups from
this catalogue for a square of $\sim5.5^{\circ}\times5.5^{\circ}$
around the position of each group defined by \cite{osmond04}.  This
area is chosen to match the HI observations.  However, we extended the
search regions around the NGC 1332, NGC 1407 and NGC 4636 groups.  NGC
1332 and NGC 1407 occupy the same region of sky and a circle of radius
$15^{\circ}$ was examined around these galaxies in
\cite{brough05}.  NGC 4636 lies south of the Virgo cluster centre and it was
necessary to extend the study to $15^{\circ}\times15^{\circ}$ to
determine a boundary with the Virgo cluster.  In all the datasets we
limited the data to recession velocities $500 < v< 5000$ km s$^{-1}$,
with the exception of NGC 1332 and NGC 1407 which are limited to
$v<2500$km s$^{-1}$.  The lower limit was chosen to avoid Galactic
confusion.  The search regions and numbers of galaxies found are
summarised in Table~\ref{group_data}.  The 6dFGS database also
provides 2MASS $K_s$-band magnitudes where available.  We have used
the 2MASS $K_s$ magnitudes within the 20th magnitude isophote
(henceforth denoted as $K$).  As 2MASS is $>99$ per cent complete to
$m_K\sim13.1$
\citep{jarrett00} we assume that those galaxies without $K$ magnitudes are
fainter than the 2MASS magnitude limit.

\subsection{NED}

We have 3 groups in the Northern sky for which 6dFGS data is not
available (NGC 524, NGC 4636 and NGC 7714).  The 6dFGS is also not yet
complete.  We therefore supplemented the 6dFGS data with sources with
known recession velocities from the NASA/IPAC Extragalactic Database
(NED) in the same position and velocity range.  This added an extra
1220 unique galaxies, also detailed in Table~\ref{group_data}.  The
total number of galaxes from NED and 6dFGS is 1735.

NED is a heterogeneous data resource.  We illustrated in
\cite{brough05} that using NED our sample is complete in photometry and velocity
to a minimum of $K\leq11$ mag, and that including data fainter than
this has little effect, within the errors, on the results we obtain.


\subsection{HI}

Kilborn et al. (in preparation) details new galaxies found in the
Parkes HI maps of these 16 groups and also the velocities found for
previously catalogued galaxies without known velocities.  We use these
new data in our dynamical analysis and the numbers for each region are
summarised in Table~\ref{group_data}.  These galaxies are all fainter
than the 2MASS apparent-magnitude limit ($m_K\sim13.1$;
\citealt{jarrett00}).

\subsection{Magnitudes}

To avoid the effects of peculiar velocities which are significant at
recession velocities less than $2000$ km s$^{-1}$ \citep{marinoni98}
we use independent distances in order to calculate the absolute
magnitudes.  The distances are primarily from the distance moduli
($DM$) from surface brightness fluctuation studies by \cite{tonry01},
corrected following the work of \cite{jensen03}, indicated by `TJ' in
Table~\ref{group_data}.  The distance modulus to the NGC 1332 and NGC
1407 groups is calculated from the NGC 1407 galaxy globular cluster
luminosity function of \cite{forbes05}, denoted by `F' in
Table~\ref{group_data}.  Distances for the remaining groups were
calculated by \cite{osmond04} from their mean group velocities after
correcting for the infall into Virgo and the Great Attractor, denoted
`O' in Table~\ref{group_data}.  Absolute magnitudes are calculated as
$M=m-DM-A$. Galactic Extinction ($A$) is calculated using the
extinction maps of \cite{schlegel98} and is of the order $A_K\sim0.01$
mag and $A_B\sim0.09$ mag.

\begin{table*}
\begin{center}
\caption{Search regions, distances, and sources of galaxy data for each group.}
\label{group_data}
\begin{tabular}{lccccccccc}
\hline
Group&Min RA&Max RA&Min Dec&Max Dec&$D$&6dFGS&HI&NED&Total\\
&(J2000)&(J2000)&(J2000)&(J2000)&(Mpc)&(sources)&(sources)&(sources)&\\
\hline
NGC 524 & 1:14 & 1:34 & 06:30&12:00& 22.3 (TJ)& 0 & 0 & 33 & 33 \\
NGC 720 & 1:40 & 2:03 & -16:30&-11:00& 25.7 (TJ)& 6 & 3 & 10 & 19 \\
NGC 1052 & 2:28 & 2:52 & -11:15&-05:45& 18.0 (TJ)& 30 & 0 & 42 & 72 \\
NGC 1332 & 2:16 & 4:40 & -6:00&-36:00& 20.9 (F)& 158 & 0 & 499 & 657 \\
NGC 1407 & 2:16 & 4:40 & -6:00&-36:00& 20.9 (F)& 158 & 0 & 499 & 657 \\
NGC 1566 & 3:54 & 4:36 & -58:30&-52:30& 21.0 (O)& 9 & 2 & 18 & 29 \\
NGC 1808 & 4:55 & 5:23 & -40:30&-35:00& 17.0 (O)& 17 & 0 & 5 & 22 \\
NGC 3557 & 10:55 & 11:24 & -40:45&-34:30& 42.5 (TJ)& 36 & 3 & 17 & 56 \\
NGC 3783 & 11:10 & 12:05 & -43:00&-33:00& 36.0 (TJ)& 81 & 3 & 27 & 111 \\
NGC 3923 & 11:25 & 12:15 & -34:00&-23:00& 21.3 (O)& 92 & 2 & 29 & 123 \\
NGC 4636 & 12:24 & 13:05 & -02:00&08:00& 13.6 (TJ)& 0 & 1 & 352 & 353 \\
NGC 5044 & 13:02 & 13:27 & -19:15&-13:30& 29.0 (TJ)& 25 & 6 & 56 & 87 \\
NGC 7144 & 21:36 & 22:12 & -51:30&-46:00& 22.8 (TJ)& 1 & 3 & 15 & 19 \\
HCG 90 & 21:48 & 22:15 & -35:30&-30:30& 36.0 (O)& 39 & 0 & 66 & 105 \\
IC 1459 & 22:42 & 23:12 & -40:00&-33:30& 27.2 (TJ)& 21 & 0 & 26 & 47 \\
NGC 7714 & 23:25 & 23:47 & -01:30&04:30& 39.0 (O)& 0 & 1 & 25 & 26 \\
\hline
\end{tabular} 
\flushleft
The columns indicate (1) Group name; (2), (3), (4) and (5) outline the
Right Ascension and Declination range over which the 6dFGS and NED
searches were made; (6) distance [from \cite{tonry01} corrected
following Jensen et al. (2003; TJ), Forbes et al. (2005; F) or Osmond
\& Ponman (2004; O)]; (7) Number of sources in the group region from
the 6dFGS catalogue; (8) Number of sources in the group region from
the HI survey of Kilborn et al. (in preparation); (9) Number of
sources in the group region from NED; (10) Total number of sources in
each region.
\end{center}
\end{table*}

\section{Group Membership}
\label{members_sect}
In order to study the dynamics of these groups it is important to
determine which galaxies are associated with each other in each field.
We used the FOF percolation algorithm which finds group structures in
galaxy data based on positional and velocity information and does not
rely on any {\it a priori} assumption about the geometrical shape of
groups.  As we are examining a small range in recession velocities we
do not adopt the method used by \cite{huchra82} to compensate for the
sampling of the galaxy luminosity function as a function of the
distance of the group.

Owing to the similarity in sampling between the 2dFGRS and 6dFGS at
these recession velocities we follow the prescriptions of the 2dFGRS
Percolation-Inferred Galaxy Group (2PIGG;
\citealt{eke04}) catalogue to determine the most appropriate value of 
limiting density contrast, $\delta \rho/\rho$,  
\begin{equation}
\frac{\delta \rho}{\rho}=\frac{3}{4} \pi D_0^3\left[\int_{-\infty}^{M_{lim}}\phi(M)dM\right]^{-1}-1.
\end{equation}
The number density contour surrounding each group represents a fixed
number density enhancement relative to the mean number density.  We
assume the differential galaxy luminosity function defined by
\cite{kochanek01}, which \cite{ramella04} determine to be a good
approximation for the $K$-band groups luminosity function
($M_\star=-22.6$, $\alpha=-1.09$ and $\phi_\star=0.004$ for $H_0=70$
km s$^{-1}$ Mpc$^{-1}$).  The 2PIGG limiting density contrast $\delta
\rho/\rho=150$ then gives $D_0=0.29$ Mpc.  We also follow 2PIGG to
calculate our velocity limit, $V_0$.  The peculiar motion of galaxies
moving in a gravitational potential lengthens the group along the
line-of-sight in velocity space -- giving the `Finger of God'
effect.  If we assume that the projected spatial ($D_0$) and the
line-of-sight dimensions of a group in velocity space ($V_0$) are in
proportion, \cite{eke04} show that a ratio of 12 for $V_0$
relative to $D_0$
is the most appropriate for a linking volume, giving $V_0=347$ km
s$^{-1}$ here.

The FOF algorithm was run over the whole sample of galaxies in each
region.  We remove all groups with $N\leq3$ galaxies as these have
been shown by many surveys to be significantly more likely to be false
positives found by the FOF algorithm
(e.g. \citealt{ramella95,nolthenius97,diaferio99}).

Groups corresponding to the positions and velocities presented in
\cite{osmond04} were found in 15 out of the 16 groups.  In contrast,
no group with $>3$ members was found in the NGC 7144 field.  This
group was initially determined to consist of three galaxies (NGC 7155,
NGC 7144 and NGC 7145) by \cite{huchra82}. \cite{tully87} used a
hierarchical group finding technique and found a group consisting of
the same 3 galaxies in his Nearby Galaxies Catalogue.  \cite{maia89}
analysed the Southern Sky Redshift Survey (SSRS, \citealt{dacosta88})
with a FOF analysis and added a new group member (NGC 7151).
\cite{garcia93} examined the LEDA database using both FOF and
hierarchical group-finding techniques and added a further new group
member (ESO 236-G035).  
There is no obvious group system present in the optical data,
suggesting that previous group finders were tuned to find looser
structures.  There is no group X-ray emission in this region, only
emission associated with the NGC 7144 galaxy and \cite{osmond04} only
found two galaxies associated with this group.  Thus, contrary to
earlier claims, we do not find evidence for a group and do not discuss
this group further, leaving a sample of 15 groups.


We compared the groups defined by taking three different density
contrasts: $\delta \rho / \rho=100$ (i.e. less dense than the 2PIGG
catalogue), $\delta \rho / \rho=150$ (the 2PIGG value) and $\delta
\rho / \rho=200$ (i.e. more dense).  We find in 4 groups (NGC 1566,
NGC 1808, NGC 3557 and NGC 3783) that there is no change in group
membership and therefore none in mean velocity or velocity dispersion.
For all 15 groups we find the mean difference in group members to be
$\langle \Delta N (150-200)\rangle =2.6\pm0.9$ and $\langle \Delta N
(150-100)\rangle =-6.8\pm3.3$ i.e. on average, there are less galaxies
if the density contrast is increased and more galaxies if it is
lowered, as expected.  Comparing the velocity dispersions calculated
for the different memberships we can examine the effect of the choice
of density contrast on the calculated dynamical parameters: $\langle
\Delta \sigma_v(150-200)\rangle=25\pm17$ km s$^{-1}$ and $\langle
\Delta \sigma_v(150-100)\rangle=3\pm7$ km s$^{-1}$.  These differences
are smaller than the errors on the calculated velocity dispersions
and, within the error on the mean difference, consistent with no
difference.
We therefore conclude that the group properties defined by FOF are
robust to the choice of density contrast and present results using the
2PIGG $\delta \rho / \rho=150$ in Table~\ref{fof_groups}.  The
galaxies determined to be group members are listed in
Appendix~\ref{group_gals}.  The spatial distribution and
velocity-distance diagrams for these groups are illustrated in
Appendix~\ref{group_pics}.

\begin{table*}
\vbox to220mm{\vfil 
\caption{Landscape table goes here}
\label{fof_groups}
\vfil}
\end{table*}

\begin{figure}
\begin{center}
	\resizebox{20pc}{!}{
	\rotatebox{-90}{
	\includegraphics{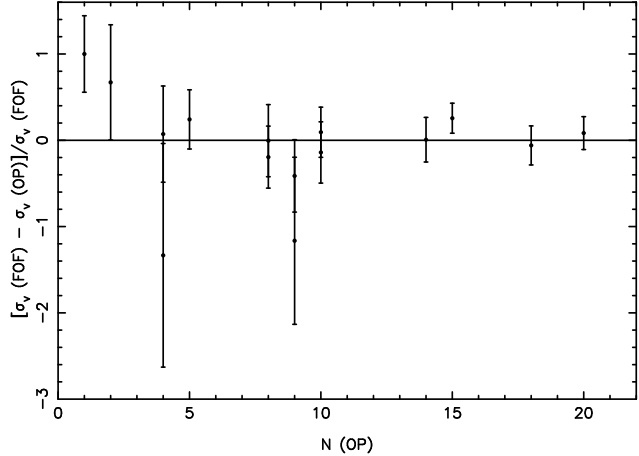} 
		}	 
	}	
  \end{center}
\caption{Comparing the velocity dispersions calculated by Osmond \& Ponman 
(2004), $\sigma_{v}(OP)$, and those calculated using the FOF
membership, $\sigma_{v} (FOF)$ with the number of members assigned by
Osmond \& Ponman, $N$ (OP).  Error bars indicate the combined
$1\sigma$ errors on the velocity dispersions.}
\label{sigma_compare}
\end{figure}

The velocity dispersions derived using the membership determined by
FOF are compared in Figure~\ref{sigma_compare} with those derived by
\cite{osmond04}.  As \cite{osmond04} only determined one member within 
$r_{500}$ for the NGC 3783 group, no velocity dispersion was measured.
As a result, this group is substantially offset in
Figure~\ref{sigma_compare}. The mean difference between the two
samples is $\langle \Delta \sigma_v\rangle=-8\pm17$ km s$^{-1}$,
i.e. consistent with no difference.

\section{Dynamical Properties of Groups}
\label{dynamics_sect}
The dynamical properties are calculated in the same way as introduced
in \cite{brough05}: For each group defined by the FOF algorithm the
luminosity-weighted centroid and mean recession velocity were
calculated.  These and the dynamical parameters calculated using the
FOF group members are summarised in Table~\ref{fof_groups}.  

%

The velocity dispersion, $\sigma_v$, was calculated using the gapper
algorithm \citep{beers90}:

\begin{equation}
\sigma_v=c \sqrt \frac{\pi}{[n(n-1)]} \sum_{i=1}^{n-1} w_i g_i,
\label{eq:sigma}
\end{equation}
where $w_i=i(n-i)$ and $g_i=z(i+1)-z(i)$.  This method is insensitive
to outliers, giving a robust estimate of $\sigma_v$ for small groups.
The corresponding errors are estimated using the jackknife algorithm.

The crossing time is calculated as a function of the Hubble time
($H_0^{-1}$) following \cite{huchra82}, as:
\begin{equation}
t_c=\frac{3~r_H}{5^{3/2}\sigma_v},
\end{equation}
where the harmonic radius, $r_H$, is independent of the velocity
dispersion and is given below.  
\begin{equation}
r_H=\pi~D~sin~\left[\frac{n(n-1)}{2 \sum_i \sum_{j>i} \theta_{ij}^{-1}}\right],
\end{equation}
where $D$ is the distance to the group given in
Table~\ref{group_data}, $n$ is the number of members of each group,
and $\theta_{ij}^{-1}$ is the angular separation of group members. The
errors on $r_H$ are estimated using the jackknife method and the
errors on $t_c$ are calulcated using standard error propagation
analysis.  \cite{nolthenius87} indicate that a crossing time $>0.09$
H$_0^{-1}$ suggests that a group has not yet had time to virialize.

The virial mass $M_V$ was calculated using the virial mass 
mass estimator of \cite{heisler85}.
\begin{equation}
M_V=\frac{3~\pi~N}{2~G}\frac{\sum_i V_i^2} {\sum_{i<j}1/R_{gc,ij}},
\label{virial_mass}
\end{equation}
where $V_i$ is the observed radial component of the velocity of the
galaxy $i$ with respect to the systemic group velocity and $R_{gc,ij}$
is its projected, group-centric, separation from other group members.
This assumes that the group is virialised.  If the group is not
virialised then the calculated virial mass will be an underestimate of
the true value.  However, \cite{mamon} show that this is not a large
effect.  We estimate the rms error on the virial mass using the
jackknife method.

\begin{figure}
\begin{center}
	\resizebox{20pc}{!}{
	\rotatebox{-90}{
	\includegraphics{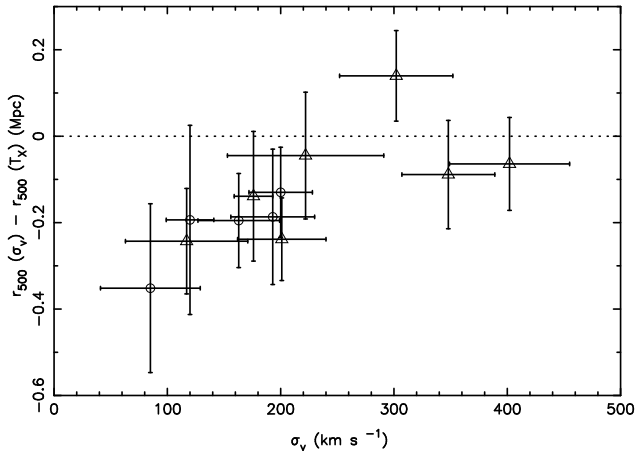} 
		}	 
	}	
  \end{center}
\caption{Comparing the $r_{500}$ radius calculated using the velocity 
dispersion, $r_{500}(\sigma_v)$, and calculated using X-ray
temperature, $r_{500} (T_X)$.  Groups with intra-group X-ray emission
(G-sample) are indicated by the triangles, groups with galaxy-halo
emission only (H-sample) by circles - the groups undetected in X-rays
do not have $T_X$ measurements.  Error bars indicate combined $1\sigma$
errors on the $r_{500}$ measurements and $1\sigma$ errors on
$\sigma_v$.  The dotted line indicates no difference between the two
methods.}
\label{r_500_compare}
\end{figure}

The radius corresponding to an overdensity of 500 times the critical
density -- $r_{500}$, provides a measure of the size of a group.  This
radius is equivalent to $\sim1.5r_{200}$, the virial radius, but is a
more robust measure of size than $r_{200}$ at the low densities of
groups.  We calculate $r_{500}$ as a function of the velocity
dispersion calculated above, following \cite{osmond04} as,
\begin{equation}
r_{500}(\sigma_v)=\frac{0.096\sigma_v}{H_0}.
\end{equation}
The rms error on $r_{500}$ is calculated using standard error
propagation.  It is also possible to calculate $r_{500}$ as a function
of the X-ray temperature, $T_X$, following \cite{evrard96} as,
\begin{equation}
\label{tx_eq}
r_{500}(T_X)=\frac{124}{H_0}\sqrt\frac{T_X}{10}.
\end{equation}
\cite{osmond04} conclude that $r_{500}(T_X)$ is the more robust of the two 
measures.  
The two methods of calculating $r_{500}$ are compared in
Figure~\ref{r_500_compare} for the 12 groups for which \cite{osmond04}
were able to measure $T_X$ values.  The mean difference
$\langle r_{500}(\sigma_v)-r_{500}(T_X)\rangle=-0.14\pm0.04$.  $r_{500}(\sigma_v)$ is
smaller than $r_{500}(T_X)$ for the lower velocity dispersion groups,
becoming more consistent at $\sigma_v>200$ km s$^{-1}$.  
This suggests that $\sigma_v$ may be underestimated for these low
velocity groups, either due to low group membership or to these groups
being less likely to be virialised than the higher velocity dispersion
groups.  However, 
as $T_X$ is only available for 12 of the 15 groups, we will use
$r_{500}(\sigma_v)$.  The conclusions drawn in
Section~\ref{composite_sect} do not change if $r_{500}(T_X)$ is used
instead.

The $K$-band luminosities of the constituent galaxies were summed to
calculate the total $K$-band luminosity of the group, $L_K(Tot)$.
Those galaxies without known $K$ magnitudes were assigned $m_K=13.5$,
this has negligible effect on the total calculated as it is dominated
by the brightest galaxies.  We have also not extrapolated the total
luminosity to infinitely small luminosities, however, \cite{girardi02}
show that this correction is only of the order of 5 per cent.  The
$1\sigma$ error on $L_K(Tot)$ is estimated from the errors on the
individual magnitudes from 2MASS.

The mass-to-light ratios were calculated by dividing the virial mass
by the sum of the $K$-band luminosities of the member galaxies.  We
calculated the rms error on the mass-to-light ratio following standard
error propagation analysis.

We obtained morphological T-types for galaxies from the
Hyper-Lyon-Meudon Extragalactic DAtabase (HyperLEDA;
\citealt{paturel97}).  T-types are numerical codes chosen to correspond to 
morphological galaxy type as defined in the Second Reference Catalogue
(RC2; e.g. \citealt{corwin77}).  The correspondance with Hubble type
is given in more detail in \cite{paturel97}.  In summary, T-types of
$-5\leq$ T-type $\leq0$ correspond to E to S0a galaxies whilst
$0<$ T-type $\leq10$ correspond to Sa to Irr galaxy types.  We limited
these data to those galaxies with magnitudes brighter than the
apparent-magnitude limit of the 2MASS catalogue ($m_K\sim13.1$;
\citealt{jarrett00}) at the distance of our furthest group (NGC 3557;
$M_K\leq-20$) in order to sample the same part of the luminosity
function in each group.  The early-type fraction, $f_{early}$, was
calculated as the fraction of galaxies in each group with $M_K\leq-20$
and T-type $\leq0.0$.  The errors were calculated from the poisson
errors on these values.

The extent of the group is the maximum projected spatial distance
between a group member and the centre of the group.

\section{Individual Group Properties}
\label{indiv_sect}
The area of sky around each group in which we collected data are all
shown in Appendix~\ref{group_pics}, which also illustrates the extent
of the group and hte group members defined by FOF.  In this Section we
briefly discuss the dynamical properties of each group, based on the
values given in Table~\ref{fof_groups}.  It is difficult to make a
statement as to whether the groups are virialized or not.  However, we
do assume that groups are dynamically mature if they have bright X-ray
emission, above-average velocity dispersion, above-average virial
mass, early-type brightest group galaxies (BGGs) and low crossing
times (mean values for our sample are given in
Table~\ref{fof_groups}).






\subsection{NGC 524 Group} 

This group 
is in the Northern hemisphere so does not have 6dFGS data available.
The group exhibits only galaxy-scale X-ray emission but has an
early-type BGG (NGC 524) only $29$ kpc and 22 km s$^{-1}$ from the FOF
centroid, average velocity dispersion, average mass, crossing time,
and a higher than average early-type fraction.  These properties
suggest that this group is dynamically mature.

\subsection{NGC 720 Group}
\label{sect_n720}

\cite{osmond04} find this group to have extended intra-group X-ray 
emission.  The brightest galaxy in the group, NGC 720, is an
early-type galaxy and is spatially coincident with the centre of the
group (within $21$ kpc).  However, it is offset in velocity by 230 km
s$^{-1}$, nearly twice the velocity dispersion.  The other 5 group
members are all $>2$ magnitudes fainter than NGC 720 itself,
suggesting that this could be a fossil group.  The X-ray luminosity of
the group is just below the fossil group definition of
log$_{10}L_X>10^{42}h_{50}^{-2}$ erg s$^{-1}$ \citep{jones03}.  The
group has a low mass, however, it also has a high fraction of
early-type galaxies consistent with the suggestion that this could be
a fossil group.

\subsection{NGC 1052 Group}
\label{sect_n1052}
This group 
shows galaxy-scale X-ray emission associated with the BGG -- NGC 1052.
NGC 1052 is an elliptical galaxy, offset both kinematically and
spatially from the group centre defined by the FOF algorithm.  The
group consists of 29 members, however its velocity dispersion and
virial mass are relatively low and it has a low early-type fraction.
Spatially the group is quite extended and irregular, with NGC 1052
lying in a sub-clump which is offset to the group centroid.  These
properties all suggest that this is a dynamically immature group just
coming together for the first time.

\subsection{NGC 1332 Group}

\cite{brough05} concluded, on the basis of its centrally coincident bright
lenticular galaxy, few members, average velocity dispersion, low
crossing time and high early-type fraction that this is a low-mass,
compact, dynamically mature group with a galaxy population similar to
that of a much more massive group.


\subsection{NGC 1407 Group}

\cite{brough05} concluded on the basis of its high mass-to-light ratio, high 
early-type fraction, symmetric intra-group X-ray emission, bright
central elliptical galaxy and short crossing time that this structure
is a dynamically mature group.  This has since been corroborated by
\cite{trentham06} who came to similar conclusions using deeper
observations.

\subsection{NGC 1566 Group}

This group 
is an overdensity within the very loose grouping of galaxies in the
direction of the Dorado constellation.  This grouping is part of the
Fornax wall that stretches up through the Fornax cluster to the NGC
1407 and NGC 1332 groups.  The group defined by \cite{osmond04} is
centred on NGC 1566 and its galaxy-scale X-ray emission.  \cite{n1566}
found 26 members in the group, selected on the basis of cuts in
velocity and distance from the position and velocity of NGC 1566, and
calculate a velocity dispersion of $282\pm30$ km s$^{-1}$.  The FOF
algorithm does not find NGC 1566 to be a member of a group in this
field.  The BGG is an early-type galaxy -- NGC 1553 and is only $29$
kpc and 37 km s$^{-1}$ from the FOF centroid.  
In contrast to \cite{n1566}, the group defined by the FOF algorithm
only contains 4 members and has a very low velocity dispersion, and
low mass, however, surprisingly it also has a high fraction of
early-type galaxies.  These properties generally suggest that this
group is not dynamically mature.

\subsection{NGC 1808 Group}
\label{sect_n1808}
\cite{osmond04} do not detect X-ray emission $>3\sigma$ background 
and define this group to be undetected in X-rays.  Its late-type BGG
(NGC 1792), few members, low velocity dispersion, low mass and high
crossing time indicate that this group is dynamically immature.

\subsection{NGC 3557 Group}

This group was first found by \cite{klemola69} and confirmed by
\cite{garcia93}. \cite{zabludoff00} describe the group as being 
marginally X-ray detected 
with a lower number density of galaxies than their other, X-ray
detected, groups.  However they also find it to have a large number of
member galaxies (22) and a high velocity dispersion ($282\pm50$ km
s$^{-1}$).  They conclude that this group has a much lower
dwarf-to-giant galaxy ratio than their other, more X-ray luminous,
groups, however it is not possible to verify that here as our data are
not deep enough.  \cite{osmond04} find the group to have intra-group
X-ray emission.  
We find the group to have 14 members and a high velocity dispersion,
consistent with that measured by \cite{zabludoff00}.  The early-type
BGG (NGC 3557) is only $21$ kpc and 171 km s$^{-1}$ from the FOF
centroid.  In addition to the group's high mass, low crossing time,
and high early-type fraction these properties are all consistent with
NGC 3557 being a dynamically mature group.

\subsection{NGC 3783 Group}
\label{sect_n3783}

\cite{osmond04} find this group to have group-scale X-ray emission.  
However, the X-rays only extend to a radius of 69 kpc, and it has a
very low X-ray luminosity.  \cite{n3783} suggest that this group may
be a very young group.  We find its late-type BGG (NGC 3783) $360$ kpc
and 27 km s$^{-1}$ from the FOF centroid. The low velocity dispersion,
mass and low early-type fraction of this group
suggest that this group is dynamically immature, in agreement with
\cite{n3783}.

\subsection{NGC 3923 Group}

This group 
shows galaxy-scale X-ray emission, centred on the BGG, NGC 3923, which
is an early-type galaxy lying close ($85$ kpc and 22 km s$^{-1}$) to
the FOF centroid.  The group consists of many members, and has an
average velocity dispersion and mass, but a low early-type fraction,
suggesting that this group is dynamically immature.

\subsection{NGC 4636 Group}

This group is also called the NGC 4343 group.  
The centre of the group defined by \cite{osmond04} is 3 degrees (0.7
Mpc) south of the centre of the Virgo cluster, however the NGC 4636
group exhibits X-ray luminous intra-group X-ray emission that is
distinct from the emission from Virgo.  The group does not have 6dFGS
data, however, detailed analysis of the region, in particular by SDSS
\citep{abazajian03}, means that it has a much higher number of galaxies 
with velocities than do other groups studied, but few with
$K$-magnitudes.  Using the same FOF technique as used for the other
groups therefore fails in this region.  The extra depth of sources
results in confusion between this group and the Virgo cluster and a
FOF analysis with all galaxies determines a group that is
indistinguishable from Virgo.

For this region we therefore cut the data by the apparent magnitude
limit of 2MASS ($K\leq13.1$ mag) to perform the FOF analysis.  This
defines a group that is distinct from the Virgo cluster, as suggested
by its X-ray emission.  The early-type galaxy NGC 4636 is the
brightest galaxy in the group, close to the FOF centroid (within $87$
kpc and 38 km s$^{-1}$) and, given the apparent-magnitude cut, has a
large number of members, average velocity dispersion and high mass.
On balance, these properties suggest that this group is dynamically
mature.

\subsection{NGC 5044 Group}

\cite{osmond04} show this group to have group-scale X-ray emission and 
it is the most X-ray luminous group in our sample -- putting it on the
border of cluster-mass scales.  Its large size means that it has been
the subject of much study.  Most recently, \cite{cellone05} have
analysed the group in detail, finding a group of 27 members with a
velocity dispersion $\sigma_v=431$ km s$^{-1}$.

The large elliptical, NGC 5044, is the brightest galaxy in the group
and is $91$ kpc and 156 km s$^{-1}$ from the FOF centroid.  Our
velocity dispersion is consistent with that of \cite{cellone05}.  This
is a classically dynamically mature group with many
members, high velocity dispersion, high mass, high early-type fraction and
a low crossing time.

\subsection{HCG 90}

This group was first found by \cite{klemola69} and then defined as a
Hickson Compact Group (HCG) by \cite{hickson82}.  \cite{maia89}
established that this group extends well beyond the compact nucleus
examined by \cite{hickson82}.  \cite{zabludoff98} found that this is
in fact a massive group consisting of 16 members with a high velocity
dispersion ($293\pm36$ km s$^{-1}$) and extended X-rays.
The three central galaxies (NGC 7173, NGC 7174, and NGC 7176) appear
to be interacting, although they are dominated in luminosity by the
elliptical NGC 7176 which appears to be undisturbed.
\cite{longo94} illustrate that NGC 7176 and NGC 7174 are not actually
interacting and that this is in fact a projection effect.  We find the
elliptical galaxy NGC 7176 to be the BGG.  It lies only $33$ kpc and
84 km s$^{-1}$ from the FOF centroid.  We find the group to consist of
many members, and it has a low crossing time and an average velocity
dispersion, however it has a low early-type fraction suggesting that
it might still be dynamically immature.

\subsection{IC 1459 Group}

\cite{osmond04} find this group to have extended intra-group X-ray 
emission.  We find the group to have an early-type BGG -- IC 1459 is
$32$ kpc and 153 km s$^{-1}$ from the FOF centroid.  The group has few
members and a high velocity dispersion, high virial mass, high
mass-to-light ratio but a lower than average early-type fraction.
These properties generally suggest that this group is a dynamically
mature group structure.

\subsection{NGC 7714 Group}

\cite{osmond04} find the X-ray emission of this group to be $<3\sigma$ 
background, it is therefore, undetected in X-rays.  This group is in
the Northern hemisphere and does not therefore have 6dF data.  The
group defined by FOF does not contain NGC 7714.  
The BGG (NGC 7716) is a late-type galaxy, only $10$ kpc and 26 km
s$^{-1}$ from the FOF centroid.  The group has very few members (4,
and only 1 with a measured $K$-magnitude), very low velocity
dispersion, very low mass, and no early-type galaxies.  This all
suggests that this group is not dynamically mature.

\subsection{Summary}

The properties of these groups depend strongly on
their X-ray luminosities.  The groups with the highest X-ray
luminosities have the highest masses and early-type BGGs that lie
close to the FOF defined centroid (Section~\ref{section_bggs}). They
also have the most members, highest velocity dispersions, and low
crossing times and high early-type fractions.  Although these groups
generally have extended intra-group X-ray emission, there are some
showing only galaxy halo X-ray emission.


In contrast, those groups with the lowest and undetected X-ray
luminosities have low masses and velocity dispersions and they
generally have fewer members than those groups with the highest X-ray
luminosities.  The properties of the NGC 3783 group are consistent
with these groups, despite its marginally extended X-ray emission.

These differences suggest that the X-ray luminosity of these groups is
more closely related to their dynamical properties than whether or not
the groups show extended or halo X-ray emission, as defined by
\cite{osmond04}.

\section{Statistics}

In the relationships presented below we calculated Kendall's rank
correlation probabilities for each group parameter pair using the
IRAF/STSDAS/STATISTICS package routine.  Kendall's rank correlation
was used because it is more reliable for samples where $N<30$ than the
Spearman rank correlation.  It is also non-parametric --
\cite{helsdon00} show that the scatter in the properties of groups is
intrinsic rather than statistical so it is appropriate not to take the
statistical errors into account in the correlation.  

Two of the groups only have upper limits for their X-ray luminosities.
The survival analysis tasks available in IRAF can take this into
account in the correlation.  Survival analysis tasks are discussed in
more detail in \cite{osullivan01}.  In summary, survival analysis
assumes that the upper limits provided hold only limited information
regarding the true values of the X-ray luminosities of these groups.
We take the uncensored parameter as the independent variable and the
censored parameter (i.e. X-ray luminosity) as the dependent parameter.

We also use the Buckley-James algorithm available in the same package
to fit straight lines to our data.  The Buckley-James method is also
non-parametric, using the Kaplan-Meier estimator for the residuals to
calculate the regression, and can also take censored data into
account.  As for the correlation, we take the uncensored parameter as
the independent variable and the censored parameter (i.e. X-ray
luminosity) as the dependent parameter.

\section{Scaling Relations}
\label{scaling_sect}
Diffuse X-rays are emitted by a hot plasma trapped in the
gravitational potential of a galaxy, group or cluster.  The X-ray
luminosities of groups of galaxies are therefore a measure of the hot
gas in the system and it is interesting to determine how the gas
properties correlate with the properties of the galaxies that share
the same potential well.

Assuming that the evolution of groups and clusters is solely due to
their collapse under gravity, groups are simply scaled down clusters.
This `self-similar' model predicts that X-ray luminosity is
proportional to the velocity dispersion of member galaxies:
$L_{X}(Bol)\propto \sigma_v^4$ (e.g. \citealt{quintana82}).  Previous
research on clusters (e.g. \citealt{mahdavi01,hilton05}) has resulted
in $L_{X}(Bol)\propto \sigma_v^{4.8\pm0.7}$ \citep{hilton05}.
Although this does not rule out simple, self-similar evolution, it is
also consistent with feedback processes (i.e. heating by
non-gravitational processes) being significant.

\begin{figure}
\begin{center}
    \resizebox{20pc}{!}{ \rotatebox{-90}{
    \includegraphics{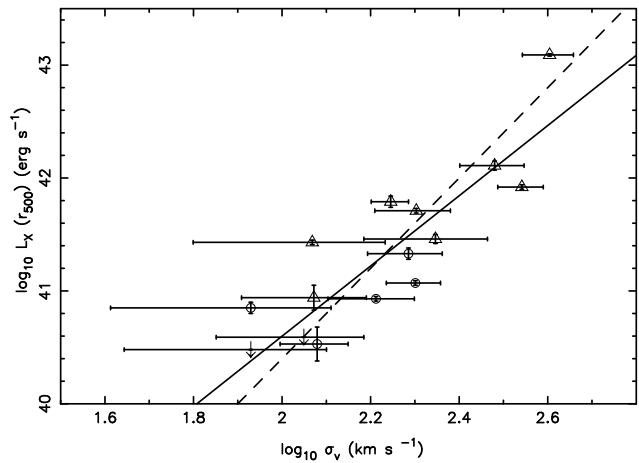} } }
    \end{center}
\caption{The relationship between $L_X(r_{500})$ and $\sigma_v$.  Groups 
with intra-group X-ray emission (G-sample) are indicated by the
triangles, groups with galaxy-halo emission only (H-sample) by circles
and groups undetected in X-rays (U-sample) by their upper-limit.
Error bars indicate $1\sigma$ errors.  The dashed line indicates the
self-similar prediction whilst the solid line indicates the regression
line fit given by Equation~\ref{eq:lx_sigma}.}
\label{sigma_lx}
\end{figure}

One of the main observational signatures of non-gravitational
processes is a steepening of the X-ray relationships in the group
regime (e.g. \citealt{borgani04}).  Hence, it is also important to
examine the relationships followed by groups.  Previous analysis of
groups reveals that the $L_X-\sigma_v$ relation does not steepen in
group-sized systems, with a growing consensus that groups are
consistent with the cluster relationship
(\citealt{mulchaey98,helsdon00,mahdavi01} and T. Ponman, private
communication 2005). We observe a 99.96 per cent correlation between
these properties (Figure~\ref{sigma_lx}) and, using the Buckley-James
algorithm described above, find a relationship for all 15 groups of
\begin{equation}
{\rm log_{10}}L_X(r_{500})=3.11^{\pm0.59}{\rm log_{10}}\sigma_v +34.38,
\label{eq:lx_sigma}
\end{equation}
with a standard deviation on the regression of 0.40.  
This is close to the prediction of self-similarity.  However, 
\cite{osmond04} discuss the biases in their $L_X$ calculation which
lead to a flattening of this relationship.  As we use their X-ray
luminosities these biases are also present in this sample, leading to
uncertainty in the true value of this relationship.

\begin{figure}
\begin{center}

    \resizebox{20pc}{!}{
	\rotatebox{-90}{
    	\includegraphics{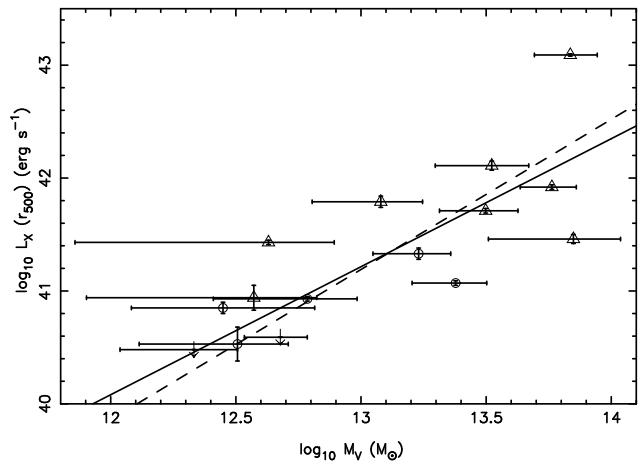} 
	}	 
	}
  \end{center}
\caption{The relationship between the X-ray luminosity $L_X(r_{500})$ and 
the virial mass, $M_V$.  G-sample galaxies are indicated by triangles,
H-sample by circles and U-sample groups by upper limits.  Error bars
indicate $1\sigma$ errors.  The dashed line indicates the self-similar
prediction whilst the solid line indicates the regression line fit
given by Equation~\ref{eq:lx_mv}.
}
\label{lx_mv}
\end{figure}

Self-similarity predicts that the X-ray emitting gas is bound to
groups with a mass proportional to the virial mass -- $M_V\propto
T_X^{3/2}$ \citep{borgani04}.  $L_X\propto T_X^2$, therefore
$L_X\propto M_V^{4/3}$.  Figure~\ref{lx_mv} illustrates the close
relationship we observe between these parameters -- correlated at the
99.91 per cent level.  Using the Buckley-James algorithm described
above we find a relationship for all 15 groups of
\begin{equation}
{\rm log_{10}}L_X(r_{500})=1.13^{\pm0.27} {\rm log_{10}}M_V +26.48,
\label{eq:lx_mv}
\end{equation}
with a standard deviation on the regression of 0.48.  
This is consistent with the prediction of self-similarity.

Early-type fractions in clusters are observed to correlate strongly
with $L_X$ \citep{edge91}.  In groups, \cite{zabludoff98} found a
strong correlation of early-type fraction with velocity dispersion.
They also found {\it no} early-type galaxies in the 3 groups without
observed X-ray emission and early-type fractions up to those observed
in clusters ($f_{early}\sim0.55-0.65$; \citealt{whitmore93}) in their
9 groups with observed X-ray emission ($0.25<f_{early}<0.55$).  We
also find three of our groups (NGC 1808, NGC 3783 and NGC 7714) to
have no early-type galaxies.  There is no X-ray emission associated
with the NGC 1808 or NGC 7714 groups, similar to the
\cite{zabludoff98} groups with no early-type galaxies, however the NGC
3783 group shows marginally extended X-ray emission, characteristic of
intra-group gas.  This group is shown in Section~\ref{sect_n3783} and
by \cite{n3783} to be a young group in the process of forming,
consistent with this low early-type fraction.

\begin{figure}
\begin{center}

    \resizebox{20pc}{!}{
	\rotatebox{-90}{
    	\includegraphics{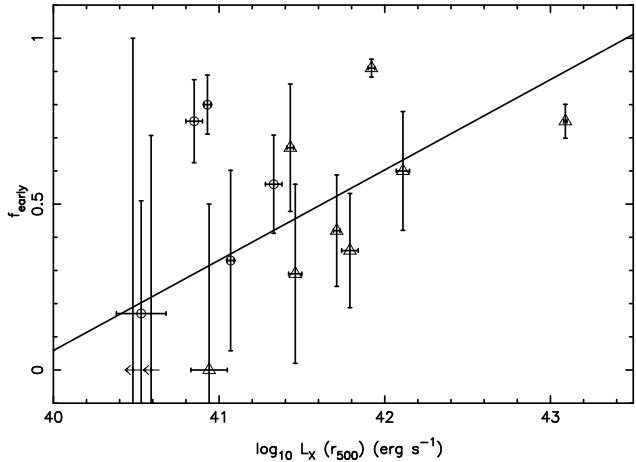} 
	}	 
	}
  \end{center}
\caption{The fraction of early-type galaxies, $f_{early}$, in each group 
with respect to the group X-ray luminosity $L_X(r_{500})$.  G-sample
galaxies are indicated by triangles, H-sample by open circles and
U-sample groups by upper limits.  Error bars indicate $1\sigma$ errors
on the X-ray luminosities and poisson errors on the early-type
fractions.  The solid line indicates the regression line fit given by
Equation~\ref{eq:lx_fsp}.}
\label{lx_fsp}
\end{figure}

\begin{figure}
\begin{center}

    \resizebox{20pc}{!}
	{ \rotatebox{-90}{
    \includegraphics{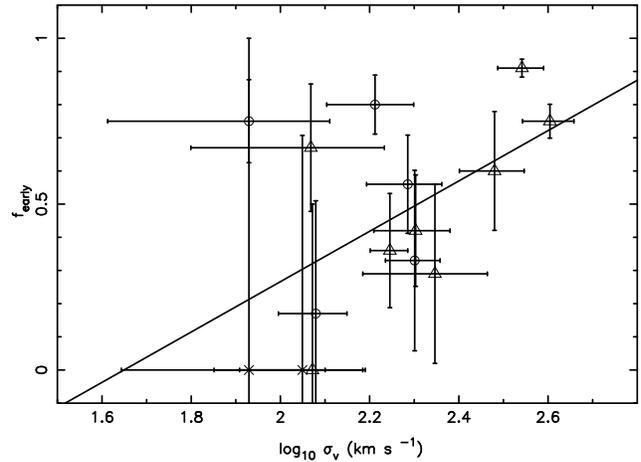} 
} 
}
    \end{center}
\caption{The relationship between the fraction of early-types, 
$f_{early}$, in each group and the velocity dispersion, $\sigma_v$.
G-sample galaxies are indicated by triangles, H-sample by open circles
and U-sample groups by crosses.  Error bars indicate $1\sigma$ errors
on the velocity dispersions and poisson errors on the early-type
fractions.  The solid line indicates the regression line fit given by
Equation~\ref{eq:sig_fsp}.}
\label{sigma_fsp}
\end{figure}

\cite{osmond04} and \cite{wilman05} find weaker trends with early-type 
fraction. 
We find the early-type fraction to be correlated with $L_X$ (95.87 per
cent correlation, Figure~\ref{lx_fsp}) and $\sigma_v$ (90.98 per cent
correlation; Figure~\ref{sigma_fsp}).  Fitting to these data using the
Buckley-James algorithm we find:
\begin{equation}
f_{early}=0.27^{\pm0.11} {\rm log_{10}}L_X(r_{500}) -10.8,
\label{eq:lx_fsp}
\end{equation}
and,
\begin{equation}
f_{early}=0.76^{\pm0.35} {\rm log_{10}}\sigma_v -1.2.
\label{eq:sig_fsp}
\end{equation}
These relationships are illustrated in Figures~\ref{lx_fsp}
and~\ref{sigma_fsp}.

In Section~\ref{indiv_sect} we concluded that the X-ray luminosity of
the groups provides more information on the optical properties of
these groups than the extent of their X-ray emission.  This conclusion
is supported here, with X-ray luminosity highly correlated with
velocity dispersion, virial mass, and the morphologies of the group
members.

\subsection{Mass-to-Light Ratio}

The ratio of mass-to-light in a system represents a proxy for the
efficiency with which stars are formed within haloes of different
masses.  Previous studies have used group dynamical masses
(e.g. \citealt{carlberg01,girardi02,eke04b}), comparison of the
theoretical mass function with the observed luminosity function
\citep{marinoni02}, mass measurements from weak lensing
(e.g. \citealt{parker05}) and the fundamental plane \citep{zaritsky06}
to examine mass-to-light ratios.  There is a growing consensus that
mass-to-light ratios reach some minimum at a system mass of
$\sim10^{12}M_{\odot}$ (e.g. \citealt{marinoni02,eke04b,zaritsky06})
and rise steadily through group-sized systems
(\citealt{carlberg01,girardi02,marinoni02,eke04b,parker05,zaritsky06}).
However there is some disagreement on cluster mass scales with
\cite{sanderson03} and \cite{eke04b} observing no relationship of
mass-to-light ratio with cluster mass whilst \cite{girardi02} and
\cite{zaritsky06} do.

\begin{figure}
\begin{center}

    \resizebox{20pc}{!}{
	\rotatebox{-90}{
    	\includegraphics{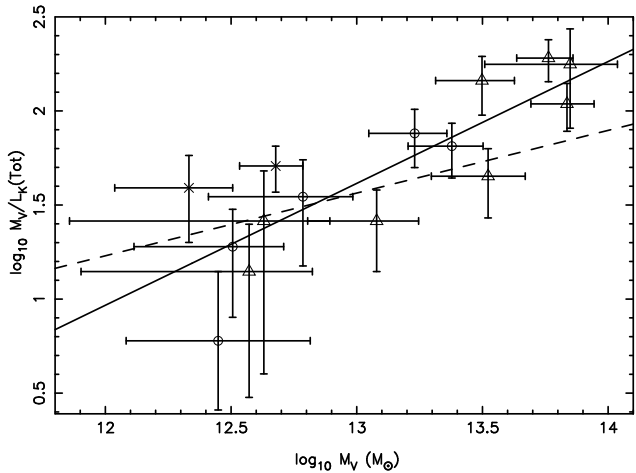}
	}	 
	}
  \end{center}
\caption{The relationship between the mass-to-light ratio and the 
virial mass of the group.  G-sample galaxies are indicated by
triangles, H-sample by circles and U-sample groups by crosses.  The
error bars give the $1\sigma$ errors.  The solid line indicates the
regression line fit to all 15 groups given by Equation~\ref{eq:ml_mv},
whilst the dashed line is the relationship of Marinoni et al. (2002).}
\label{m2l_mv}
\end{figure}

Using the virial masses and total K-band luminosities calculated for
our groups, we show the relationship of the mass-to-light ratio with
mass in Figure~\ref{m2l_mv}.  The mass-to-light ratio clearly
increases with system mass.  Using the Buckley-James algorithm we
find a relationship of
\begin{equation}
{\rm log_{10}}(M_V/L_K)=0.65^{\pm0.12} {\rm log_{10}}M_V -6.8,
\label{eq:ml_mv}
\end{equation}
This is similar to the relationship determined by \cite{eke04b} and
\cite{parker05} but steeper than the $M/L_B\propto M^{0.33}$
measured by \cite{marinoni02}.

\begin{figure}
\begin{center}

    \resizebox{20pc}{!}{
	\rotatebox{-90}{
    	\includegraphics{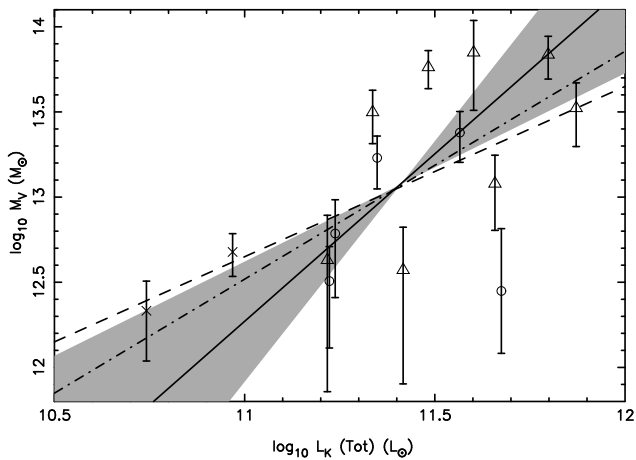}
	}	 
	}
  \end{center}
\caption{The relationship between the virial mass of the group and its total 
luminosity.  G-sample galaxies are indicated by triangles, H-sample by
circles and U-sample groups by crosses.  The error bars give $1\sigma$
errors on $M_V$, the $1\sigma$ errors on the total $K$-band
luminosities are of the order of the size of the points and are not
therefore plotted.  The dashed line indicates the simple $M_V\propto
L_K$ line, the dot-dashed line the relationship of Girardi et
al. (2002), whilst the shaded areas indicate the direct and inverse
fits and the solid line gives the bisecting fit
(Equation~\ref{eq:lk_mv}).}
\label{lk_mv}
\end{figure}

However, \cite{girardi02} and \cite{eke04b} indicate the pitfalls of
working with correlated quantities.  We therefore follow
\cite{girardi02} and examine the more robust relationship between mass
and light, presented in Figure~\ref{lk_mv}.  The scatter in this
relationship is evident.  We therefore fit both $M_V\propto L_K$ and
the inverse, $L_K\propto M_V$, and use the difference between these
fits as an estimate of the error on the fit.  Using the Buckley-James
algorithm we find a bisecting relationship of,
\begin{equation}
{\rm log_{10}}M_V=2.0^{\pm0.9} {\rm log_{10}}L_K (Tot)-9.4.
\label{eq:lk_mv}
\end{equation}
In Figure~\ref{lk_mv} we also indicate the $M_V\propto L_K$
relationship and the \cite{girardi02} fit to their data:
$M_V\propto L_B^{1.338\pm0.033}$.  The slope fitted here is consistent
with that of \cite{girardi02} and with mass increasing faster than
luminosity.

\section{Brightest Group Galaxies}
\label{section_bggs}
Hierarchical structure formation predicts that the galaxy at the
centre of a dark-matter halo will continue growing as it accretes
other galaxies and that it will grow at the expense of other galaxies.
It is, therefore, expected to be the brightest, most massive galaxy in
the halo at all times.  In this paradigm the bright elliptical
galaxies found at the centres of clusters (Brightest Cluster Galaxies;
BCGs; e.g. \citealt{brough02,bcg05}) form in the group environment and
as the groups merge to form clusters, newly accreted massive galaxies
will sink to the centre of the potential well by dynamical friction
and merge with the central galaxy.  Groups with extended X-ray
emission are frequently observed with bright, early-type galaxies at
the centre of their X-ray emission, similar to the BCGs observed at
the centres of clusters \citep{zabludoff98,osmond04}.  If the
hierarchical picture of structure formation is correct then these
brightest group galaxies (BGGs) can be seen as evolutionary tracers of
the system \citep{zabludoff98,brough02,lin04}.

Here, the BGGs were selected as the galaxy with the brightest
$K$-magnitude within $\pm2\sigma_v$ of the mean group velocity.  The
BGGs are detailed in Table~\ref{bggs}.  Of 8 groups with extended
X-ray emission in this sample, 7 have early-type BGGs.  The exception
is the NGC 3783 group, which is shown in Section~\ref{sect_n3783} and
by \cite{n3783} to have properties which are consistent with this
group still being in the process of relaxation.

Comparing the magnitude distribution of early- and late-type BGGs we
found that the early-type BGGs are more luminous than the late-type
BGGs, as also seen by \cite{osmond04}.  A Kolmogorov-Smirnov (KS) test
gives a probability of only 3.0 per cent that the early and late-type
BGGs are drawn from the same parent population.

\begin{table*}
\begin{center}
\caption{Properties of Brightest Group Galaxies (BGGs).}
\label{bggs}
\begin{tabular}{lccccccccc}
\hline
Group&BGG&$M_K$&$L_{K}(BGG)$&$|R-<R>|$&$\frac{|R-<R>|}{r_{500}}$&$v-{\bar{v}}$&$\frac{v-{\bar{v}}}{\sigma_v}$&$M_{K,1}-M_{K,2}$&Type\\
&&(mag)&$(10^{11}L_{\odot})$&(kpc)&&(km s$^{-1}$)&&(mag)&\\
\hline
NGC 524 & NGC 0524&-24.53$\pm$0.01&  1.33$\pm$0.02& 29&0.11& -22&-0.11&1.73$\pm$0.03&E\\
NGC 720 & NGC 0720&-24.66$\pm$0.02&  1.50$\pm$0.02& 21&0.13& 230& 1.97&2.95$\pm$0.04&E\\
NGC 1052& NGC 1052&-23.78$\pm$0.01&  0.67$\pm$0.01&144&0.87& 185& 1.54&0.51$\pm$0.02&E\\
NGC 1332& NGC 1332&-24.48$\pm$0.02&  1.27$\pm$0.02& 42&0.19&  50& 0.31&1.70$\pm$0.03&E\\
NGC 1407& NGC 1407&-24.77$\pm$0.02&  1.66$\pm$0.03& 95&0.20&  92& 0.26&1.43$\pm$0.03&E\\
NGC 1566& NGC 1553&-25.26$\pm$0.02&  2.61$\pm$0.04& 29&0.25& -37&-0.44&0.53$\pm$0.02&E\\
NGC 1808& NGC 1792&-24.07$\pm$0.02&  0.87$\pm$0.01& 25&0.16& 149& 1.33&3.25$\pm$0.07&L\\
NGC 3557& NGC 3557&-25.90$\pm$0.02&  4.70$\pm$0.07& 21&0.05& 171& 0.57&1.73$\pm$0.02&E\\
NGC 3783& NGC 3783&-24.12$\pm$0.02&  0.91$\pm$0.02&360&2.24&  -9&-0.08&0.07$\pm$0.02&L\\
NGC 3923& NGC 3923&-25.02$\pm$0.02&  2.09$\pm$0.03& 85&0.31& -22&-0.11&1.12$\pm$0.02&E\\
NGC 4636& NGC 4636&-24.05$\pm$0.02&  0.86$\pm$0.01& 87&0.32&  38& 0.19&0.84$\pm$0.03&E\\
NGC 5044& NGC 5044&-24.49$\pm$0.02&  1.28$\pm$0.02& 91&0.16& 156& 0.39&0.76$\pm$0.03&E\\
HCG 90  & NGC 7176&-24.70$\pm$0.02&  1.56$\pm$0.03& 33&0.14& -84&-0.48&0.30$\pm$0.03&E\\
IC 1459 & IC 1459 &-25.24$\pm$0.02&  2.56$\pm$0.04& 28&0.09&-107&-0.48&1.98$\pm$0.03&E\\
NGC 7714& NGC 7716&-23.51$\pm$0.03&  0.52$\pm$0.01& 10&0.09& -26&-0.31&4.04$\pm$0.10&L\\
\hline Mean Values &&$-24.57\pm0.16$&3.19$\pm$0.50&&&&&$1.53\pm0.30$&\\

\hline
\end{tabular} 
\flushleft
The columns indicate (1) Group name; (2) BGG name; (3) Absolute
$K$-band magnitude of BGG with $1\sigma$ error; (4) Luminosity of BGG
with $1\sigma$ error; (5) Offset of BGG with respect to
luminosity-weighted group centroid; (6) Offset of BGG with respect to
luminosity-weighted group centroid scaled by group $r_{500}$ radius;
(7) Offset of BGG velocity with respect to mean group velocity; (8)
Offset of BGG velocity with respect to mean group velocity, scaled by
group velocity dispersion; (9) Magnitude difference between first and
second ranked galaxies with $1\sigma$ error; (10) Morphology of BGG: E
for early-type (T-type $\leq0.0$) and L (T-type $>0.0$) for late-type
galaxies.  The final row gives the means and error on the mean
($\sigma/\sqrt{N}$) of these quantities.
\end{center}
\end{table*}


The hierarchical structure formation paradigm predicts that the
brightest galaxy in a halo will lie at rest with respect to the
potential well.  It is therefore interesting to examine whether BGGs
lie at rest with respect to the spatial and velocity centroids of
their groups.  Previous studies have found that 90 per cent of BCGs
lie within $0.38r_{200}$ of their cluster centroid \citep{lin04}.
\cite{zabludoff90} and \cite{oegerle01} find significant departures in
the velocities of BCGs from the mean cluster velocity but within the mean
velocity dispersion of their clusters.  

In the group environment, \cite{mulchaey98} found that the BGGs of
their X-ray detected groups lie within $5-10h^{-1}$ kpc of the X-ray
peak of their groups.

\begin{figure}
\begin{center}

    \resizebox{20pc}{!}{
	\rotatebox{-90}{
    	\includegraphics{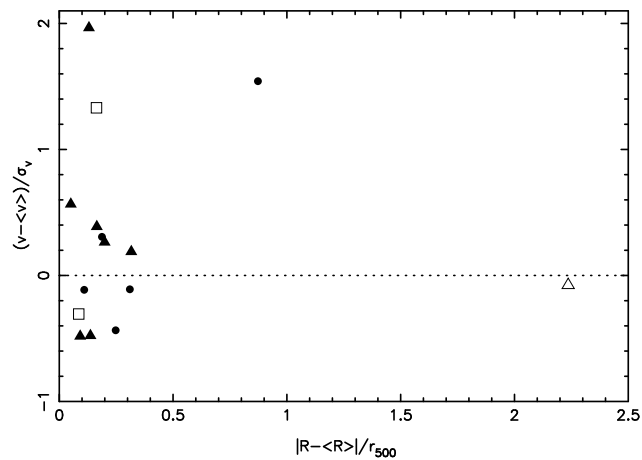} 
	}	 
	}
  \end{center}
\caption{The relationship between the offset of the BGG in position and 
velocity with respect to the luminosity-weighted centroid and the mean
velocity determined by FOF.  The symbols indicate early-type (closed
points) and late-type (open point) BGGs, with those in G-sample groups
as triangles, in H-sample groups as circles and those in U-sample groups
are marked by squares.}
\label{fig_offset_voffset}
\end{figure}

We examine the offset of the BGG from the group luminosity-weighted
centroid and mean group velocity, scaled by the size and velocity
dispersion of the group. (Figure~\ref{fig_offset_voffset}).  There are
four clear outlying BGGs in Figure~\ref{fig_offset_voffset}.  These
BGGs are all in groups showing signs of being dynamically immature:
NGC 3783 is a late-type galaxy lying close to the FOF centroid in
velocity ($0.1\sigma_v$), but it is offset by $2.2r_{500}$
spatially. Despite this it is still the closest galaxy to the FOF
centroid in this group.  This group was determined by
\cite{n3783} and in Section~\ref{sect_n3783} to still be in the
process of forming.  NGC 1792 is also a late-type galaxy and is offset
by $1.3\sigma_v$ in velocity with respect to its group centroid.  This
group (NGC 1808) was determined in Section~\ref{sect_n1808} to be
dynamically immature. NGC 1052 is an early-type galaxy but it lies at
both significant radius ($0.9r_{500}$) and velocity ($1.6\sigma_v$)
from the FOF centroid.  This group only has X-rays associated with NGC
1052 itself and the group was shown in Section~\ref{sect_n1052} to be
relaxing for the first time.  Spatially, NGC 720 lies close to the FOF
centroid, but is offset by nearly $2\sigma_v$ in velocity.  If this
group is a fossil group then it should have reached dynamical
equilibrium.  The observation that the BGG is offset in velocity by
$2\sigma_v$ therefore suggests that this might not be a fossil group.

All BGGs in dynamically mature groups lie within a group-centric
radius of $\sim0.35r_{500}$ and $\pm0.6\sigma_v$ in velocity of their
FOF centroid.  These groups all have high X-ray luminosities,
consistent with the sample of \cite{mulchaey98}.  We conclude that it
is only safe to assume that the brightest galaxy in the group lies at
rest with respect to its potential well once the group has come into
equilibrium.

\begin{figure}
\begin{center}

    \resizebox{20pc}{!}{
	\rotatebox{-90}{
    	\includegraphics{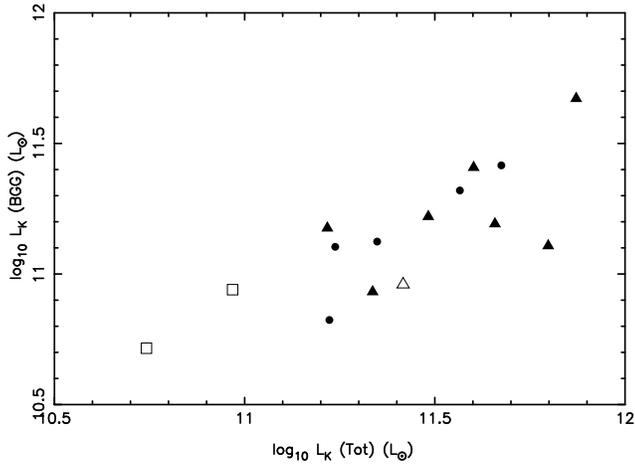} 
	}	 
	}
  \end{center}
\caption{The relationship between the $K$-band luminosity of the BGG, 
$L_{K}(BGG)$, and the total $K$-band luminosity of the group,
$L_{K}(tot)$. 
The symbols indicate early-type (closed points) and late-type (open
point) BGGs, with those in G-sample groups as triangles, in H-sample
groups as circles and those in U-sample groups are marked by squares.
The $1\sigma$ errors on the BGG and total $K$-band luminosities are of
the order of the size of the points and are not therefore plotted.}
\label{fig_lbgg_ltot}
\end{figure}

\begin{figure}
\begin{center}

    \resizebox{20pc}{!}{
	\rotatebox{-90}{
    	\includegraphics{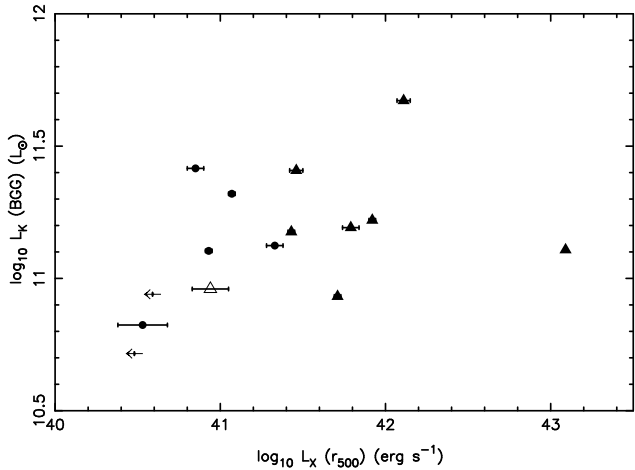} 
	}	 
	}
  \end{center}
\caption{The relationship between the $K$-band luminosity of the BGG, 
$L_{K}(BGG)$, and the X-ray luminosity, $L_X(r_{500})$ of its host
group.  The symbols indicate early-type (closed points) and late-type
(open point) BGGs, with those in G-sample groups as triangles, in
H-sample groups as circles and those in U-sample groups (both
late-types) are marked by upper-limits.  The error bars indicate
$1\sigma$ errors on the X-ray luminosities.  The $1\sigma$ errors on
the $K$-band luminosity of the BGG are of the order of the size of the
points and are not therefore plotted.}
\label{fig_lbgg_lx}
\end{figure}

\cite{brough02} found that the $K$-band aperture magnitudes of BCGs 
at low redshifts ($z<0.1$) are only weakly correlated with the X-ray
luminosity of their host cluster.  
In contrast, \cite{osmond04} find a $6\sigma$ correlation between BGG
luminosity and $L_{B}(Tot)$ and a $2.5\sigma$ correlation with group
X-ray luminosity.  Here we find a 99.87 per cent correlation with
total $K$-band luminosity (Figure~\ref{fig_lbgg_ltot}) and a 95.82 per
cent correlation with X-ray luminosity (Figure~\ref{fig_lbgg_lx}).
The fact that the BGG luminosity is more correlated with total group
luminosity reflects the dependence of $L_{K}(Tot)$ on $L_{K}(BGG)$.
Therefore, we can state that, to first order, there is a decrease in
the dependence of the luminosity of the central galaxy on its host
halo mas between groups and clusters.  This has been predicted
analytically by \cite{cooray05} who show that above a critical halo
mass of $\sim1-6\times10^{13}M_{\odot}$, the timescale on which
dynamical friction induces orbital decay in the accreted galaxies
exceeds the age of the dark-matter halo.  As a result of this the
relationship between central galaxy luminosity and host halo mass
turns over at this critical mass, albeit with some scatter.  The
groups studied here lie close to this critical mass range and lie
within the scatter.  Therefore, in a future paper we will compare the
properties of all the GEMS BGGs with those of the BCGs of
\cite{brough02} in order to explore a significantly wider halo mass
range and fully investigate this prediction.



\begin{figure}
\begin{center}

    \resizebox{20pc}{!}{
	\rotatebox{-90}{
    	\includegraphics{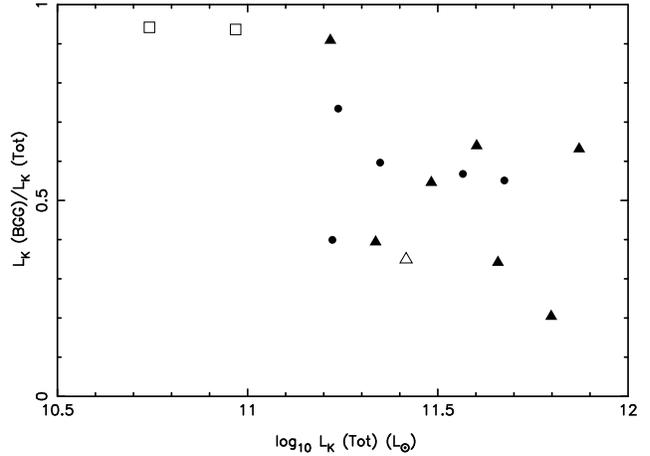} 
	}	 
	}
  \end{center}
\caption{The relationship between the ratio of the K-band luminosity of 
the BGG, $L_{K}(BGG)$, and the total K-band luminosity of the group,
$L_{K}(Tot)$ and $L_{K}(Tot)$. The symbols indicate early-type (closed
points) and late-type (open point) BGGs, with those in G-sample groups
as triangles, in H-sample groups as circles and those in U-sample groups
are marked by squares.  The $1\sigma$ errors on the BGG and total
K-band luminosities are of the order of the size of the points and are
not plotted.}
\label{fig_lratio_ltot}
\end{figure}

A result of the prediction by \cite{cooray05} is that the fraction of
total group light in the BGG will decrease with increasing mass of the
group.  \cite{lin04} observe this relationship on cluster mass scales,
with BCGs constituting 40--50 per cent of total light at poor-cluster
mass scales, down to 5 per cent at rich-cluster mass scales.  We
observe a continuation of that effect here to very poor groups where
the brightest galaxy makes up $>90$ per cent of the total group
luminosity (Figure~\ref{fig_lratio_ltot}). 

In the hierarchical structure formation paradigm, the total luminosity
of the group is expected to rise as the group accretes galaxies, such
that the correlations between group mass, total $K$-band luminosity
and X-ray luminosity are retained.  The correlation of the luminosity
of the BGG with the total luminosity of the group
(Figure~\ref{fig_lbgg_ltot}) then means that the BGGs themselves must
have also accreted galaxies.  However, the dynamical time required for
the accreted galaxies to fall in and merge with the central galaxy
increases with increasing group mass.  Above the critical mass
predicted by \cite{cooray05} the accreted galaxies have not had time
to merge with the central galaxy, such that the fraction of group
light in the BGG falls with increasing total group luminosity.

The accretion of galaxies by the group, that are not cannibalized by
the BGG, will also affect the dominance of the BGG over the second
brightest galaxy in the group.  We only observe a weak correlation
with total light of galaxies in the system (90.79 per cent),
consistent with \cite{osmond04}, \cite{lin04} and
\cite{miles04}.  However, the mean offset $\langle
M_{K,1}-M_{K,2}\rangle=1.53\pm0.30$ mag, is significantly higher than
\cite{lin04} who find $\langle M_{K,1}-M_{K,2}\rangle=0.66\pm0.48$ mag
in their sample of 93 BCGs.  The increase in domination of the BGG
into the group environment
is consistent with \cite{loh05} who find the dominance of BCGs to be
more prominent in group-like environments than in cluster-like
environments.  
The observation that BGGs are more dominant in the group environment
than in the cluster environment adds evidence to the merging picture
in which larger groups have accreted more galaxies but the BGG has yet
to absorb the accreted members.  

\subsection{Summary}

The BGGs of all the dynamically mature groups are early-type galaxies
and lie within a group-centric radius of $0.3r_{500}$ and
$\pm0.6\sigma_v$ in velocity of the FOF centroid.  They are
significantly brighter than the late-type BGGs.

The luminosity of the BGGs increases with increasing total $K$-band
luminosity and X-ray luminosity of the group. However, the fraction of
group light in the BGG and the dominance of the BGG over the second
brightest galaxy fall with increasing total group luminosity.  These
properties are all consistent with the paradigm in which BGGs grow by
mergers at early times in group evolution while the group continues to
grow by accreting infalling galaxies.

\section{A Composite Group}
\label{composite_sect}
The numbers of galaxies in each of our groups are few.  Therefore, in
order to study the group-wide properties statistically it is necessary
to stack the galaxies in each group to construct a composite group.
To sample the same portion of the luminosity function of each group we
cut the sample by absolute magnitude, based on the apparent-magnitude
limit of 2MASS (i.e. $m_K<13.1$; \citealt{jarrett00}) at the distance
of the furthest group (NGC 3557), $M_K\leq-20$ mag.  This creates a
composite group with 113 galaxies.

Owing to the size and number of regions studied it was unfeasible to
obtain new photometric data.  We therefore used HyperLEDA to obtain
total $B$-band magnitudes and morphological T-types for the absolute
magnitude-limited sample.  This resulted in 112 galaxies with both
$B$-band magnitudes and T-types.

We also used the dataset to construct a field sample.  Galaxies within
$\pm2\sigma_v$ of the group defined by FOF, which were not defined to
be members of that group {\it or} other groups in the same field, were
selected to be field galaxies.  The velocity range was chosen such
that galaxies in the same velocity range of the group would be at the
same distance so that accurate absolute magnitudes could be
calculated.  Above the absolute magnitude limit this gives 161 field
galaxies, of which 157 have both $B$-band magnitudes and T-types.

\subsection{Velocity Dispersion Profile}

The line-of-sight velocity dispersion as a function of the projected
group-centric distance provides information on the velocity anisotropy
of galaxy orbits.  Clusters generally have falling velocity dispersion
profiles with radius, consistent with isotropic galaxy orbits
(e.g. \citealt{carlberg97,girardi02,lokas03}).  However, at group
scales \cite{carlberg01} observed a rise in the projected velocity
dispersion with radius.  Using this to examine the mass-to-light
profile of their groups they concluded that galaxies contract within
their dark matter haloes by dynamical friction.  In contrast,
\cite{zabludoff98,girardi02} and \cite{parker05} observed falling velocity 
dispersion profiles in their group samples, which are consistent with
clusters.

\begin{figure}
\begin{center}

    \resizebox{20pc}{!}{
	\rotatebox{-90}{
    	\includegraphics{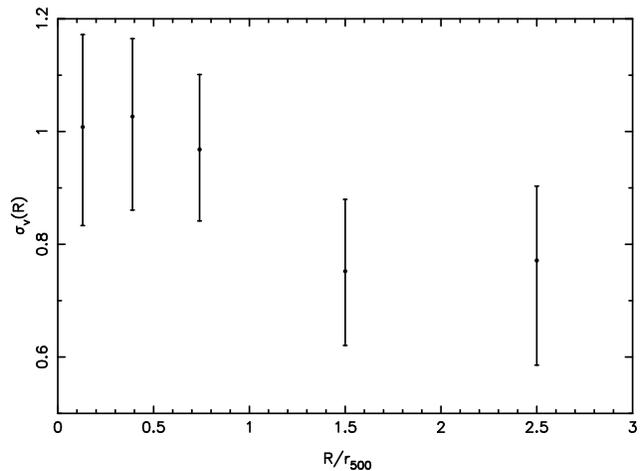} 
	}	 
	}
  \end{center}
\caption{The line-of-sight velocity dispersion, $\sigma_v(R)$ as a function 
of the scaled projected group-centric distance, $R/r_{500}$, of our
composite group.  The error bars indicate $1\sigma$ errors on the
velocity dispersions.}
\label{fig_voffset_rad}
\end{figure}

We have scaled our galaxy data by their group velocity dispersion and
group size (as measured by their $r_{500}$ radius), binned the
galaxies into equal number bins and applied the gapper algorithm
(Equation~\ref{eq:sigma}) to calculate robust velocity dispersions.
The corresponding errors are estimated using the jackknife algorithm.
The velocity dispersion profile of our composite group, illustrated in
Figure~\ref{fig_voffset_rad}, extends to a radius $>2r_{500}$, much
further than previous studies.  We observe that the velocity
dispersion profile falls with radius.  The large error bars mean that
it is not possible to determine any possible velocity anisotropy of
the galaxy orbits with these data, however they are consistent with
isotropic orbits.


\subsection{Magnitudes}

\cite{miles04} obtained deep $B$ and $R$-band photometry of 25 GEMS groups 
and found a difference in the luminosity function (LF) of galaxies in
groups with X-ray luminosities log$_{10}L_X(Bol)<41.7$ erg s$^{-1}$ in
comparison to the LF of galaxies in groups with X-ray luminosities
greater than this.  They found a dip at $-19<M_B<-17$ ($-20.5<M_K<
-23$) in the LF of galaxies in groups with low X-ray luminosities,
mainly associated with the early-type galaxies.  They concluded that
the dip is a result of current rapid dynamical evolution in the low
X-ray luminosity groups.


Our sample has advantages over that of \cite{miles04} in that all of
our galaxies are spectroscopically confirmed as group members, our
stacked group extends to larger radii ($2r_{500}$ in comparison to the
$0.3r_{500}$ of \cite{miles04}) and we also have a field sample with
which to compare our group properties.  However, we only study 15
groups to $M_K=-20$ so the data is too shallow to conclusively examine
the LF of this sample, or further divide the sample by galaxy
morphology.

\begin{figure}
\begin{center}
    \resizebox{20pc}{!}{
	\rotatebox{-90}{
    	\includegraphics{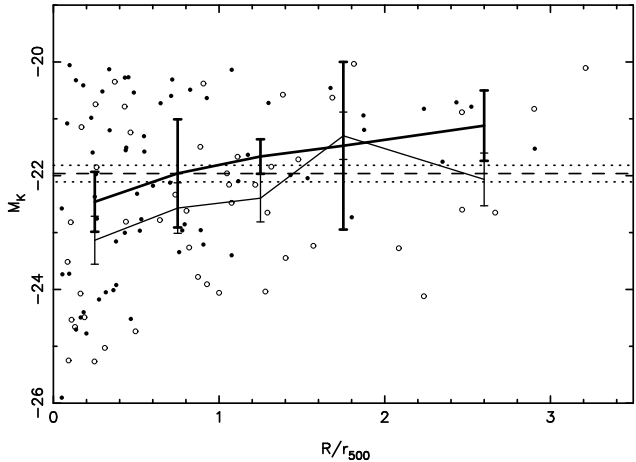}
	}	 
	}
  \end{center}
\caption{The relationship of $M_K$ with scaled group radius.  The solid 
points indicate galaxies in high-$L_X$ groups
(log$_{10}L_X(r_{500})>41.7$ erg s$^{-1}$) and open points indicate
galaxies in low-$L_X$ groups (log$_{10}L_X(r_{500})<41.7$ erg
s$^{-1}$).  The solid line indicates the mean luminosity at each
radius for high-$L_X$ groups (thick black line) and low-$L_X$ groups
(thin black line), whilst the error bars show the error on the mean
magnitude. The dashed line shows the mean luminosity in our field
sample and the dotted lines indicate the error on this mean value.}
\label{fig_mag_radius}
\end{figure}

We examine the radial dependence of the magnitudes of the galaxies in
our sample in Figure~\ref{fig_mag_radius}.  We divide the sample into
high and low-$L_X$ groups by the X-ray luminosity at which
\cite{miles04} observe a separation in the galaxy populations: 
log$_{10}L_X(r_{500})=41.7$ erg s$^{-1}$. 

Our composite group indicates that the mean galaxy magnitude fades
with distance from the group centre.  We also see that the mean
magnitudes of the galaxies in high X-ray luminosity groups are fainter
than those in low X-ray luminosity groups.  This is in contrast to
\cite{miles04} who observed that the total B-band luminosity within
a radius of $0.3r_{500}$ is lower and more concentrated in low-$L_X$
groups.  However, our result is consistent with observations that the
ratio of dwarf-to-giant galaxies increases with group mass
\citep{ferguson91,zabludoff00,wilman05,cellone05}.  We explicitly examine 
this by calculating the number of galaxies with $M\leq M^{\star}$
(`giants'; $M^{\star}=-22.6$; \citealt{kochanek01}) and $M>M^{\star}$
(`dwarfs') in high- and low X-ray luminosity groups.  We find
dwarf-to-giant ratios of $1.71\pm0.27$ in high-$L_X$ groups and
$0.96\pm0.2$ in low X-ray luminosity groups, where the errors given
are the poisson errors on these quantities.  Therefore, the fact that
the mean magnitudes in high X-ray luminosity groups are fainter than
those in low X-ray luminosity groups is an indication of the higher
dwarf-to-giant ratio in this environment.  We suggest that the
magnitude difference observed by \cite{miles04} is a result of the
more numerous bright galaxies within $R<0.3r_{500}$ in high-$L_X$
groups compared to low-$L_X$ groups (Figure~\ref{fig_mag_radius}).

\subsection{Colours}

\begin{figure}
\begin{center}

    \resizebox{20pc}{!}{
	\rotatebox{-90}{
    	\includegraphics{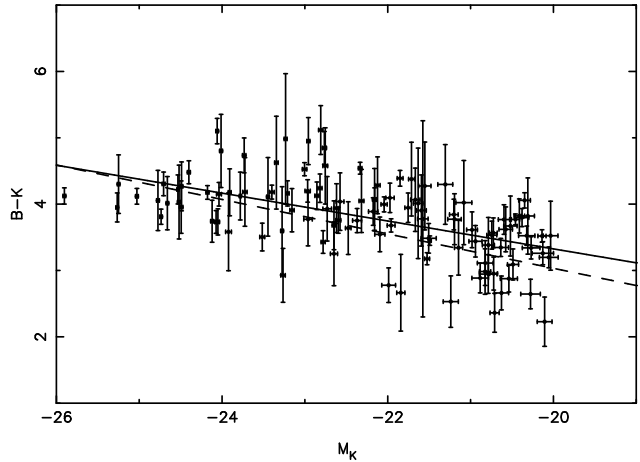} 
	}	 
	}
  \end{center}
\caption{$B-K$ vs $M_K$ colour-magnitude diagram. The solid line is a 
least-squares fit to the composite group data (solid points) given in
Equation~\ref{eq:colour_mag} whilst the dashed line is a least-squares
fit to the field data given in Equation~\ref{eq:colour_mag_field}.
Error bars indicate the errors on the 2MASS magnitudes and the
combined errors on the colours.}
\label{fig_col_mag}
\end{figure}

Galaxy colour has been shown to depend on both luminosity and
environment: The highest luminosity galaxies are the reddest
(e.g. \citealt{faber73,visvanathan77,balogh04sdss,baldry04,blanton05})
and there are a higher fraction of red galaxies in the densest
environments
(e.g. \citealt{oemler74,butcher84,girardi03,balogh04sdss,tanaka04}).
Previous analyses of the group environment have found that the colours
of galaxies in groups are redder than those in the field
\citep{girardi03,tovmassian04}.  In order to examine the relationship
between colour and environment, we separate out the effects of
luminosity by correcting the colours of the galaxies to the colour
they would have at a specific magnitude based on the slope of the
colour-magnitude relation (c.f. \citealt{kodama01,tanaka05}).  We
therefore fit a colour-magnitude relation using the Buckley-James
algorithm described above.  The best-fit straight line to all 113
absolute-magnitude limited galaxies is described by:
\begin{equation}
B-K=-0.21^{\pm0.03}M_K-0.88,
\label{eq:colour_mag}
\end{equation}
with an rms scatter $\sigma=0.49$ mag, and is shown in
Figure~\ref{fig_col_mag}.  We also fit a colour-magnitude relation to
the field galaxies and find that this is consistent within the errors:
\begin{equation}
B-K=-0.26^{\pm0.03}M_K-2.2.
\label{eq:colour_mag_field}
\end{equation}

\begin{figure}
\begin{center}

    \resizebox{20pc}{!}{
	\rotatebox{-90}{
    	\includegraphics{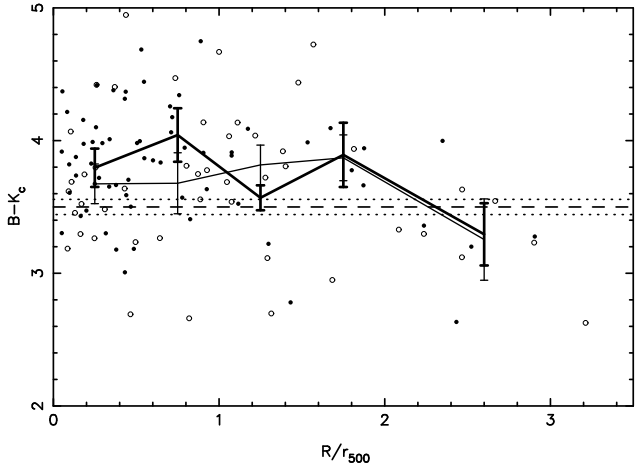} 
	}	 
	}
  \end{center}
\caption{Relationship of $B-K$ colour with scaled group radius 
($R/r_{500}$).  The colours are normalised to $M_K=-22$ mag as
described in the text.  The solid points indicate galaxies in
high-$L_X$ groups (log$_{10}L_X(r_{500})>41.7$ erg s$^{-1}$) and open
points indicate galaxies in low-$L_X$ groups
(log$_{10}L_X(r_{500})<41.7$ erg s$^{-1}$).  The solid line indicates
the mean colour at each radius for high-$L_X$ groups (thick black
line) and low-$L_X$ groups (thin black line), whilst the error bars
show the error on the mean colour.  The dashed line indicates the mean
colour of field galaxies, also normalised to $M_K=-22$ mag, whilst the
dotted lines indicate the error on the mean of the field colours.}
\label{fig_col_radius}
\end{figure}

We correct the colours of the galaxies to a magnitude of $M_K=-22$ mag
based on Equation~\ref{eq:colour_mag},
i.e. $B-K_c=(B-K)-0.21(M_K+22)$. Examining the relationship of the
normalised colours with scaled radius (Figure~\ref{fig_col_radius}) we
observe a weak correlation of colour with radius (correlated at the 86
per cent level) such that galaxies further out are bluer than those in
the centre of groups.  
A KS test indicates that the colours of the galaxies in the high-$L_X$
environment are unlikely to be drawn from the same parent population
as those galaxies in the low-$L_X$ groups at the 95.4 per cent level.
Comparing the two group populations to the field, the colours of the
galaxies in the high-$L_X$ environment are unlikely to have been drawn
from the same population as those in the field at the $99.8$ per cent
level.  In contrast, the colours of the galaxies in the low-$L_X$
groups are unlikely to have been drawn from the same population as the
field at only the $57$ per cent level.  It appears that the colours of
galaxies in high X-ray luminosity groups are significantly redder than
those in low X-ray luminosity groups, which are similar to those of
galaxies in the field.

\subsection{Morphology}

The morphology-density relationship, and its close equivalent the
morphology-radius relationship have been known to exist in clusters
for years (e.g. \citealt{dressler80,whitmore93}).

In the group environment, \cite{tran01} and \cite{wilman05} have shown
that the fraction of early-type galaxies in groups is higher than the
field and that fraction decreases from the group core to the outer
regions consistent with the situation in clusters.  However,
\cite{helsdon_morph} find that, although groups show similar spiral
fractions to clusters within a virial radius, the group
morphology-density relation is offset from the cluster relation.  They
conclude that this is a result of the projection of the 3D density
onto the line of sight and a higher merging rate in the group
environment.



\begin{figure}
\begin{center}
    \resizebox{20pc}{!}{
	\rotatebox{-90}{
	\includegraphics{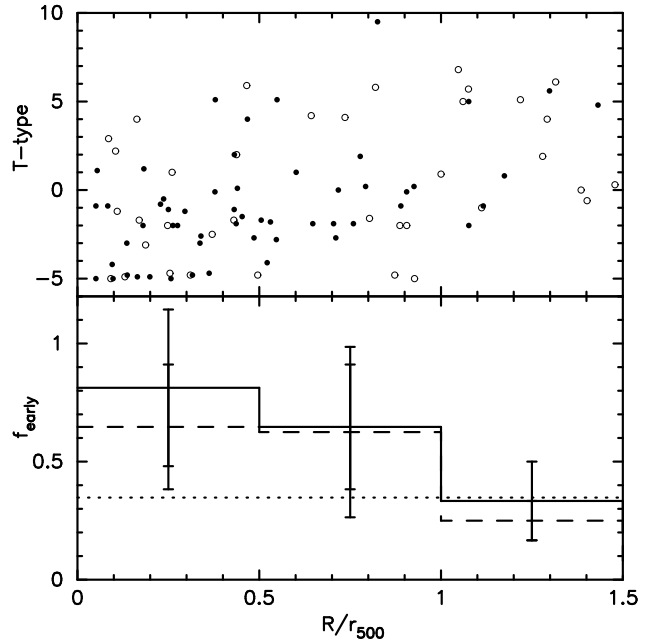} 		
	}	 
	}
  \end{center}
\caption{The upper plot indicates the morphological T-types with scaled 
group-radius, solid points are galaxies in high X-ray luminosity
groups (log$_{10}L_X(r_{500})>41.7$ erg s$^{-1}$), open points in low
X-ray luminosity groups (log$_{10}L_X(r_{500})<41.7$ erg s$^{-1}$).
The lower plot shows the early-type fraction ($f_{early}$; proportion
of galaxies with T-type $\leq0.0$) with scaled group radius for galaxies
in high-$L_X$ groups (solid line), low-$L_X$ groups (dashed line) and
the field (dotted line).  The error bars indicate poisson errors on
each bin.}
\label{fig_histo_ttype_rad}
\end{figure}

In Figure~\ref{fig_histo_ttype_rad} we show the distribution of the
morphologies with scaled group-centric radius. We observe that both
high- and low X-ray luminosity groups have significantly higher
fractions of early-type galaxies than the field value within a radius
of $r_{500}$.  These fractions become equivalent to the field at radii
larger than this.  A KS test confirms this, indicating that the
morphologies of the galaxies in groups within $r_{500}$ are not drawn
from the same population as those in the field at the $>99.999$ per
cent level.

We can also compare our group early-type fractions to those in
clusters if we assume that passive galaxies are equivalent to
early-type galaxies.  Within a radius of $r_{200}$ ($\sim1.5r_{500}$)
\cite{hilton05} find a fraction of passive galaxies of $0.76\pm0.02$
in high X-ray luminosity clusters and $0.64\pm0.02$ for low-$L_X$
clusters. For our groups we find an early-type fraction within a
$1.5r_{500}$ of $0.71\pm0.04$ for high X-ray luminosity groups and
$0.51\pm0.08$ for low-$L_X$ groups, where the errors are poisson
errors.  There is a clear difference between the high- and low-X-ray
luminosity groups and a KS test confirms that the morphologies of
galaxies in high-$L_X$ groups are not drawn from the same population
as those in low-$L_X$ groups at the $97.34$ per cent level.  In
comparison to the clusters we find that the low-$L_X$ groups have a
lower early-type fraction than clusters and high-$L_X$ groups have
similar early-type fractions to the cluster environment. However, we
bear in mind the work of \cite{helsdon_morph} and add the caveat that
these groups may not follow the same morphology-density relationship
as clusters.

\begin{figure}
\begin{center}
    \resizebox{20pc}{!}{
	\rotatebox{-90}{
	\includegraphics{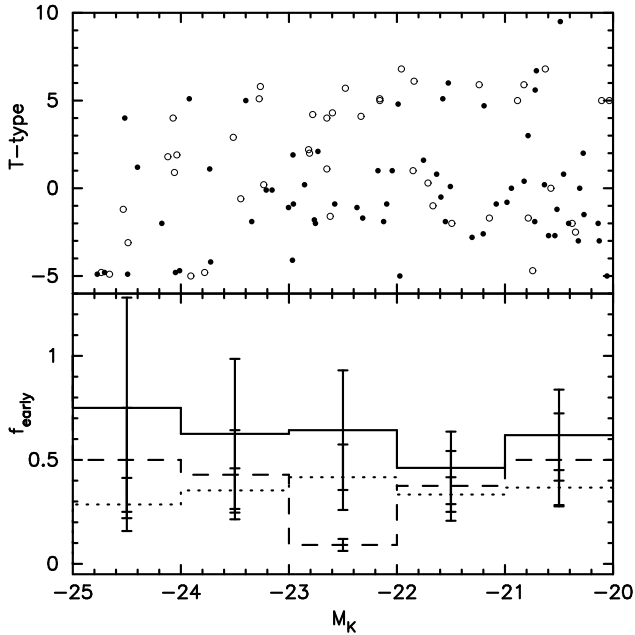}
	}	 
	}
  \end{center}
\caption{The upper plot shows the distribution of morphological T-types 
with absolute magnitude, $M_K$, solid points are galaxies in high
X-ray luminosity groups (log$_{10}L_X(r_{500})>41.7$ erg s$^{-1}$),
open points in low X-ray luminosity groups
(log$_{10}L_X(r_{500})<41.7$ erg s$^{-1}$).  The lower plot shows the
early-type fraction ($f_{early}$; proportion of galaxies with T-type
$\leq0.0$) for galaxies in high-$L_X$ groups (solid line), low-$L_X$
groups (dashed line) and the field (dotted line).  The error bars
indicate possion errors on each bin.}
\label{fig_histo_ttype_mag}
\end{figure}

\cite{wilman05} observe an enhancement of the fraction of passive galaxies 
in group environments in comparison to the field at all magnitudes.
Figure~\ref{fig_histo_ttype_mag} demonstrates that galaxies in high
X-ray luminosity groups have higher early-type fractions than the
field at all magnitudes, whereas the galaxies in low X-ray luminosity
groups are more consistent with the field.  A KS test confirms that
the morphologies of galaxies in high-$L_X$ groups are not drawn from
the same population as those in the field at $>99.999$ per cent level.
In contrast the galaxies in the low-$L_X$ groups are consistent with
not being drawn from the same population as those in the field at only
the $58$ per cent level.

In summary, groups have higher fractions of early-type galaxies than the
field at radii less than $r_{500}$ and similar fractions to the field
at radii beyond this.  The morphologies of galaxies in low X-ray
luminosity groups are similar to those in the field whilst the
morphologies of galaxies in the inner regions of high X-ray luminosity
groups are similar to those in clusters.

\section{Discussion and Conclusions}
\label{concl_sect}

We have examined the properties of 16 galaxy groups from the GEMS
sample \citep{osmond04,forbes06} which have additional wide-field HI
observations (Kilborn et al. in preparation).  Using galaxy positions
and recession velocities from the 6dFGS DR2, NED and new recession
velocities and positions from the HI observations, we determine group
membership using a FOF algorithm.  We show that the group properties
we derive from that membership are robust to the choice of limiting
number density contrast.  However, we do not find a group at the
position of the NGC 7144 group (determined by \citealt{osmond04}) at
any limiting density contrast.  This region only has X-ray emission
from the halo of NGC 7144 itself.  We, therefore, do not include this
group in the rest of our analysis.

We examined the dynamical parameters of the remaining 15 groups,
finding that groups with higher X-ray luminosities are more likely to
show extended intra-group X-ray emission and bright early-type BGGs
located near to the group centroid, than those groups with lower X-ray
luminosities.  Our analysis suggests that the X-ray luminosities of
these groups are more closely related to their dynamical properties
than whether the groups show intra-group, galaxy halo or no detectable
X-ray emission as defined by \cite{osmond04}.

Investigating the scaling relations followed by the galaxy groups we
find that groups with higher X-ray luminosities have higher velocity
dispersions and masses than those with lower X-ray luminosities:
$L_X(r_{500})\propto\sigma_v^{3.11\pm0.59}$, and $L_X(r_{500})\propto
M_V^{1.13\pm0.27}$.  These relationships are consistent with the
predictions of self-similarity (e.g. \citealt{borgani04}), i.e. groups
following the same scaling relations as clusters.  We also find the
virial mass to be proportional to the total $K$-band luminosity in the
system: $M_V\propto L_K(Tot)^{2.0\pm0.9}$, indicating that the mass of
the groups is increasing faster than their luminosity. This increase
in the mass-to-light ratio is unlikely to be an effect of the stellar
populations of galaxies as the $K$-band is more closely related to
galaxy mass than other wavelengths (Proctor et al. in preparation).  
The fraction of early-type galaxies in the groups are correlated with
both velocity dispersion and X-ray luminosity.

We study the properties of the BGGs and their relationship with their
host group.  We find that the BGGs of the dynamically mature groups
are early-type galaxies that lie close to the spatial and velocity
centroid of the group.  In contrast, we find both early- and late-type
BGGs in the dynamically immature groups and these lie at a range of
spatial and velocity separations from their group centroids.  The
early-type BGGs are significantly brighter than the late-type BGGs.
We observe that the luminosities of all the BGGs increase with
increasing total $K$-band luminosity, X-ray luminosity and velocity
dispersion of their groups.  However, the fraction of group light in the
BGG, and the dominance of the BGG over the second brightest galaxy,
fall with increasing total group luminosity.  These properties are all
consistent with the paradigm in which the group grows rapidly by
accreting infalling galaxies but beyond a critical group mass the time
for infalling galaxies to merge with the central galaxy is longer than
a Hubble time, such that the growth of the BGG with respect to that of
the group slows.

In order to analyse the properties of the groups' constituent galaxies
we stack the members of the 15 groups to create a composite group.  In
the composite group we observe that galaxies are, in the mean,
fainter, bluer and morphologically later-type galaxies with increasing
radius from the group centroid.  We divide the composite group sample
by the X-ray luminosity (log$_{10}L_X(r_{500})=41.7$ erg s$^{-1}$)
which \cite{miles04} found to divide the properties of 25 GEMS groups.
Galaxies in groups with an X-ray luminosity brighter than this are
redder with a higher giant-to-dwarf ratio, and a higher early-type
fraction (close to that observed in clusters) than galaxies in
low-$L_X$ groups.  We also observe that the colours of galaxies and
early-type fractions in low X-ray luminosity groups are more closely
related to those of galaxies in the field, than are galaxies in
high-$L_X$ groups.

The examination of the composite group suggests that the properties of
galaxies in the centres of low-mass groups are already different to
those in the field, to account for the radial relationships
observed. The galaxies in the centres of the groups are more likely to
be
bright, red, early-type galaxies than the galaxies in our field
sample,
even in the dynamically younger groups that do not have large velocity
dispersions and are not generally observed to display halos of hot
X-ray emitting gas. This suggests that multiple, high-speed
galaxy-galaxy encounters (`harassment'; \citealt{moore96}) and ram
pressure stripping by a dense intra-group medium
(e.g. \citealt{quilis00}) cannot be playing a role in creating the
observed differences in these low-mass groups.

Possible mechanisms for these differences include strangulation and
merging.  Strangulation stems from current theories of galaxy
formation that suggest that isolated galaxies continuously draw gas
from a hot diffuse reservoir in their dark matter halo with which to
maintain star formation (e.g. \citealt{larson80,cole00}).  Access to
this reservoir may be halted by the galaxy falling into a group,
thereby quenching its star formation.  The galaxy will then slowly
fade and redden over $\sim1$Gyr.  However, the reservoirs are too cold
and diffuse to be observed (e.g. \citealt{benson00}).  In contrast,
galaxy mergers occur over much shorter timescales, $\sim100$ Myr: Once
a group has formed by gravitational collapse, the system relaxes
thereafter by two-body interactions, with dynamical friction causing
galaxies to fall into the centre of the group and decelerate. This
deceleration is proportional to the inverse of the difference in
velocity between the galaxies and is therefore more likely in
low-velocity dispersion groups (although there is also a dependence on
the number of members a group has). Galaxy-galaxy mergers are
therefore more likely in the low velocity dispersion groups.

The differences observed between field and group galaxies are stronger
in higher mass groups, such that the properties of galaxies in this
environment are already close to those in clusters.  The dynamical
analysis of all the groups suggests that the higher mass groups are
more dynamically mature.  We therefore conclude that the higher
early-type fractions and dwarf-to-giant ratios and redder galaxies in
higher mass groups are a result of galaxy-galaxy mergers at earlier
epochs in smaller mass groups which have since merged to become the
group we observe today, whilst, due to their lower velocity
dispersions, the less massive groups are still undergoing mergers
today.  However, the properties of all the galaxies in the groups are
unlikely to be due to mergers.  We therefore conclude that
strangulation, or some further mechanism may play some role and look
to our second paper (Kilborn et al. in preparation) examining the
neutral hydrogen in these groups to further understand the mechanisms
acting in this environment.

\section*{Acknowledgments}

We would like to thank Chris Power, Gary Mamon, Trevor Ponman, Rob
Proctor and Simon Ellis for helpful discussions.  We would also like
to thank the anonymous referee for his/her positive comments. This
publication makes use of data products from the Two Micron All Sky
Survey (2MASS) which is a joint project of the University of
Massachusetts and the Infrared Processing and Analysis
Center/California Institute of Technology, funded by the National
Aeronautics and Space Administration and the National Science
Foundation.  This research has made use of the NASA/IPAC Extragalactic
Database (NED) which is operated by the Jet Propulsion Laboratory,
California Institute of Technology, under contract with the National
Aeronautics and Space Administration.  It has also made use of the
HyperLEDA database.

\appendix

\section[]{Group Members}
\label{group_gals}

\begin{table*}
\begin{center}
\caption{Details of the galaxies in each group.  $R_{gc}$ is the 
group-centric radius of each galaxy from the centre calculated by the
FOF algorithm.  New galaxies found in HI are named
GEMS-groupname-number. Dashes indicate where information is not available.}
\label{group_mems1}
\begin{tabular}{llcccccc}
\hline
\hline 
Galaxy Name&6dFGS ID&RA (J2000)&Dec (J2000)&$\bar{v}$ (km s$^{-1}$)&$m_K$ (mag)&$R_{gc}$ (Mpc)&T-type\\
\hline
\hline 
NGC 524&&&&&&&\\
\hline
UGC 00896                      &  -                  &   1:21:27.40 &   9:10:47.30 & 2042 & 13.50$\pm$0.111 & 0.325 &   8.0\\
NGC 0489                       &  -                  &   1:21:53.90 &   9:12:23.60 & 2507 &  9.61$\pm$0.179 & 0.281 &   5.0\\
CGCG 411-038                   &  -                  &   1:22:31.82 &   9:16:53.20 & 2636 & 13.43$\pm$0.287 & 0.213 &   0.0\\
MCG +01-04-042                 &  -                  &   1:22:54.30 &   8:51:17.00 & 2219 & 13.50$\pm$0.184 & 0.297 &   0.0\\
NGC 0502                       &  -                  &   1:22:55.54 &   9: 2:57.10 & 2489 & 10.28$\pm$0.069 & 0.235 &  -2.0\\
NGC 0509                       &  -                  &   1:23:24.09 &   9:26: 0.80 & 2274 & 10.99$\pm$0.410 & 0.114 &  -1.7\\
NGC 0516                       &  -                  &   1:24: 8.07 &   9:33: 6.10 & 2432 & 10.63$\pm$0.410 & 0.045 &  -1.7\\
IC 0101                        &  -                  &   1:24: 8.55 &   9:55:49.90 & 2404 & 12.13$\pm$0.132 & 0.173 &   3.0\\
NGC 0518                       &  -                  &   1:24:17.64 &   9:19:51.40 & 2725 &  9.92$\pm$0.113 & 0.069 &   1.0\\
NGC 0522                       &  -                  &   1:24:45.85 &   9:59:40.70 & 2725 &  9.44$\pm$0.086 & 0.195 &   4.1\\
NGC 0524                       &  -                  &   1:24:47.72 &   9:32:19.80 & 2379 &  7.24$\pm$0.218 & 0.029 &  -1.2\\
NGC 0532                       &  -                  &   1:25:17.34 &   9:15:50.80 & 2361 &  8.96$\pm$0.367 & 0.116 &   2.0\\
IC 0114                        &  -                  &   1:26:22.58 &   9:54:35.80 & 2275 & 11.39$\pm$0.108 & 0.240 &  -2.0\\
UGC 01019                      &  -                  &   1:26:38.94 &  10:17:14.00 & 2182 & 13.50$\pm$0.100 & 0.369 &   8.9\\
LEDA 093841                    &  -                  &   1:27:37.30 &   8:50:24.00 & 2435 & 13.50$\pm$0.100 & 0.394 &   0.0\\
UGC 01050                      &  -                  &   1:28:12.80 &  10:26: 2.00 & 2396 & 13.50$\pm$0.001 & 0.510 &   8.0\\
\hline 
NGC 720&&&&&&&\\
\hline
KUG 0150-138                   &  -                  &   1:52:36.06 & -13:34:39.70 & 1374 & 13.50$\pm$0.225 & 0.106 &   0.0\\
2MASX J01524752-1416211        & 6dF J0152475-141621 &   1:52:47.53 & -14:16:21.40 & 1448 & 11.48$\pm$0.328 & 0.223 &   0.0\\
NGC 0720                       & 6dF J0153005-134419 &   1:53: 0.49 & -13:44:19.00 & 1663 &  7.40$\pm$0.401 & 0.021 &  -4.9\\
2MASX J01535632-1350125        & -                   &   1:53:56.32 & -13:50:12.60 & 1496 & 12.40$\pm$0.217 & 0.100 &   0.0\\
MCG -02-05-072                 & -                   &   1:54: 3.15 & -14:15:10.70 & 1423 & 10.34$\pm$0.555 & 0.238 &   0.3\\
6dF J0154050-135421            & 6dF J0154050-135421 &   1:54: 5.04 & -13:54:21.20 & 1321 & 13.50$\pm$0.100 & 0.126 &  99.9\\
\hline
NGC 1052&&&&&&&\\ 
\hline
$[$RC3$]$ 0231.5-0635              & -                   &   2:33:57.04 &  -6:21:36.20 & 1410 & 13.50$\pm$0.100 & 0.766 &  99.9\\
USGC S092 NED09                & -                   &   2:35:28.79 &  -7: 8:59.00 & 1532 & 13.50$\pm$0.100 & 0.535 &  99.9\\
6dF J0235320-070936            & 6dF J0235320-070936 &   2:35:32.00 &  -7: 9:36.30 & 1554 & 13.50$\pm$0.100 & 0.530 &  99.9\\
NGC 0991                       & -                   &   2:35:32.69 &  -7: 9:16.00 & 1532 & 11.18$\pm$0.361 & 0.530 &   5.0\\
NGC 1022                       & -                   &   2:38:32.70 &  -6:40:38.70 & 1453 &  8.64$\pm$0.367 & 0.440 &   1.1\\
SDSS J023848.50-080257.7       & -                   &   2:38:48.50 &  -8: 2:57.70 & 1665 & 13.50$\pm$0.100 & 0.240 &  99.9\\
NGC 1035                       & -                   &   2:39:29.09 &  -8: 7:58.60 & 1241 &  9.13$\pm$0.465 & 0.201 &   5.1\\
6dF J0239299-080821            & 6dF J0239299-080821 &   2:39:29.92 &  -8: 8:21.10 & 1393 & 13.50$\pm$0.100 & 0.201 &  99.9\\
NGC 1042                       & 6dF J0240240-082601 &   2:40:23.97 &  -8:26: 1.00 & 1411 &  9.45$\pm$0.576 & 0.217 &   6.1\\
UGCA 038                       & -                   &   2:40:30.19 &  -6: 6:23.00 & 1327 & 13.50$\pm$0.100 & 0.549 &   9.0\\
NGC 1047                       & -                   &   2:40:32.84 &  -8: 8:51.60 & 1340 & 11.38$\pm$0.173 & 0.137 &  -0.7\\
NGC 0961                       & -                   &   2:41: 2.46 &  -6:56: 9.10 & 1295 & 13.50$\pm$0.100 & 0.286 &  99.9\\
NGC 1052                       & 6dF J0241048-081521 &   2:41: 4.80 &  -8:15:21.00 & 1591 &  7.51$\pm$0.360 & 0.144 &  -4.8\\
SDSS J024120.22-071706.0       & -                   &   2:41:20.22 &  -7:17: 6.00 & 1663 & 13.50$\pm$0.100 & 0.173 &  99.9\\
$[$VC94$]$ 023858-0820.4           & -                   &   2:41:25.53 &  -8: 7:36.70 & 1412 & 13.50$\pm$0.100 & 0.097 &  99.9\\
SDSS J024129.37-072046.0       & -                   &   2:41:29.37 &  -7:20:46.00 & 1145 & 13.50$\pm$0.100 & 0.153 &  99.9\\
2MASX J02413514-0810243        & 6dF J0241351-081025 &   2:41:35.11 &  -8:10:24.60 & 1556 & 13.15$\pm$0.100 & 0.109 &   0.0\\
SDSS J024149.95-075530.1       & -                   &   2:41:49.96 &  -7:55:30.00 & 1372 & 13.50$\pm$0.100 & 0.031 &  99.9\\
SDSS J024246.84-073230.3       & -                   &   2:42:46.84 &  -7:32:30.40 & 1344 & 13.50$\pm$0.100 & 0.122 &  99.9\\
MCG -01-08-001                 & -                   &   2:43:42.80 &  -6:39: 5.00 & 1410 & 13.50$\pm$0.100 & 0.401 &   9.8\\
NGC 1084                       & -                   &   2:45:59.93 &  -7:34:43.10 & 1407 &  8.01$\pm$0.665 & 0.344 &   5.1\\
SHOC 137                       & 6dF J0248158-081724 &   2:48:15.83 &  -8:17:23.90 & 1405 & 13.50$\pm$0.100 & 0.533 &  99.9\\
SDSS J024839.95-074848.3       & -                   &   2:48:39.96 &  -7:48:48.30 & 1465 & 13.50$\pm$0.100 & 0.545 &  99.9\\
SHOC 138a                      & -                   &   2:49: 9.32 &  -7:50:27.40 & 1288 & 13.50$\pm$0.100 & 0.583 &  99.9\\
NGC 1110                       & -                   &   2:49: 9.57 &  -7:50:15.20 & 1333 & 13.50$\pm$0.273 & 0.583 &   8.8\\
SHOC 138b                      & -                   &   2:49:10.79 &  -7:49:24.50 & 1368 & 13.50$\pm$0.100 & 0.585 &  99.9\\
SDSS J024911.16-082828.7       & -                   &   2:49:11.16 &  -8:28:28.80 & 1430 & 13.50$\pm$0.100 & 0.620 &  99.9\\
SDSS J024913.41-080653.2       & -                   &   2:49:13.42 &  -8: 6:53.20 & 1369 & 13.50$\pm$0.100 & 0.595 &  99.9\\
2MASX J02400428-0744217        & 6dF J0240043-074422 &   2:40: 4.28 &  -7:44:22.00 & 1309 & 13.09$\pm$0.100 & 0.133 &   0.0\\
\hline 
\end{tabular} 
\end{center}
\end{table*}

\begin{table*}
\begin{center}
\caption{Continued.}
\label{group_mems2}
\begin{tabular}{llcccccc}
\hline
Galaxy Name&6dFGS ID&RA (J2000)&Dec (J2000)&$\bar{v}$ (km s$^{-1}$)&$m_K$ (mag)&$R_{gc}$ (Mpc)&T-type\\
\hline
\hline 
NGC 1332&&&&&&&\\ 
\hline
\hline 
NGC 1315                       & 6dF J0323066-212231 &   3:23: 6.60 & -21:22:30.70 & 1596 &  9.94$\pm$0.474 & 0.249 &  -1.0\\
NGC 1325                       & 6dF J0324256-213238 &   3:24:25.57 & -21:32:38.30 & 1589 &  8.83$\pm$0.169 & 0.144 &   4.2\\
NGC 1325A                      & -                   &   3:24:48.50 & -21:20:10.00 & 1333 & 14.03$\pm$0.517 & 0.095 &   6.6\\
ESO 548- G 011                 & -                   &   3:24:55.30 & -21:47: 2.00 & 1453 & 13.50$\pm$0.104 & 0.173 &   8.4\\
2MASX J03255262-2117204        & 6dF J0325526-211721 &   3:25:52.62 & -21:17:20.60 & 1427 & 11.77$\pm$0.100 & 0.029 &   0.0\\
NGC 1332                       & -                   &   3:26:17.32 & -21:20: 7.30 & 1524 &  7.12$\pm$0.368 & 0.042 &  -3.1\\
NGC 1331                       & 6dF J0326283-212120 &   3:26:28.34 & -21:21:20.30 & 1241 & 10.87$\pm$0.202 & 0.057 &  -4.7\\
2MASX J03263135-2113003        & -                   &   3:26:31.31 & -21:13: 0.60 & 1548 & 11.26$\pm$0.091 & 0.083 &  -2.5\\
2MASX J03273556-2113417        & 6dF J0327356-211341 &   3:27:35.57 & -21:13:41.40 & 1744 & 12.12$\pm$0.100 & 0.167 &  99.9\\
6dF J0327422-214159            & 6dF J0327422-214159 &   3:27:42.16 & -21:41:58.60 & 1294 & 13.50$\pm$0.100 & 0.208 &  99.9\\
\hline 
NGC 1407&&&&&&&\\
\hline
NGC 1383                       & 6dF J0337392-182022 &   3:37:39.24 & -18:20:22.10 & 2008 &  9.51$\pm$0.427 & 0.336 &  -1.9\\
NGC 1390                       & -                   &   3:37:52.17 & -19: 0:30.10 & 1207 & 11.70$\pm$0.209 & 0.350 &   1.2\\
NGC 1393                       & -                   &   3:38:38.58 & -18:25:40.70 & 2127 &  9.31$\pm$0.475 & 0.241 &  -1.7\\
ESO 548- G 063                 & 6dF J0339348-200053 &   3:39:34.78 & -20: 0:53.30 & 2047 & 12.18$\pm$0.162 & 0.555 &   3.7\\
ESO 548- G 064                 & 6dF J0340001-192535 &   3:40: 0.08 & -19:25:34.70 & 1873 & 11.03$\pm$0.212 & 0.339 &  -2.7\\
ESO 548- G 065                 & -                   &   3:40: 2.70 & -19:21:59.80 & 1221 & 13.21$\pm$0.138 & 0.317 &   1.3\\
IC 0343                        & -                   &   3:40: 7.14 & -18:26:36.50 & 1841 & 10.65$\pm$0.262 & 0.109 &  -0.8\\
NGC 1407                       & -                   &   3:40:11.90 & -18:34:49.40 & 1779 &  6.86$\pm$0.453 & 0.095 &  -4.9\\
2MASX J03401592-1904544        & 6dF J0340159-190454 &   3:40:15.93 & -19: 4:54.40 & 1613 & 12.26$\pm$0.297 & 0.213 &  -3.5\\
ESO 548- G 068                 & -                   &   3:40:19.17 & -18:55:53.40 & 1693 & 10.43$\pm$0.225 & 0.162 &  -2.6\\
2MASX J03404323-1838431        & 6dF J0340432-183843 &   3:40:43.23 & -18:38:43.10 & 1373 & 12.36$\pm$0.170 & 0.059 &  -1.6\\
2MASX J03405272-1828410        & 6dF J0340527-182841 &   3:40:52.73 & -18:28:40.80 & 1678 & 12.70$\pm$0.084 & 0.042 &  -1.4\\
ESO 548- G 072                 & -                   &   3:41: 0.28 & -19:27:19.40 & 2034 & 13.50$\pm$0.124 & 0.331 &   5.0\\
ESO 548- G 073                 & 6dF J0341044-190540 &   3:41: 4.41 & -19: 5:40.00 &  989 & 12.91$\pm$0.212 & 0.199 &   3.3\\
IC 0345                        & 6dF J0341091-181851 &   3:41: 9.13 & -18:18:50.90 & 1244 & 11.22$\pm$0.249 & 0.086 &  -2.0\\
ESO 548- G 076                 & 6dF J0341318-195419 &   3:41:31.81 & -19:54:18.50 & 1544 & 11.88$\pm$0.362 & 0.495 &  -1.3\\
IC 0346                        & -                   &   3:41:44.66 & -18:16: 1.10 & 2013 & 10.04$\pm$0.533 & 0.113 &  -0.5\\
6dF J0341498-193453            & 6dF J0341498-193453 &   3:41:49.82 & -19:34:52.50 & 1913 & 13.50$\pm$0.100 & 0.380 &  99.9\\
ESO 548- G 079                 & 6dF J0341561-185343 &   3:41:56.08 & -18:53:42.60 & 2029 & 11.11$\pm$0.445 & 0.141 &  -1.2\\
ESO 549- G 002                 & -                   &   3:42:57.34 & -19: 1:12.40 & 1111 & 13.50$\pm$0.323 & 0.232 &   9.5\\
APMBGC 549+118-079             & -                   &   3:44: 2.46 & -18:28:18.30 & 1979 & 13.50$\pm$0.100 & 0.257 &  99.9\\
ESO 549- G 007                 & 6dF J0344115-191910 &   3:44:11.48 & -19:19: 9.90 & 1478 & 13.50$\pm$0.121 & 0.389 &  -1.4\\
NGC 1440                       & -                   &   3:45: 2.91 & -18:15:57.70 & 1597 &  8.28$\pm$0.699 & 0.362 &  -1.9\\
NGC 1452                       & -                   &   3:45:22.31 & -18:38: 1.10 & 1737 &  8.77$\pm$0.206 & 0.378 &   0.2\\
\hline 
NGC 1566&&&&&&&\\
\hline
NGC 1546                       & 6dF J0414364-560340 &   4:14:36.38 & -56: 3:39.51 & 1238 &  8.17$\pm$0.592 & 0.164 &  -0.6\\
NGC 1549                       & 6dF J0415451-553532 &   4:15:45.13 & -55:35:32.10 & 1202 &  6.88$\pm$0.113 & 0.058 &  -4.8\\
NGC 1553                       & 6dF J0416105-554648 &   4:16:10.47 & -55:46:48.00 & 1172 &  6.35$\pm$0.218 & 0.029 &  -2.0\\
IC 2058                        & -                   &   4:17:54.35 & -55:55:58.40 & 1379 & 10.99$\pm$0.240 & 0.197 &   6.8\\
\hline 
NGC 1808&&&&&&&\\
\hline
NGC 1792                       & 6dF J0505144-375851 &   5: 5:14.41 & -37:58:50.50 & 1176 &  7.09$\pm$0.170 & 0.025 &   4.0\\
2MASX J05061389-3803154        & 6dF J0506139-380316 &   5: 6:13.89 & -38: 3:15.60 &  855 & 11.72$\pm$0.382 & 0.065 &   0.0\\
NGC 1808:$[$AB70$]$ C              & 6dF J0507423-373046 &   5: 7:42.34 & -37:30:46.10 &  969 & 13.50$\pm$0.100 & 0.205 &  99.9\\
ESO 305- G 009                 & -                   &   5: 8: 7.62 & -38:18:33.51 & 1021 & 13.50$\pm$0.902 & 0.225 &   8.0\\
2MASX J05081153-3657351        & 6dF J0508115-365735 &   5: 8:11.51 & -36:57:35.30 & 1073 & 12.92$\pm$0.100 & 0.350 &   0.0\\
NGC 1827                       & 6dF J0510046-365737 &   5:10: 4.11 & -36:57:34.90 & 1037 & 10.34$\pm$0.336 & 0.444 &   5.9\\
\hline 
\end{tabular} 
\end{center}
\end{table*}

\begin{table*}
\begin{center}
\caption{Continued.}
\label{group_mems3}
\begin{tabular}{llcccccc}
\hline
\hline 
Galaxy Name&6dFGS ID&RA (J2000)&Dec (J2000)&$\bar{v}$ (km s$^{-1}$)&$m_K$ (mag)&$R_{gc}$ (Mpc)&T-type\\
\hline
\hline 
NGC 3557&&&&&&&\\
\hline
ESO 377- G 012                 & 6dF J1108198-373726 &  11: 8:19.76 & -37:37:26.10 & 3486 & 10.22$\pm$0.398 & 0.322 &   1.9\\
NGC 3557:$[$ZM2000$]$ 0038         & -                   &  11: 8:50.20 & -37:22:39.00 & 3062 & 13.50$\pm$0.100 & 0.239 &  99.9\\
NGC 3557:$[$ZM2000$]$ 0097         & -                   &  11: 9: 6.50 & -37:13: 4.00 & 2751 & 13.50$\pm$0.100 & 0.275 &  99.9\\
NGC 3557:$[$ZM2000$]$ 0047         & -                   &  11: 9: 8.40 & -37:43:32.00 & 3146 & 13.50$\pm$0.100 & 0.225 &  99.9\\
NGC 3557:$[$ZM2000$]$ 0016         & -                   &  11: 9:10.80 & -37:23:59.00 & 3183 & 13.50$\pm$0.100 & 0.175 &  99.9\\
NGC 3557:$[$ZM2000$]$ 0025         & -                   &  11: 9:21.80 & -37:27:48.00 & 2772 & 13.50$\pm$0.100 & 0.125 &  99.9\\
NGC 3557:$[$ZM2000$]$ 0049         & -                   &  11: 9:27.70 & -37:38:43.00 & 2640 & 13.50$\pm$0.100 & 0.141 &  99.9\\
NGC 3557B                      & 6dF J1109321-372059 &  11: 9:32.13 & -37:20:58.70 & 2937 &  9.17$\pm$0.551 & 0.150 &  -4.7\\
2MASX J11093481-3737266        & 6dF J1109355-373729 &  11: 9:35.57 & -37:37:28.90 & 2902 & 13.55$\pm$0.100 & 0.113 &  99.9\\
NGC 3557                       & 6dF J1109577-373221 &  11: 9:57.65 & -37:32:21.00 & 3031 &  7.28$\pm$0.121 & 0.021 &  -5.0\\
NGC 3557:$[$ZM2000$]$ 0017         & -                   &  11:10:13.60 & -37:24:55.01 & 2447 & 13.50$\pm$0.100 & 0.084 &  99.9\\
NGC 3564                       & 6dF J1110364-373251 &  11:10:36.38 & -37:32:51.30 & 2779 &  9.00$\pm$0.101 & 0.114 &  -2.0\\
NGC 3568                       & 6dF J1110486-372652 &  11:10:48.57 & -37:26:52.30 & 2440 &  9.26$\pm$0.585 & 0.157 &   5.1\\
NGC 3557:$[$ZM2000$]$ 0032         & -                   &  11:11:42.70 & -37:32:10.00 & 2623 & 13.50$\pm$0.100 & 0.317 &  99.9\\
\hline 
NGC 3783&&&&&&&\\
\hline
ESO 320- G 004                 & 6dF J1134436-381503 &  11:34:43.62 & -38:15: 3.20 & 2829 & 11.94$\pm$0.200 & 0.397 &   5.0\\
2MASX J11351493-3755309        & 6dF J1135149-375531 &  11:35:14.92 & -37:55:30.70 & 2823 & 13.15$\pm$0.100 & 0.250 &   0.0\\
NGC 3742                       & 6dF J1135325-375723 &  11:35:32.51 & -37:57:23.01 & 2715 &  8.79$\pm$0.175 & 0.206 &   1.9\\
AM 1133-374                    & -                   &  11:35:45.70 & -38: 1:19.99 & 2870 & 13.50$\pm$0.100 & 0.185 &  99.9\\
NGC 3749                       & 6dF J1135532-375951 &  11:35:53.21 & -37:59:50.50 & 2742 &  8.77$\pm$0.191 & 0.161 &   0.9\\
ESO 320- G 013                 & 6dF J1137199-380551 &  11:37:19.86 & -38: 5:51.21 & 3018 & 13.50$\pm$0.118 & 0.145 &   3.0\\
6dF J1138589-380042            & 6dF J1138589-380042 &  11:38:58.91 & -38: 0:41.90 & 2685 & 13.50$\pm$0.100 & 0.344 &  99.9\\
NGC 3783                       & 6dF J1139017-374419 &  11:39: 1.72 & -37:44:18.90 & 2817 &  8.71$\pm$0.316 & 0.360 &   1.8\\
GEMS N3783-8                   &                     &  11:38: 1.80 & -37:57: 0.10 & 2983 & 13.50$\pm$0.100 & 0.190 &  99.9\\
\hline
NGC 3923&&&&&&&\\ 
\hline
ESO 440- G 004                 & -                   &  11:45:41.88 & -28:21:59.50 & 1842 & 13.50$\pm$0.090 & 0.445 &   8.0\\
ESO 504- G 014                 & -                   &  11:46:23.53 & -27:15: 4.30 & 1645 & 13.50$\pm$0.106 & 0.638 &   0.0\\
NGC 3885                       & 6dF J1146465-275520 &  11:46:46.49 & -27:55:19.80 & 2094 &  8.44$\pm$0.983 & 0.430 &   0.2\\
6dF J1146465-280737            & 6dF J1146465-280737 &  11:46:46.50 & -28: 7:37.20 & 2157 & 13.50$\pm$0.100 & 0.386 &  99.9\\
6dF J1147555-281157            & 6dF J1147555-281157 &  11:47:55.51 & -28:11:56.70 & 1414 & 13.50$\pm$0.100 & 0.282 &  99.9\\
6dF J1148198-290400            & 6dF J1148198-290400 &  11:48:19.82 & -29: 4: 0.30 & 1894 & 13.50$\pm$0.100 & 0.242 &  99.9\\
UGCA 247                       & 6dF J1148456-281734 &  11:48:45.62 & -28:17:34.90 & 1978 & 13.50$\pm$0.340 & 0.200 &   6.8\\
ESO 504- G 017                 & 6dF J1148464-272245 &  11:48:46.31 & -27:22:45.00 & 1874 & 11.64$\pm$0.512 & 0.497 &   5.0\\
ESO 440- G 012                 & 6dF J1148584-282641 &  11:48:58.42 & -28:26:40.50 & 1540 & 13.50$\pm$0.351 & 0.150 &   0.0\\
NGC 3904                       & 6dF J1149132-291636 &  11:49:13.23 & -29:16:36.90 & 1685 &  7.76$\pm$0.353 & 0.254 &  -5.0\\
ESO 440- G 014                 & 6dF J1150032-284017 &  11:50: 3.20 & -28:40:17.10 & 1888 & 13.50$\pm$0.179 & 0.028 &   0.0\\
ESO 440- G 015                 & 6dF J1150117-283041 &  11:50:11.70 & -28:30:41.20 & 1865 & 12.57$\pm$0.394 & 0.055 &   0.2\\
ESO 440- G 016                 & 6dF J1150198-283231 &  11:50:19.84 & -28:32:31.30 & 2158 & 12.07$\pm$0.344 & 0.042 &  -2.0\\
2MASX J11503040-2852202        & 6dF J1150304-285220 &  11:50:30.40 & -28:52:20.30 & 1661 & 11.76$\pm$0.100 & 0.082 &  -5.0\\
NGC 3923                       & 6dF J1151017-284821 &  11:51: 1.74 & -28:48:21.20 & 1808 &  6.64$\pm$0.117 & 0.085 &  -4.8\\
6dF J1151122-271459            & 6dF J1151122-271459 &  11:51:12.15 & -27:14:59.20 & 1675 & 13.50$\pm$0.100 & 0.529 &  99.9\\
2MASX J11513759-2847291        & 6dF J1151376-284729 &  11:51:37.61 & -28:47:29.10 & 1841 & 12.97$\pm$0.328 & 0.129 &  -5.0\\
6dF J1151533-281047            & 6dF J1151533-281047 &  11:51:53.27 & -28:10:46.80 & 1423 & 13.50$\pm$0.100 & 0.228 &  99.9\\
2MASX J11521217-2912554        & 6dF J1152122-291256 &  11:52:12.19 & -29:12:55.80 & 1823 & 13.72$\pm$0.100 & 0.270 &   0.0\\
NGC 3936                       & 6dF J1152206-265421 &  11:52:20.59 & -26:54:21.20 & 2011 &  9.07$\pm$0.193 & 0.676 &   4.3\\
ESO 440- G 023                 & 6dF J1152334-290719 &  11:52:33.40 & -29: 7:19.40 & 1912 & 13.50$\pm$0.260 & 0.269 &   3.0\\
UGCA 250                       & 6dF J1153241-283311 &  11:53:24.06 & -28:33:11.40 & 1700 &  9.71$\pm$0.088 & 0.287 &   6.8\\
2MASX J11532725-2833064        & 6dF J1153273-283306 &  11:53:27.25 & -28:33: 6.00 & 1664 & 12.20$\pm$0.100 & 0.291 &  99.9\\
ESO 504- G 024                 & -                   &  11:53:37.89 & -26:59:44.90 & 1894 & 13.50$\pm$0.123 & 0.688 &   8.9\\
ESO 504- G 025                 & -                   &  11:53:50.64 & -27:21: 0.00 & 1637 & 13.50$\pm$0.797 & 0.584 &   8.7\\
ESO 504- G 028                 & 6dF J1154544-271505 &  11:54:54.41 & -27:15: 4.70 & 2077 & 11.95$\pm$0.173 & 0.672 &   6.9\\
ESO 440- G 030                 & -                   &  11:55:25.58 & -28:44: 8.30 & 1821 & 13.50$\pm$0.267 & 0.473 &   3.0\\
FLASH J115712.00-280934.8      & -                   &  11:57:12.00 & -28: 9:34.80 & 1837 & 13.50$\pm$0.100 & 0.663 &  99.9\\
ESO 504- G 030                 & 6dF J1157149-274200 &  11:57:14.90 & -27:42: 0.30 & 1841 & 13.50$\pm$0.115 & 0.733 &   7.6\\
$[$KK2000$]$ 47                & -                   &  11:57:30.77 & -28: 7:27.20 & 2125 & 13.50$\pm$0.100 & 0.695 &  99.9\\
\hline 
\end{tabular} 
\end{center}
\end{table*}

\begin{table*}
\begin{center}
\caption{Continued.}
\label{group_mems4}
\begin{tabular}{llcccccc}
\hline
Galaxy Name&6dFGS ID&RA (J2000)&Dec (J2000)&$\bar{v}$ (km s$^{-1}$)&$m_K$ (mag)&$R_{gc}$ (Mpc)&T-type\\
\hline 
\hline 
NGC 4636&&&&&&&\\
\hline
\hline 
NGC 4544                       & -                   &  12:35:36.60 &   3: 2: 4.30 & 1154 & 10.22$\pm$0.065 & 0.460 &   0.8\\
NGC 4580                       & -                   &  12:37:48.39 &   5:22: 6.70 & 1034 &  8.92$\pm$0.218 & 0.646 &   1.6\\
NGC 4586                       & -                   &  12:38:28.40 &   4:19: 8.70 &  794 &  8.63$\pm$0.107 & 0.422 &   1.0\\
NGC 4587                       & -                   &  12:38:35.44 &   2:39:26.70 &  901 & 10.54$\pm$0.080 & 0.296 &  -2.0\\
NGC 4600                       & -                   &  12:40:22.99 &   3: 7: 3.80 &  852 &  9.95$\pm$0.220 & 0.178 &  -1.9\\
VCC 1920                       & -                   &  12:42:21.25 &   2: 3:59.30 & 1311 & 12.94$\pm$0.854 & 0.236 &  -2.0\\
NGC 4630                       & -                   &  12:42:31.16 &   3:57:37.10 &  737 & 10.19$\pm$0.224 & 0.227 &   9.5\\
NGC 4636                       & -                   &  12:42:49.87 &   2:41:16.00 &  938 &  6.63$\pm$0.196 & 0.087 &  -4.8\\
VCC 1947                       & -                   &  12:42:56.32 &   3:40:35.30 &  974 & 11.19$\pm$0.355 & 0.156 &  -5.0\\
NGC 4643                       & -                   &  12:43:20.14 &   1:58:42.10 & 1273 &  7.47$\pm$0.189 & 0.249 &  -0.1\\
NGC 4665                       & -                   &  12:45: 5.97 &   3: 3:20.60 &  785 &  7.52$\pm$0.325 & 0.104 &  -0.1\\
NGC 4688                       & -                   &  12:47:46.46 &   4:20: 9.90 &  986 & 11.67$\pm$0.638 & 0.406 &   6.0\\
CGCG 043-030                   & -                   &  12:47:59.82 &   4:41:41.30 & 1023 & 12.99$\pm$0.870 & 0.482 &   0.0\\
NGC 4701                       & -                   &  12:49:11.57 &   3:23:19.40 &  723 &  9.96$\pm$0.244 & 0.357 &   5.6\\
NGC 4713                       & -                   &  12:49:57.87 &   5:18:41.10 &  652 &  9.97$\pm$0.288 & 0.669 &   6.7\\
NGC 4765                       & -                   &  12:53:14.42 &   4:27:47.20 &  716 & 10.94$\pm$0.146 & 0.678 &   0.2\\
NGC 4808                       & -                   &  12:55:48.95 &   4:18:14.80 &  778 &  9.15$\pm$0.086 & 0.799 &   6.0\\
\hline 
NGC 5044&&&&&&&\\
\hline
2MASX J13115849-1644541        & 6dF J1311585-164454 &  13:11:58.49 & -16:44:54.10 & 2907 & 12.03$\pm$0.569 & 0.396 &   0.0\\
NGC 5010                       & -                   &  13:12:26.35 & -15:47:52.30 & 2975 &  9.38$\pm$0.354 & 0.491 &  -0.9\\
MCG -03-34-014                 & -                   &  13:12:35.43 & -17:32:32.70 & 2760 &  8.94$\pm$0.101 & 0.594 &   5.0\\
NGC 5017                       & 6dF J1312545-164557 &  13:12:54.50 & -16:45:57.00 & 2451 &  9.37$\pm$0.166 & 0.288 &  -4.1\\
MCG -03-34-019                 & 6dF J1313055-162841 &  13:13: 5.48 & -16:28:41.30 & 1989 & 10.79$\pm$0.658 & 0.241 &  -1.9\\
MCG -03-34-020                 & -                   &  13:13:12.48 & -16: 7:50.10 & 2663 & 11.03$\pm$0.594 & 0.302 &  -2.8\\
LEDA 083798                    & -                   &  13:13:28.42 & -16:18:52.90 & 2682 & 13.50$\pm$0.134 & 0.220 &   7.0\\
MCG -03-34-022                 & -                   &  13:13:32.24 & -17: 4:43.40 & 2929 & 10.16$\pm$0.200 & 0.332 &   1.0\\
6dF J1313501-173048            & 6dF J1313501-173048 &  13:13:50.08 & -17:30:47.80 & 2223 & 13.50$\pm$0.100 & 0.518 &  99.9\\
NGC 5030                       & -                   &  13:13:54.15 & -16:29:27.40 & 2535 &  9.97$\pm$0.174 & 0.138 &  -1.1\\
2MASX J13135622-1616244        & 6dF J1313562-161624 &  13:13:56.23 & -16:16:24.40 & 2445 & 12.21$\pm$0.116 & 0.186 &  -3.0\\
2MASXi J1313594-162303         & -                   &  13:13:59.52 & -16:23: 3.80 & 2411 & 13.50$\pm$0.100 & 0.145 &  99.9\\
NGC 5031                       & -                   &  13:14: 3.22 & -16: 7:23.20 & 2839 &  9.33$\pm$0.095 & 0.238 &  -1.1\\
LEDA 083813                    & -                   &  13:14: 7.40 & -16:25:35.80 & 2661 & 13.50$\pm$0.121 & 0.121 &  -5.0\\
2MASX J13141733-1626189        & -                   &  13:14:17.36 & -16:26:19.50 & 2462 & 12.51$\pm$0.121 & 0.099 &  -5.0\\
MCG -03-34-025                 & 6dF J1314304-173201 &  13:14:30.42 & -17:32: 0.90 & 2517 & 11.70$\pm$0.100 & 0.511 &   0.2\\
2MASX J13143485-1629289        & 6dF J1314349-162929 &  13:14:34.86 & -16:29:28.90 & 2397 & 12.28$\pm$0.109 & 0.054 &  -5.0\\
NGC 5035                       & -                   &  13:14:49.23 & -16:29:33.70 & 2181 &  9.76$\pm$0.431 & 0.028 &  -0.9\\
NGC 5037                       & -                   &  13:14:59.37 & -16:35:25.10 & 1887 &  8.60$\pm$0.262 & 0.030 &   1.1\\
NGC 5038                       & -                   &  13:15: 2.13 & -15:57: 6.50 & 2222 &  9.57$\pm$0.247 & 0.293 &  -1.8\\
2MASX J13150409-1623391        & 6dF J1315041-162339 &  13:15: 4.08 & -16:23:39.30 & 1977 & 12.67$\pm$0.496 & 0.070 &  -5.0\\
MCG -03-34-033                 & -                   &  13:15:17.57 & -16:29:10.20 & 3442 & 11.26$\pm$0.627 & 0.046 &  -0.9\\
NGC 5044                       & -                   &  13:15:23.97 & -16:23: 7.90 & 2704 &  7.84$\pm$0.394 & 0.091 &  -4.9\\
2MASX J13153203-1628509        & 6dF J1315320-162851 &  13:15:32.04 & -16:28:51.10 & 3249 & 12.02$\pm$0.107 & 0.075 &  -3.0\\
NGC 5046                       & -                   &  13:15:45.12 & -16:19:36.60 & 2214 & 10.36$\pm$0.216 & 0.142 &  -5.0\\
NGC 5049                       & -                   &  13:15:59.30 & -16:23:49.80 & 2744 &  9.58$\pm$0.574 & 0.145 &  -2.0\\
2MASX J13164875-1620397        & 6dF J1316488-162040 &  13:16:48.75 & -16:20:39.70 & 2619 & 12.07$\pm$0.130 & 0.250 &  -1.5\\
MCG -03-34-040                 & 6dF J1316562-163535 &  13:16:56.23 & -16:35:34.70 & 2112 & 12.90$\pm$0.108 & 0.250 &   7.8\\
NGC 5054                       & -                   &  13:16:58.49 & -16:38: 5.50 & 1741 &  7.82$\pm$0.556 & 0.258 &   4.0\\
MCG -03-34-041                 & -                   &  13:17: 6.13 & -16:15: 7.90 & 2628 & 10.76$\pm$1.477 & 0.303 &   5.1\\
LCSB S1851O                    & 6dF J1317364-163225 &  13:17:36.37 & -16:32:25.30 & 2919 & 12.17$\pm$0.100 & 0.332 &  99.9\\
GEMS N5044-1                   &                     &  13:14: 9.93 & -16:41:41.60 & 3074 & 13.50$\pm$0.100 & 0.132 &  99.9\\
\hline 
\end{tabular} 
\end{center}
\end{table*}

\begin{table*}
\begin{center}
\caption{Continued.}
\label{group_mems5}
\begin{tabular}{llcccccc}
\hline
\hline 
Galaxy Name&6dFGS ID&RA (J2000)&Dec (J2000)&$\bar{v}$ (km s$^{-1}$)&$m_K$ (mag)&$R_{gc}$ (Mpc)&T-type\\
\hline 
\hline 
HCG 90&&&&&&&\\
\hline
ESO 466- G 025                 & 6dF J2158245-321411 &  21:58:24.45 & -32:14:11.50 & 2497 & 12.00$\pm$0.271 & 0.608 &   3.0\\
2dFGRS S407Z170                & -                   &  21:59: 0.59 & -31:45: 3.20 & 2492 & 13.50$\pm$0.170 & 0.498 &   0.0\\
ESO 466- G 029                 & -                   &  21:59:15.46 & -31:13:42.70 & 2772 & 13.50$\pm$0.200 & 0.630 &   0.0\\
2dFGRS S407Z162                & -                   &  21:59:15.50 & -31:13:18.60 & 2832 & 13.50$\pm$0.100 & 0.633 &   0.0\\
NGC 7163                       & 6dF J2159204-315259 &  21:59:20.45 & -31:52:59.30 & 2737 & 10.06$\pm$0.483 & 0.434 &   2.1\\
MCG -05-52-001                 & -                   &  21:59:56.62 & -31:27:42.60 & 2540 & 11.85$\pm$0.224 & 0.451 &   0.0\\
ESO 404- G 018                 & -                   &  22: 1:10.16 & -32:34:43.69 & 2268 & 13.50$\pm$0.252 & 0.428 &   6.7\\
ESO 466- G 036                 & -                   &  22: 1:20.46 & -31:31:46.90 & 2559 & 11.16$\pm$0.831 & 0.283 &   0.8\\
2dFGRS S408Z202                & -                   &  22: 1:21.44 & -31:31:52.90 & 2457 & 13.50$\pm$0.100 & 0.281 &  99.9\\
DUKST 466-064                  & -                   &  22: 1:29.79 & -31:57:46.90 & 2760 & 13.50$\pm$0.175 & 0.095 &   0.0\\
NGC 7172                       & 6dF J2202019-315211 &  22: 2: 1.87 & -31:52:11.10 & 2557 &  8.39$\pm$0.172 & 0.044 &   1.2\\
NGC 7173                       & -                   &  22: 2: 3.19 & -31:58:25.30 & 2497 &  9.07$\pm$0.516 & 0.023 &  -4.2\\
2dFGRS S407Z097                & -                   &  22: 2: 4.81 & -31:52:13.50 & 2384 & 13.50$\pm$0.100 & 0.043 &   0.0\\
$[$PCM2000$]$ 31                   & -                   &  22: 2: 5.33 & -31:58:56.50 & 2674 & 13.50$\pm$0.100 & 0.028 &  99.9\\
NGC 7174                       & 6dF J2202065-315934 &  22: 2: 6.49 & -31:59:33.90 & 2746 & 13.50$\pm$0.826 & 0.034 &   2.5\\
IRAS  21592-3214               & -                   &  22: 2: 7.50 & -31:59:28.00 & 2778 & 13.50$\pm$0.826 & 0.034 &   2.5\\
NGC 7176                       & 6dF J2202085-315923 &  22: 2: 8.45 & -31:59:23.30 & 2503 &  8.09$\pm$0.177 & 0.033 &  -4.8\\
ESO 466- G 043                 & -                   &  22: 2:14.95 & -31:13:12.40 & 2608 & 11.60$\pm$0.375 & 0.452 &   4.7\\
ESO 466- G 044                 & -                   &  22: 2:15.88 & -31:45:23.30 & 2818 & 12.25$\pm$0.173 & 0.117 &  -2.7\\
2dFGRS S408Z175                & -                   &  22: 2:15.89 & -31:13: 6.10 & 2679 & 13.50$\pm$0.100 & 0.453 &  99.9\\
2dFGRS S407Z090                & -                   &  22: 2:16.07 & -31:57:11.80 & 2541 & 13.50$\pm$0.180 & 0.029 &   0.0\\
ESO 466- G 046                 & 6dF J2202440-315926 &  22: 2:44.01 & -31:59:26.20 & 2260 & 11.28$\pm$0.204 & 0.106 &   0.1\\
NGC 7187                       & 6dF J2202445-324811 &  22: 2:44.49 & -32:48:11.60 & 2740 & 13.50$\pm$0.425 & 0.553 &  -1.0\\
ESO 466- G 047                 & -                   &  22: 2:46.05 & -31:57:19.70 & 2556 & 13.50$\pm$0.137 & 0.107 &   5.0\\
DUKST 404-032                  & 6dF J2202502-323437 &  22: 2:50.17 & -32:34:36.80 & 2221 & 13.50$\pm$0.150 & 0.418 &   0.0\\
2dFGRS S408Z047                & -                   &  22: 3:36.51 & -32:26:54.19 & 2649 & 13.50$\pm$0.118 & 0.399 &   0.0\\
2dFGRS S408Z045                & -                   &  22: 3:41.65 & -32:44:44.40 & 2871 & 13.50$\pm$0.236 & 0.566 &   0.0\\
ESO 404- G 027                 & 6dF J2203478-321706 &  22: 3:47.84 & -32:17: 6.00 & 2611 & 10.80$\pm$0.247 & 0.345 &   4.8\\
ESO 466- G 051                 & 6dF J2203484-315721 &  22: 3:48.36 & -31:57:20.90 & 2674 & 10.70$\pm$0.254 & 0.269 &  -0.9\\
ESO 404- G 028                 & 6dF J2204152-323615 &  22: 4:15.23 & -32:36:14.80 & 2377 & 11.97$\pm$0.324 & 0.539 &   0.4\\
2dFGRS S408Z018                & -                   &  22: 5:13.33 & -32:22:32.80 & 2800 & 13.50$\pm$0.100 & 0.563 &   0.0\\
2dFGRS S409Z039                & -                   &  22: 5:21.08 & -31:59:59.70 & 2401 & 13.50$\pm$0.305 & 0.513 &   2.5\\
2dFGRS S408Z283                & -                   &  22: 5:26.54 & -31:33:34.10 & 2618 & 13.50$\pm$0.135 & 0.578 &   0.0\\
2dFGRS S409Z237                & -                   &  22: 6:18.58 & -32:10:32.00 & 2742 & 13.50$\pm$0.099 & 0.679 &   0.0\\
2dFGRS S409Z216                & -                   &  22: 7:15.58 & -32:12:33.20 & 2393 & 13.50$\pm$0.071 & 0.829 &   0.0\\
2dFGRS S409Z198                & -                   &  22: 7:53.84 & -31:56:55.20 & 2651 & 13.50$\pm$0.135 & 0.912 &   0.0\\
2dFGRS S409Z193                & -                   &  22: 8: 6.79 & -31:44:57.70 & 2486 & 13.50$\pm$0.072 & 0.953 &   0.0\\
2MASX J22090574-3147414        & -                   &  22: 9: 5.76 & -31:47:41.70 & 2381 & 12.52$\pm$0.189 & 0.104 &   2.0\\
\hline 
IC 1459&&&&&&&\\
\hline
NGC 7418                       & 6dF J2256361-370148 &  22:56:36.13 & -37: 1:47.80 & 1417 &  8.91$\pm$0.405 & 0.250 &   5.8\\
IC 5269B                       & 6dF J2256367-361459 &  22:56:36.72 & -36:14:59.09 & 1638 & 10.94$\pm$0.377 & 0.142 &   5.9\\
NGC 7418A                      & -                   &  22:56:41.15 & -36:46:21.20 & 2102 & 13.50$\pm$0.497 & 0.131 &   6.6\\
IC 5264                        & -                   &  22:56:53.04 & -36:33:14.99 & 1940 &  9.36$\pm$0.213 & 0.032 &   2.2\\
NGC 7421                       & 6dF J2256543-372050 &  22:56:54.33 & -37:20:50.70 & 1801 &  9.53$\pm$0.478 & 0.394 &   4.0\\
IC 1459                        & 6dF J2257106-362744 &  22:57:10.61 & -36:27:44.20 & 1713 &  6.93$\pm$0.441 & 0.028 &  -5.0\\
2MASX J22571092-3640103        & -                   &  22:57:10.92 & -36:40:10.40 & 1945 & 13.05$\pm$0.134 & 0.071 &  -3.5\\
ESO 406- G 031                 & 6dF J2257408-352349 &  22:57:40.76 & -35:23:49.40 & 1592 & 13.50$\pm$0.162 & 0.537 &   1.9\\
IC 5269                        & -                   &  22:57:43.66 & -36: 1:34.40 & 1967 &  9.56$\pm$0.366 & 0.245 &  -1.6\\
IC 5270                        & 6dF J2257549-355129 &  22:57:54.86 & -35:51:28.50 & 1929 &  9.70$\pm$0.396 & 0.328 &   5.7\\
\hline 
NGC 7714&&&&&&&\\
\hline
SDSS J233631.29-002943.3       &  -                  &  23:36:31.29 &   0:29:43.30 & 2495 & 13.50$\pm$0.100 & 0.125 &  99.9\\
NGC 7716                       &  -                  &  23:36:31.46 &   0:17:50.30 & 2571 &  9.46$\pm$0.207 & 0.010 &   2.9\\
SHOC 608                       &  -                  &  23:36:46.84 &   0:37:24.50 & 2633 & 13.50$\pm$0.100 & 0.216 &  99.9\\
UGC 12709                      &  -                  &  23:37:24.02 &   0:23:30.10 & 2682 & 13.50$\pm$0.129 & 0.155 &   8.3\\
\hline 
\hline 
\end{tabular} 
\end{center}
\end{table*}

\section[]{Spatial Distribution of Groups}
\label{group_pics}

\begin{figure*}
\begin{center}

    \resizebox{40pc}{!}{
     \rotatebox{-90}{
	\includegraphics{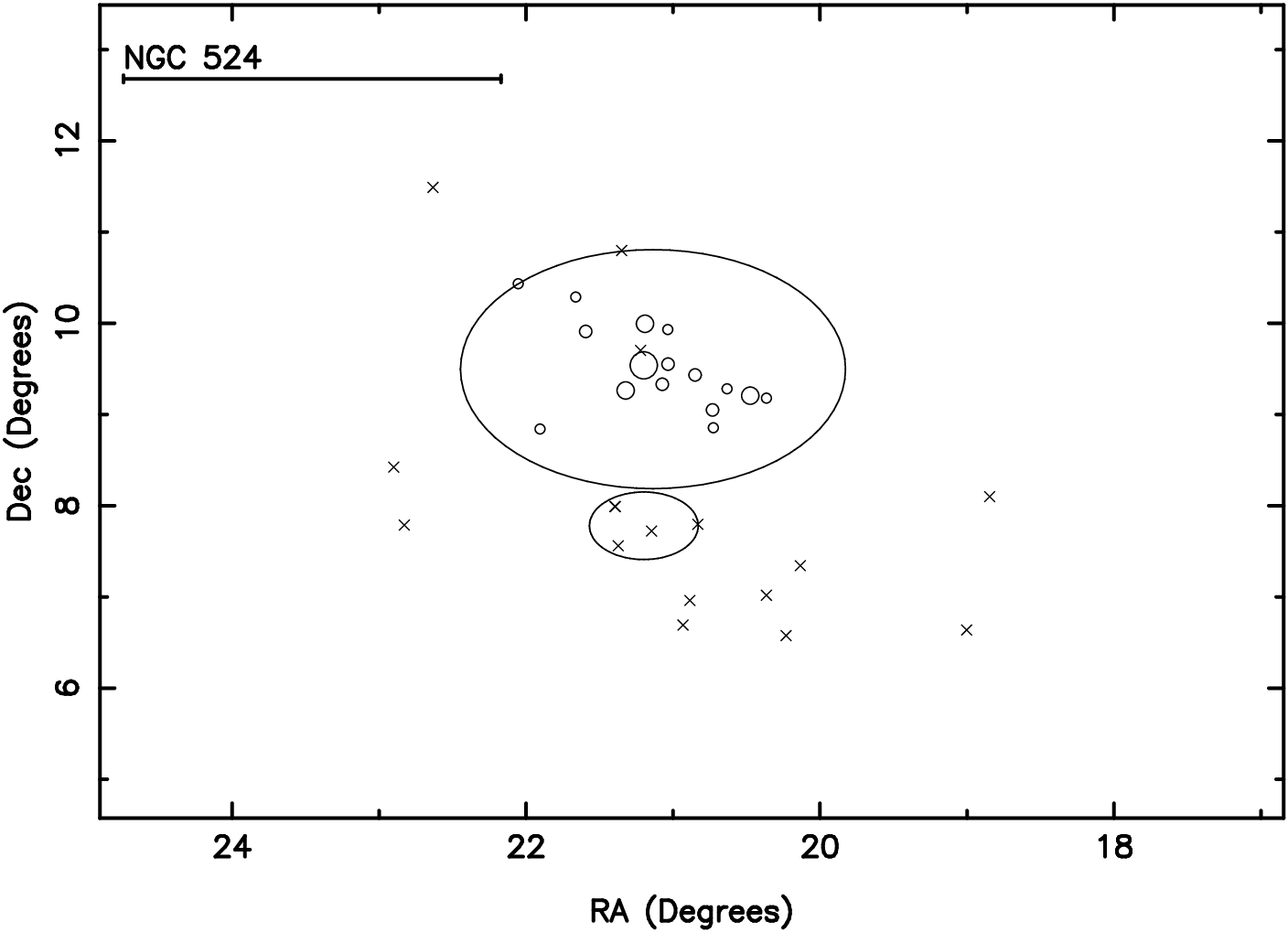} 
    }
     \rotatebox{-90}{
	\includegraphics{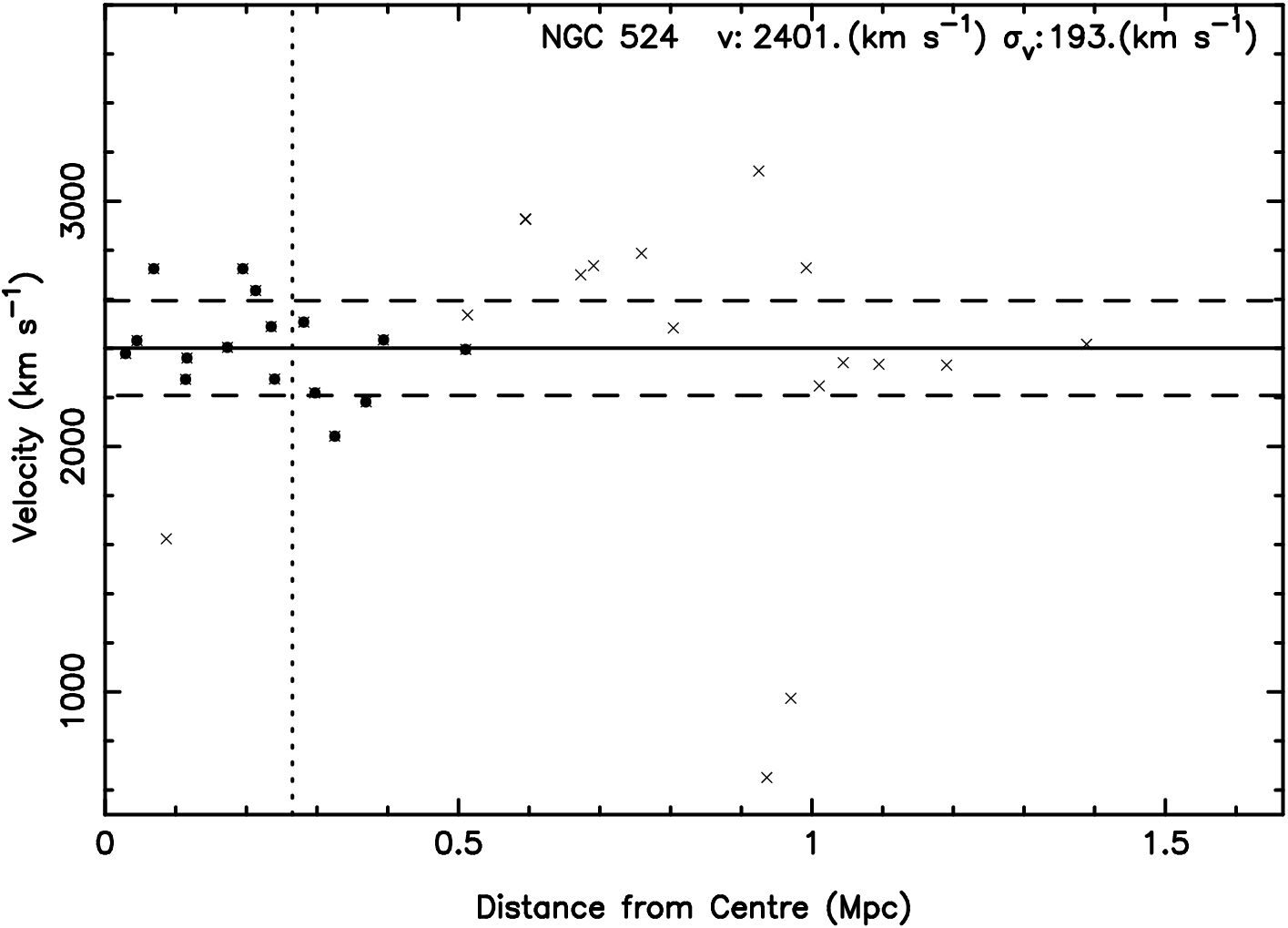} 
    }
}

     \resizebox{40pc}{!}{
     \rotatebox{-90}{
	\includegraphics{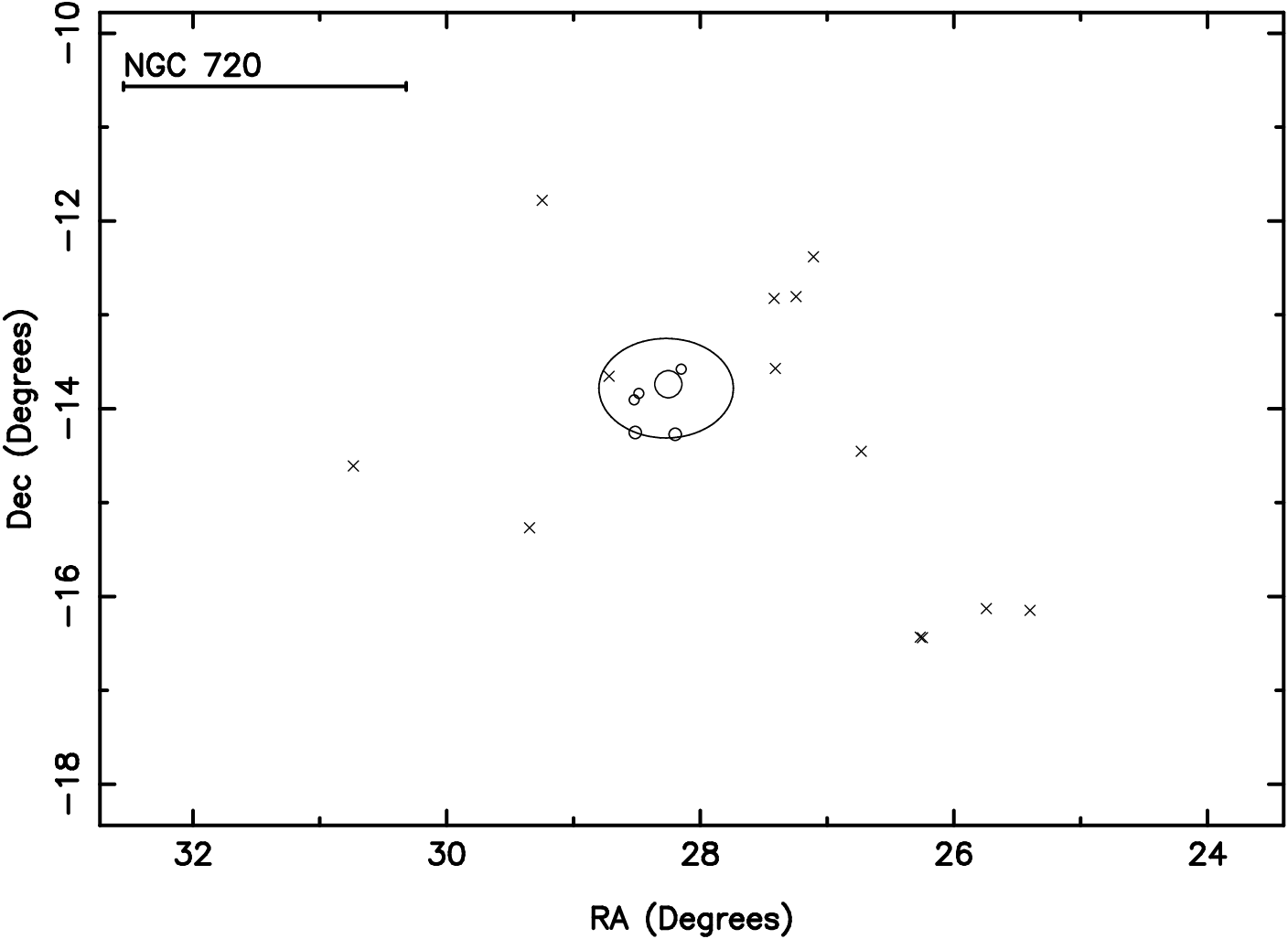}
}  
     \rotatebox{-90}{
	\includegraphics{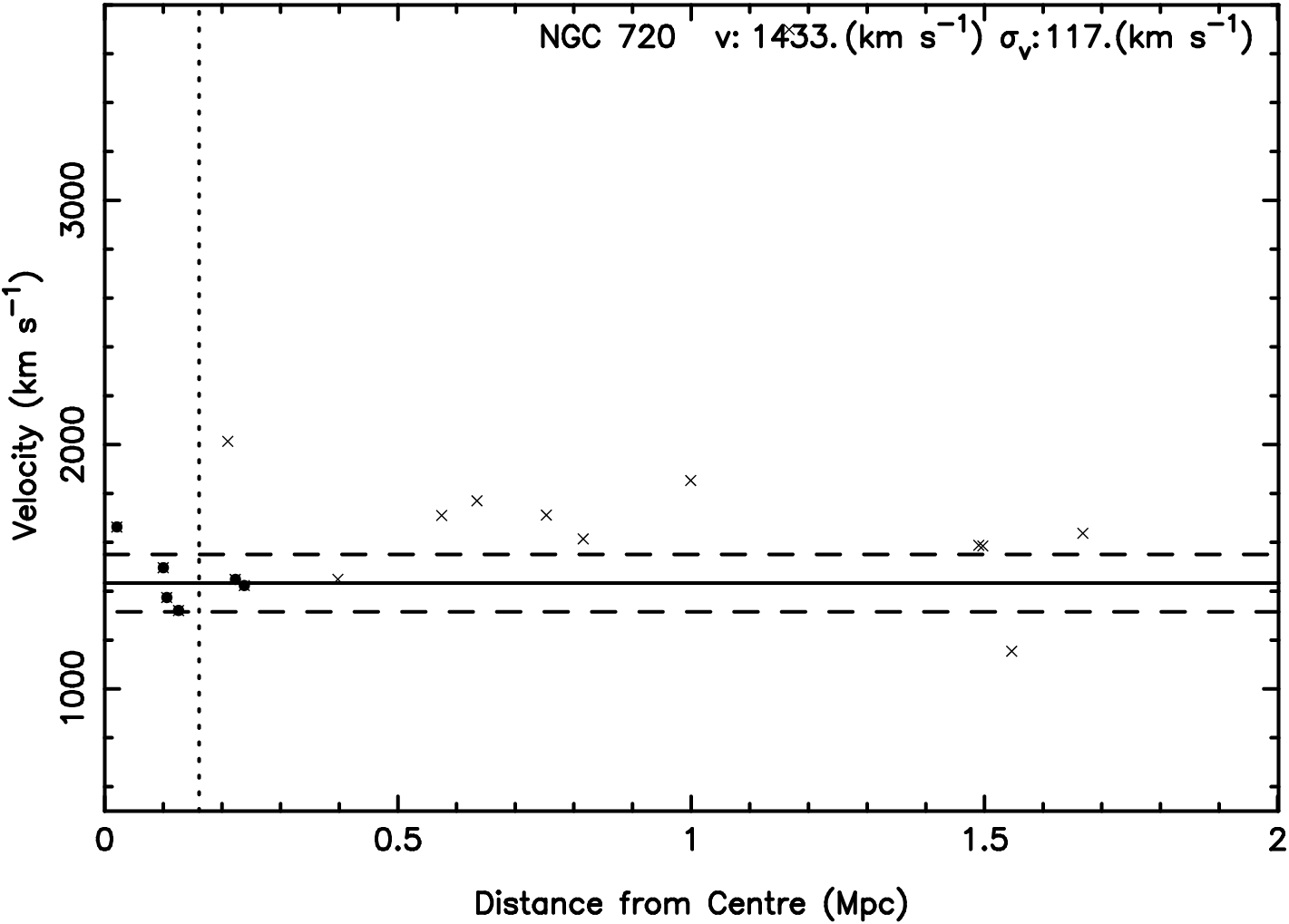}
} 
}  
   
	\resizebox{40pc}{!}{
     \rotatebox{-90}{
	\includegraphics{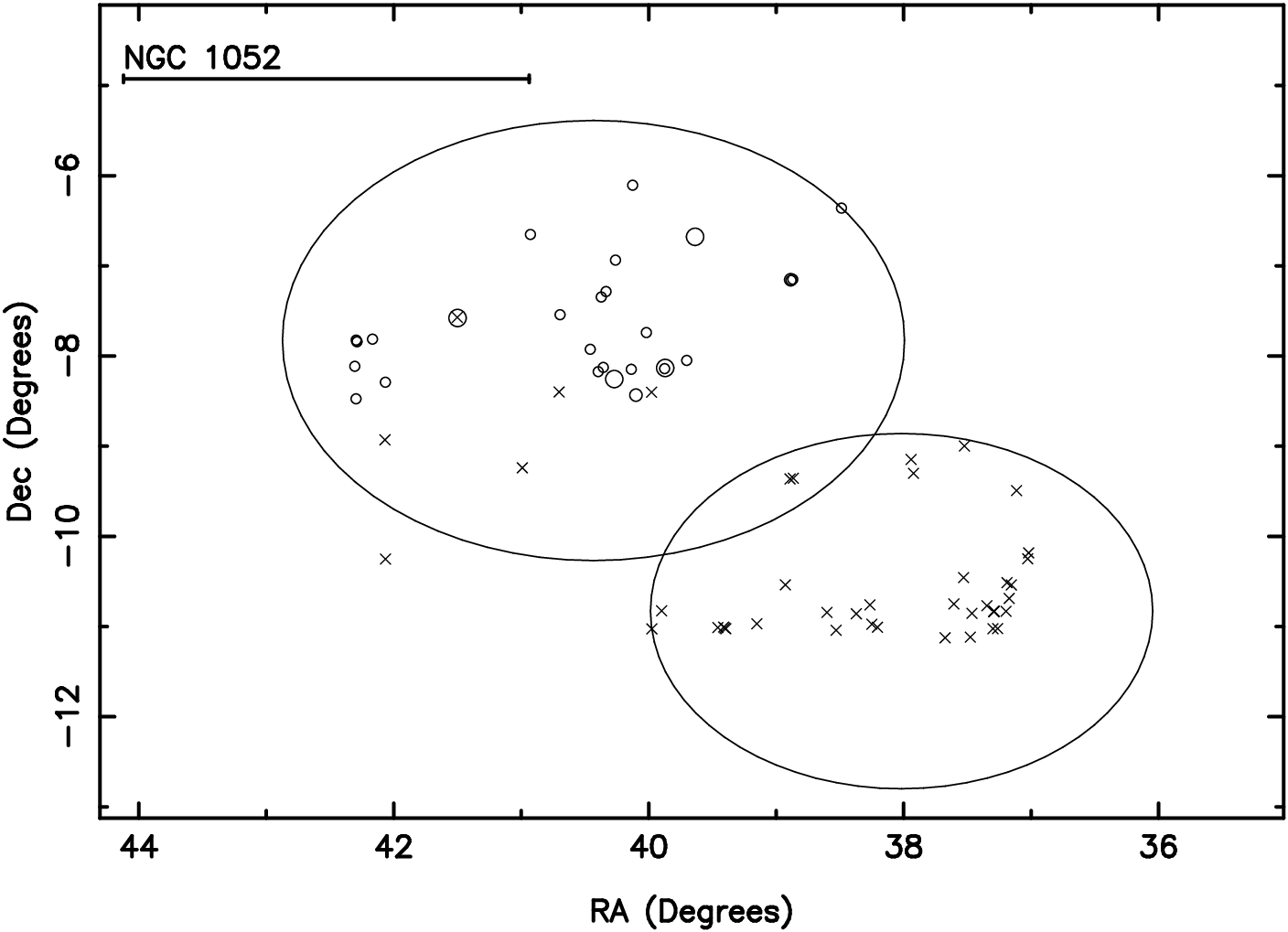} 
    }
     \rotatebox{-90}{
	\includegraphics{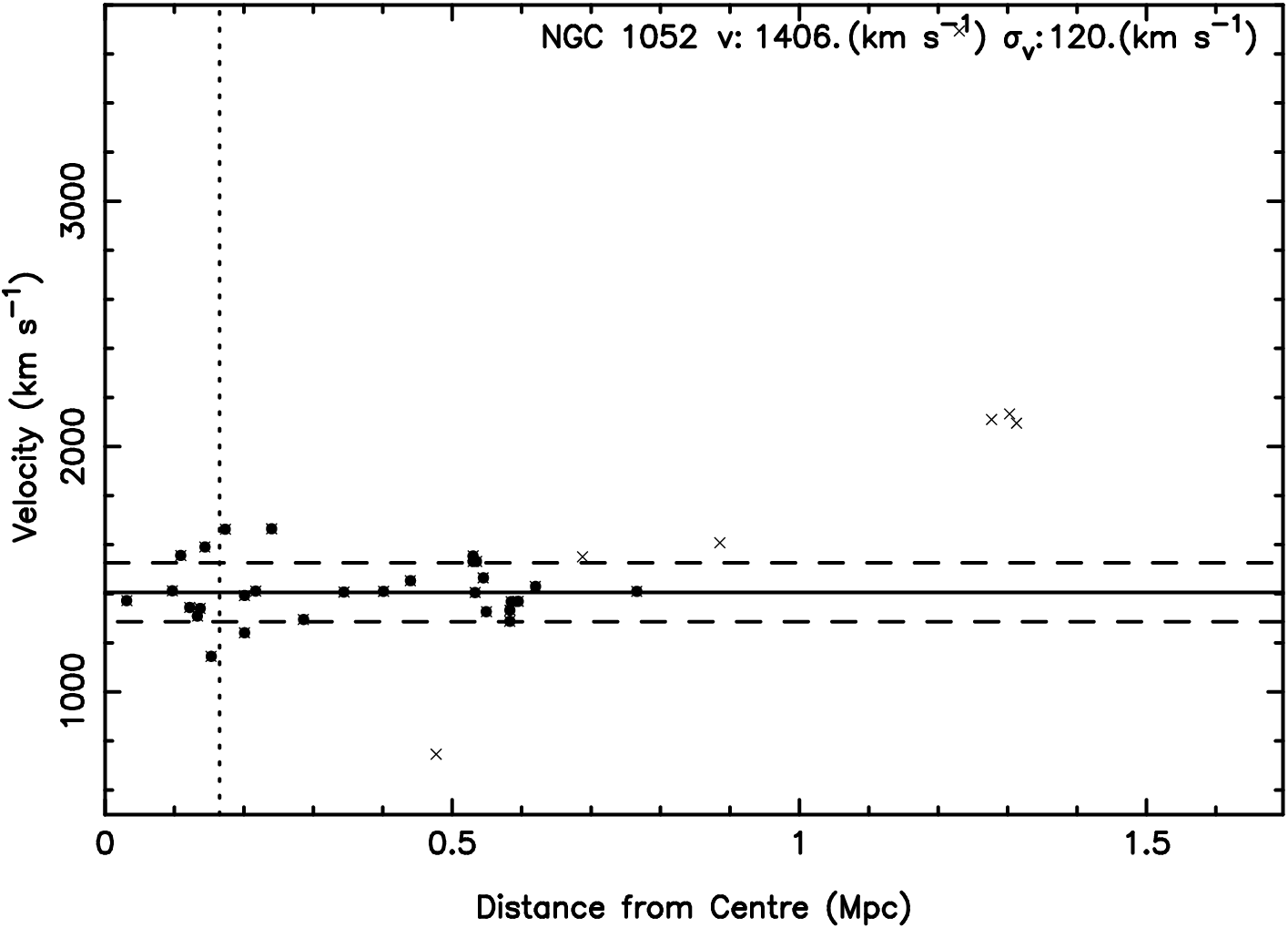} 
    }
}

  \end{center}
\caption {The left-hand panel indicates the spatial distribution of 
the galaxies in each region. The ellipses mark the maximum extent of
all groups determined by FOF.  The circles indicate group members
defined by FOF, scaled by $M_K$
.  The crosses indicate other galaxies in the same field.  The line in
the top left-hand corner indicates 1 Mpc at the distance of the group
being studied.  The right-hand panel shows the velocity-distance plots
for each group, the centroid defined by FOF.  The solid line marks the
mean velocity while the dashed lines indicate the velocity dispersion.
The vertical dotted line marks the $r_{500}$ radius from
Table~\ref{fof_groups}.  Solid points mark the group members defined
by FOF (shown as open circles in the left-hand panels) and crosses
indicate other galaxies in the same region.}
\end{figure*}

 \begin{figure*}
\begin{center}
\resizebox{40pc}{!}{
     \rotatebox{-90}{ 
	\includegraphics{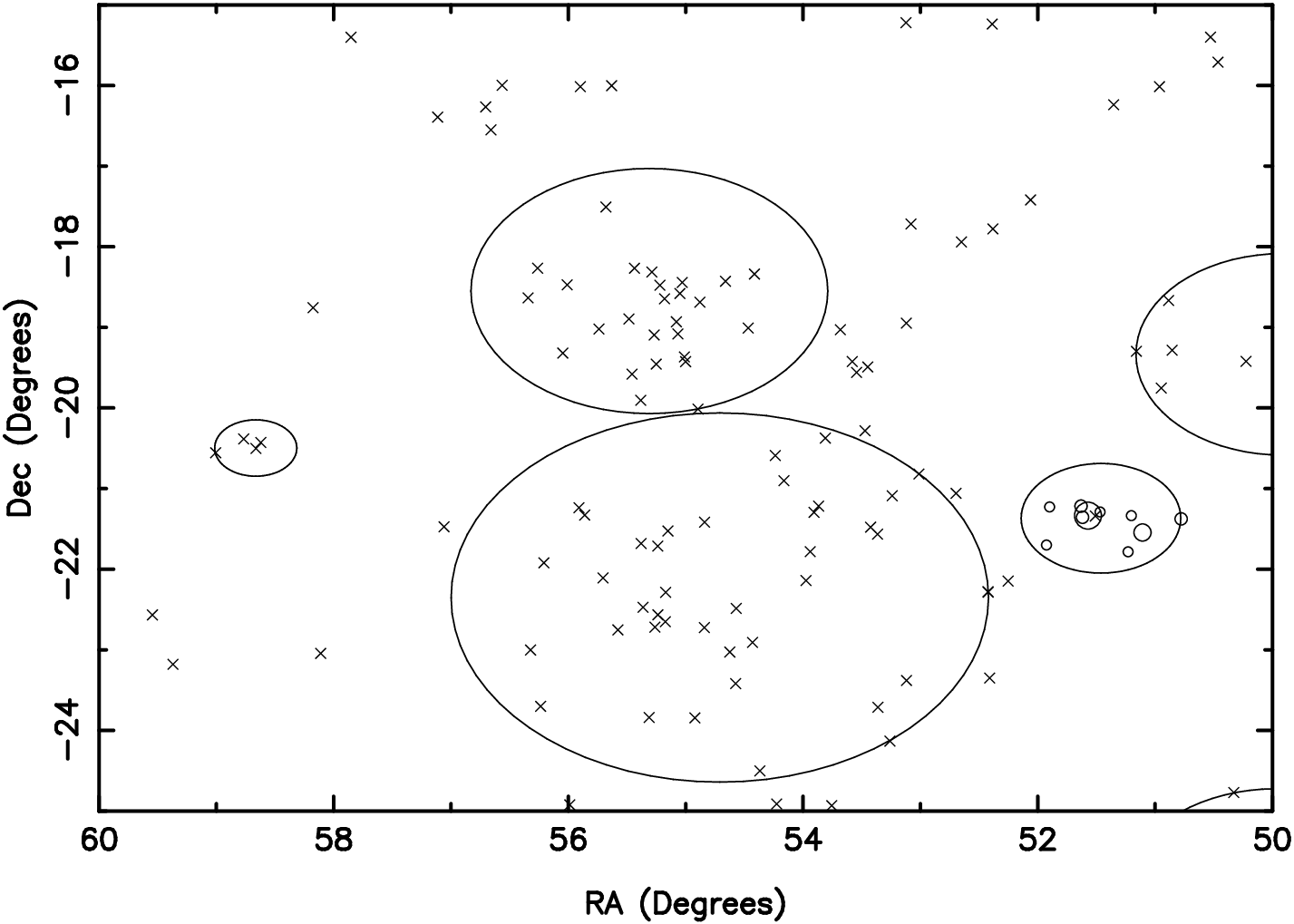} 
    }
     \rotatebox{-90}{ 
	\includegraphics{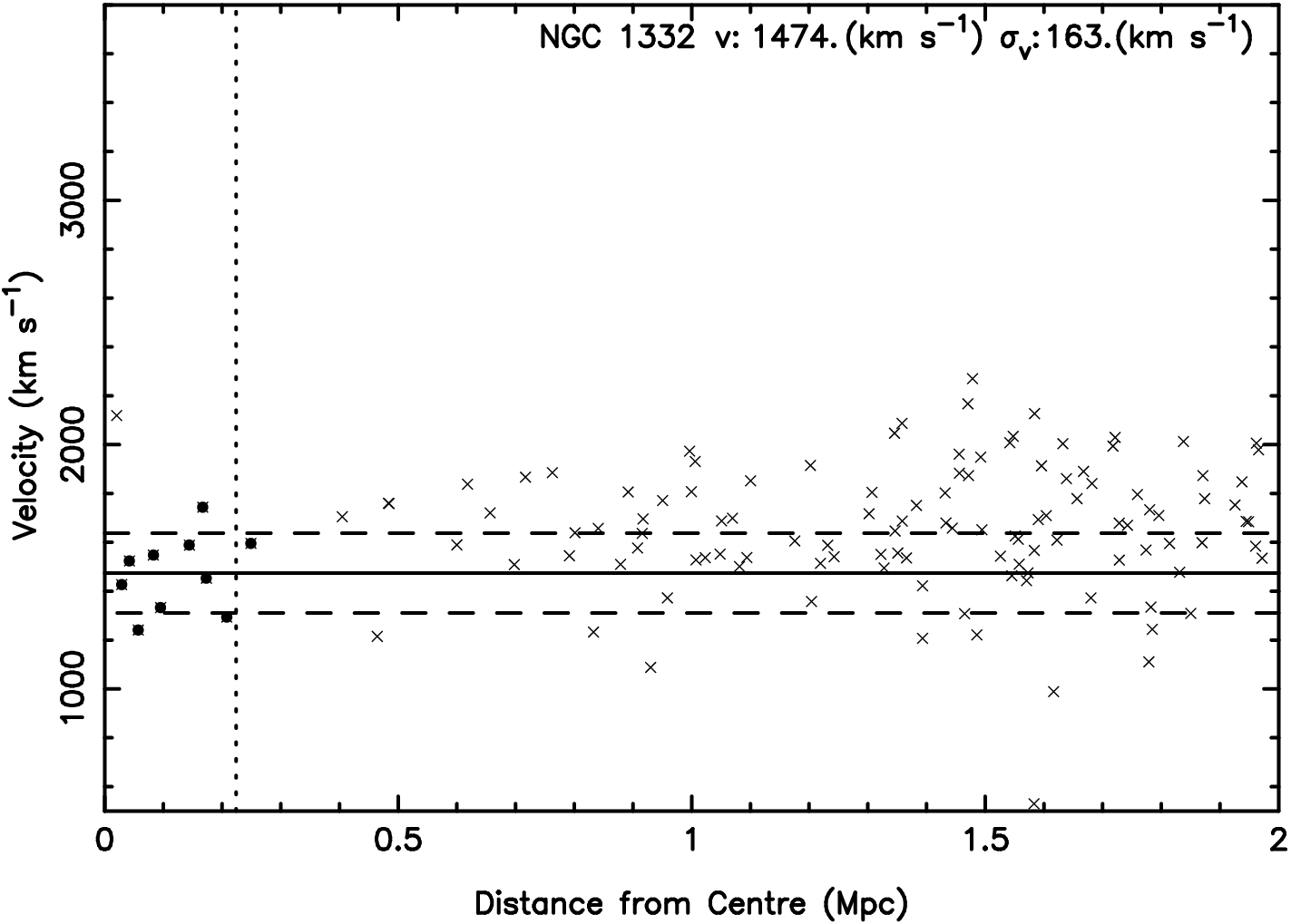} 
    }
}
     \resizebox{40pc}{!}{
     \rotatebox{-90}{
	\includegraphics{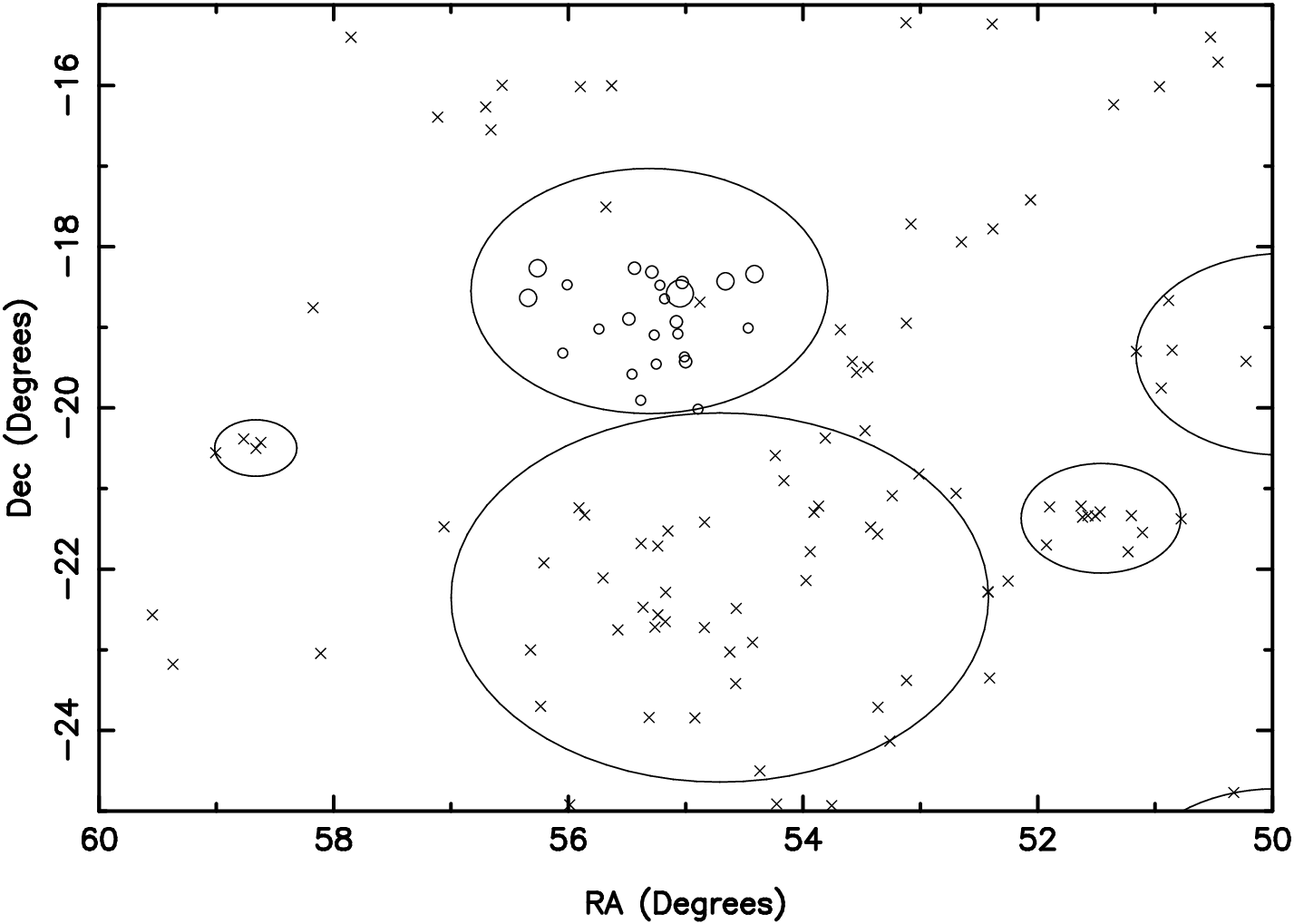}
}  
     \rotatebox{-90}{
	\includegraphics{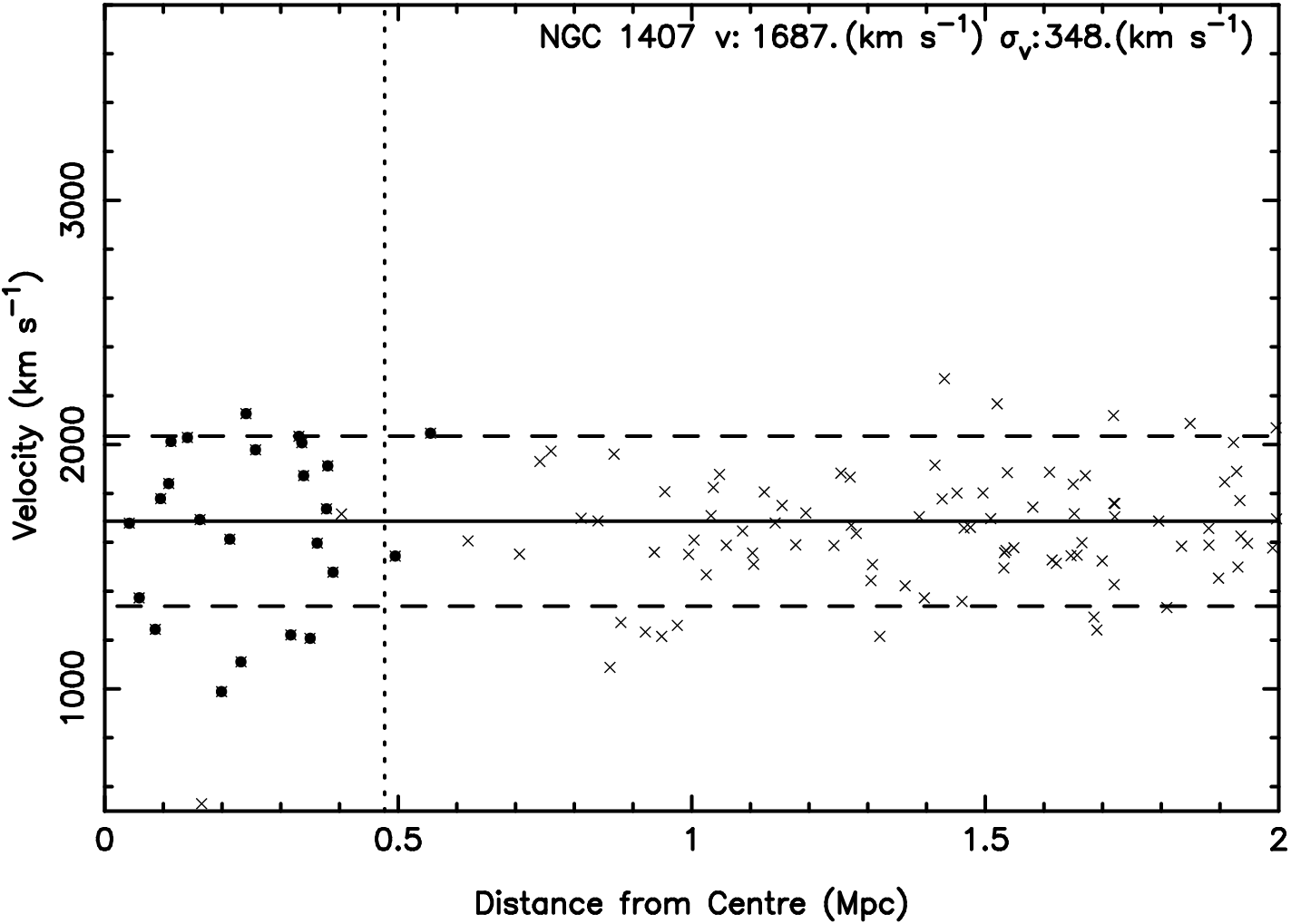}
}    
}  
    \resizebox{40pc}{!}{
     \rotatebox{-90}{
	\includegraphics{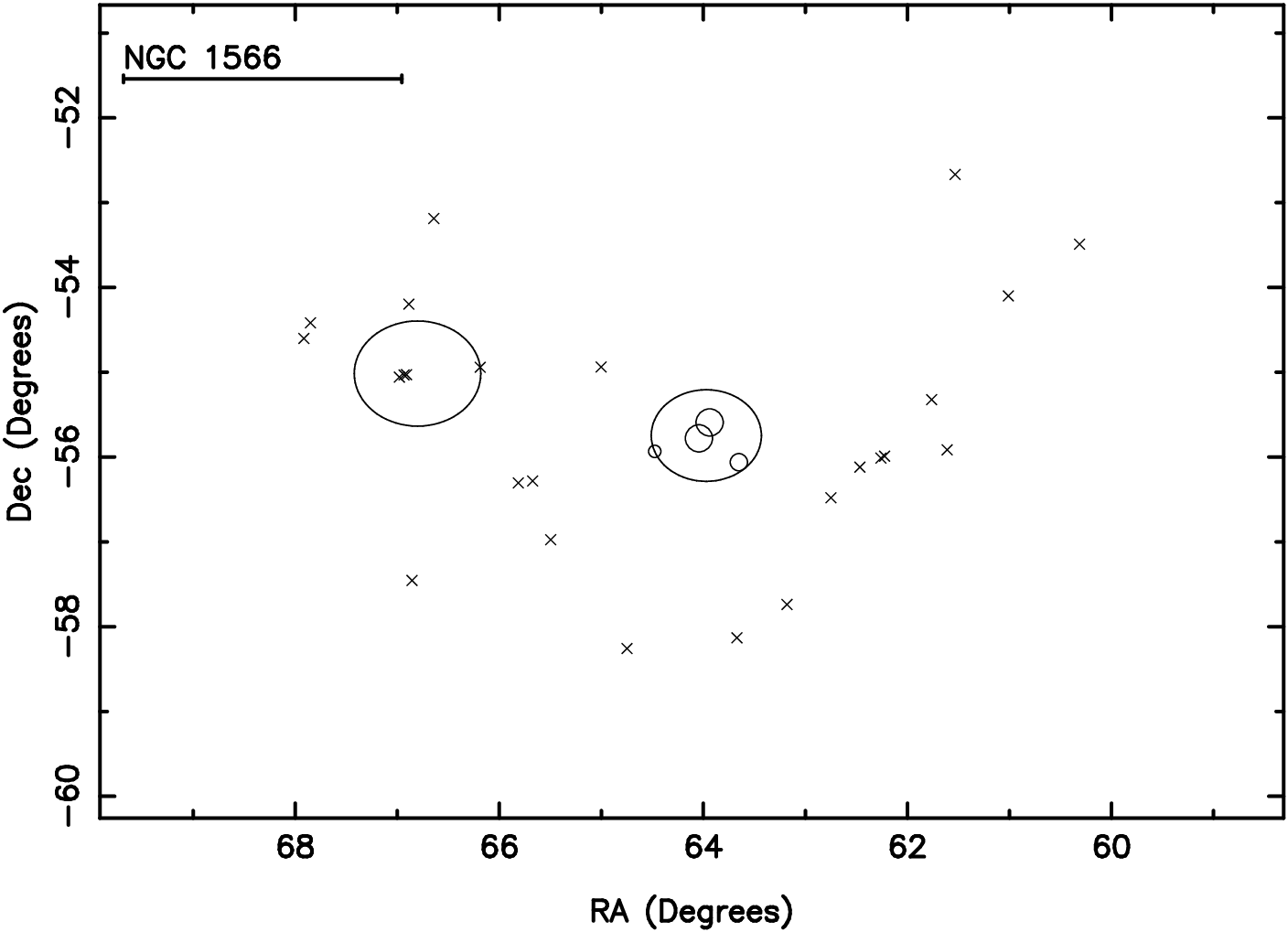} 
}
    \rotatebox{-90}{
	\includegraphics{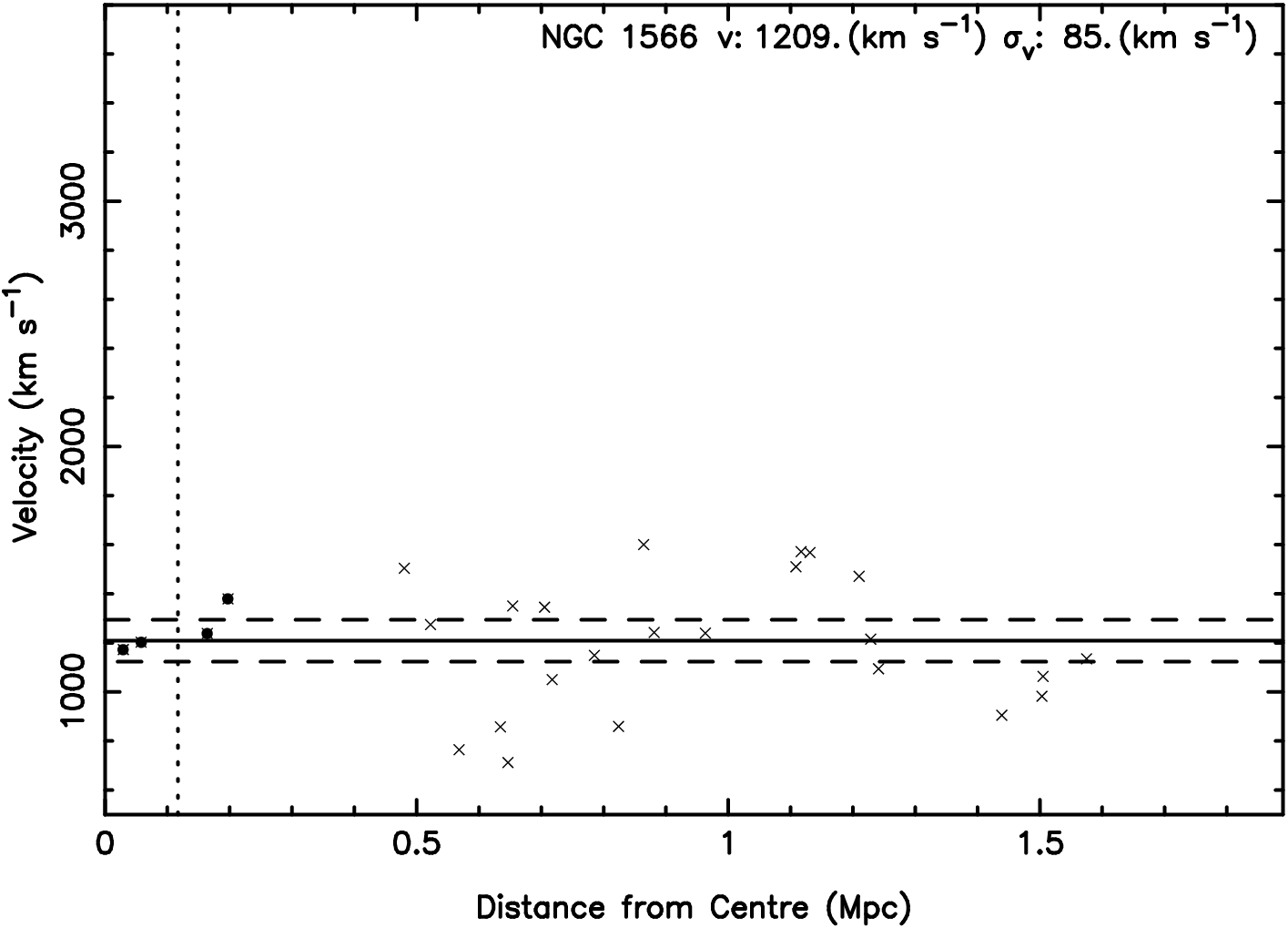} 
}
}

  \end{center}
\caption {Continued.}
\end{figure*}

\begin{figure*}
\begin{center}
    \resizebox{40pc}{!}{
     \rotatebox{-90}{
	\includegraphics{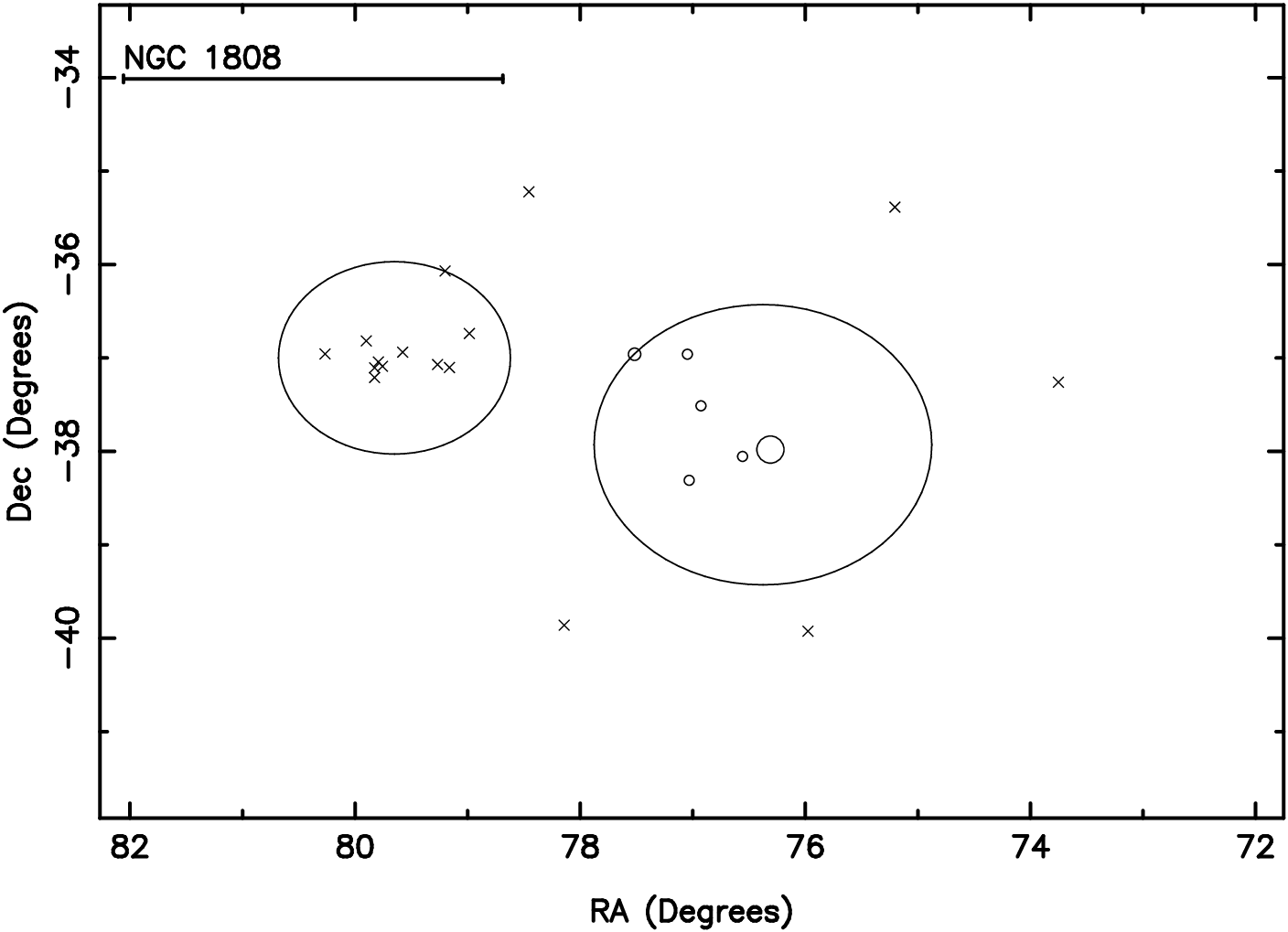} 
    }
     \rotatebox{-90}{
	\includegraphics{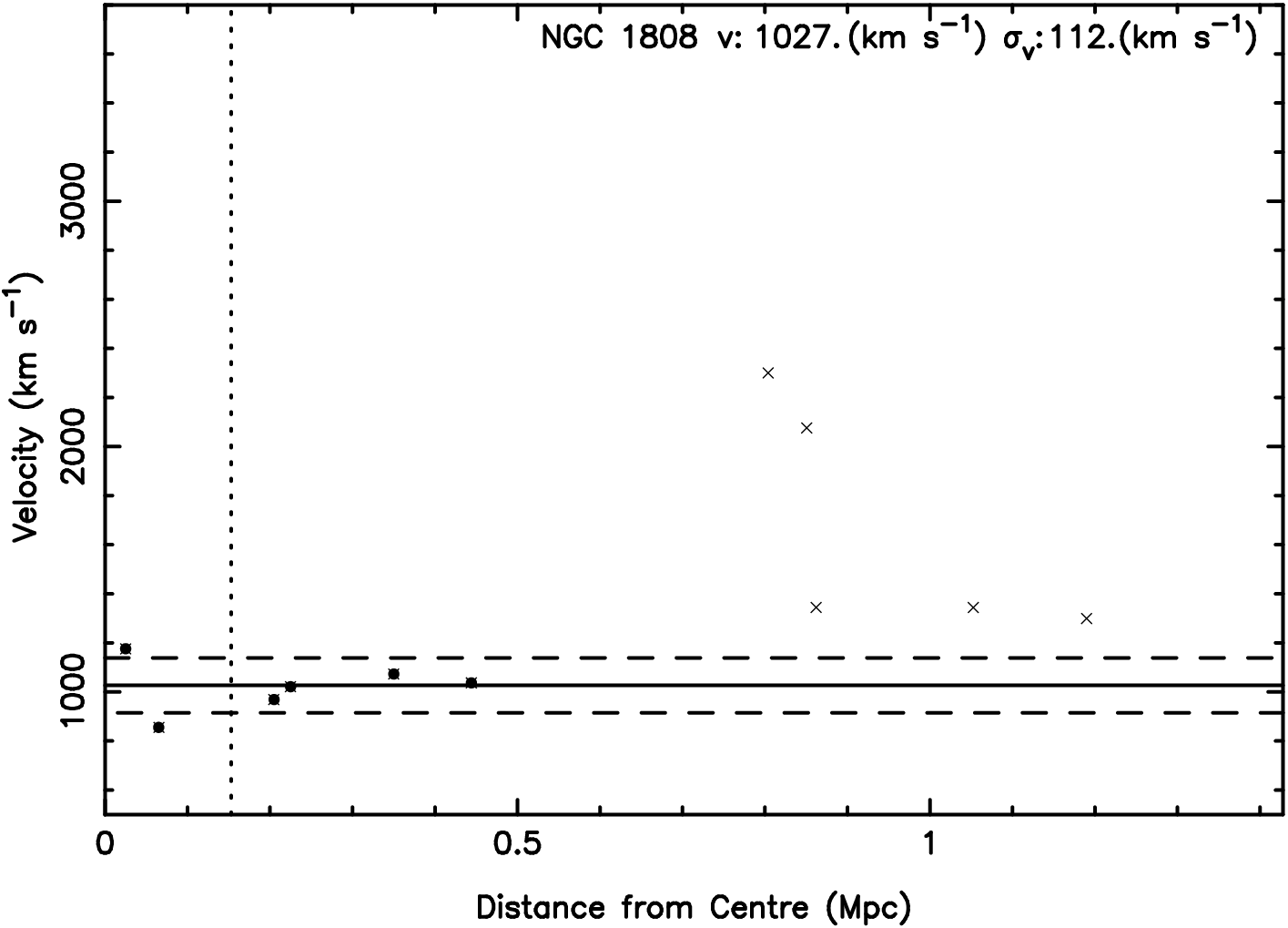} 
    }
}

    \resizebox{40pc}{!}{
     \rotatebox{-90}{
	\includegraphics{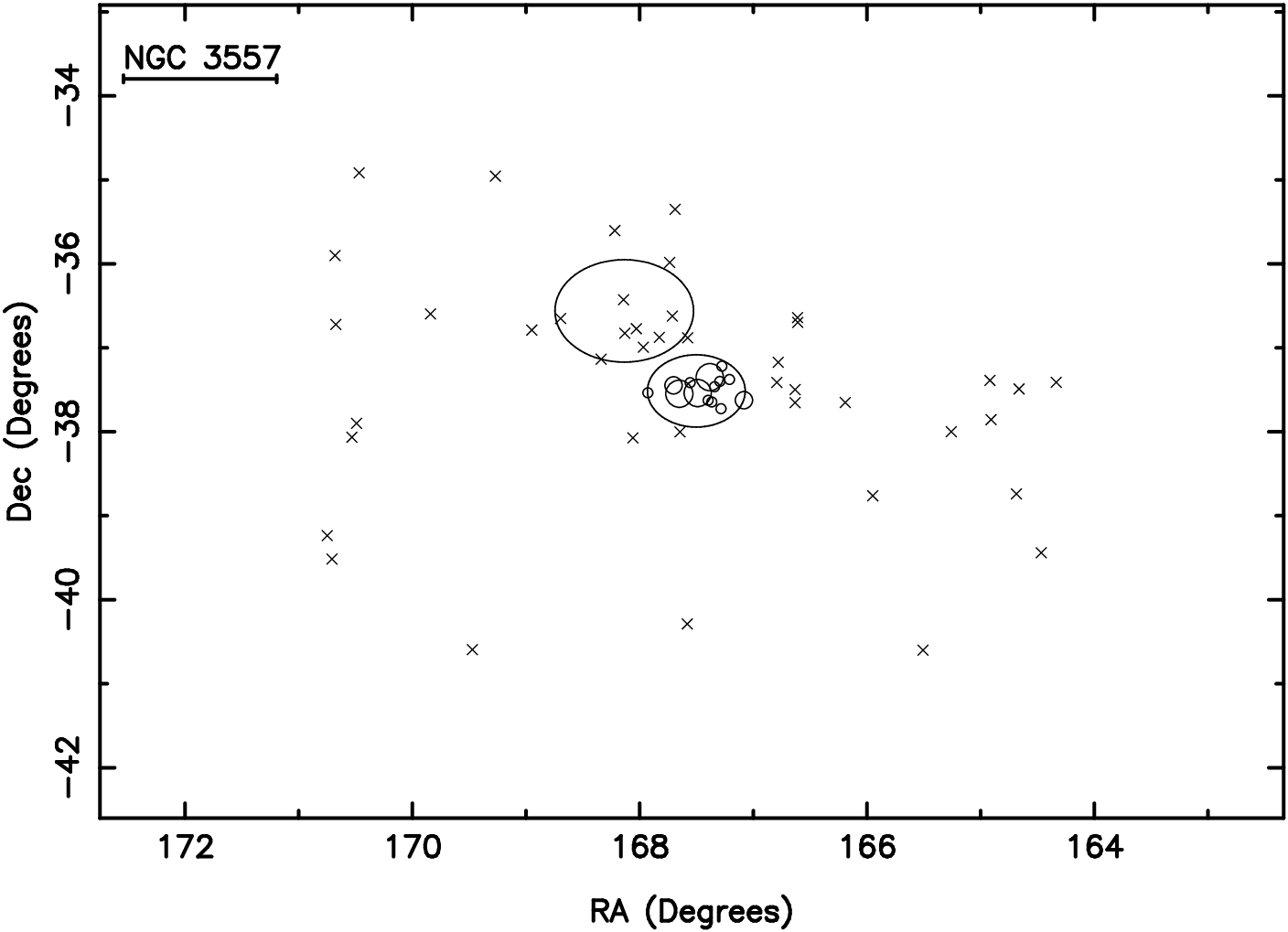}
} 
     \rotatebox{-90}{
	\includegraphics{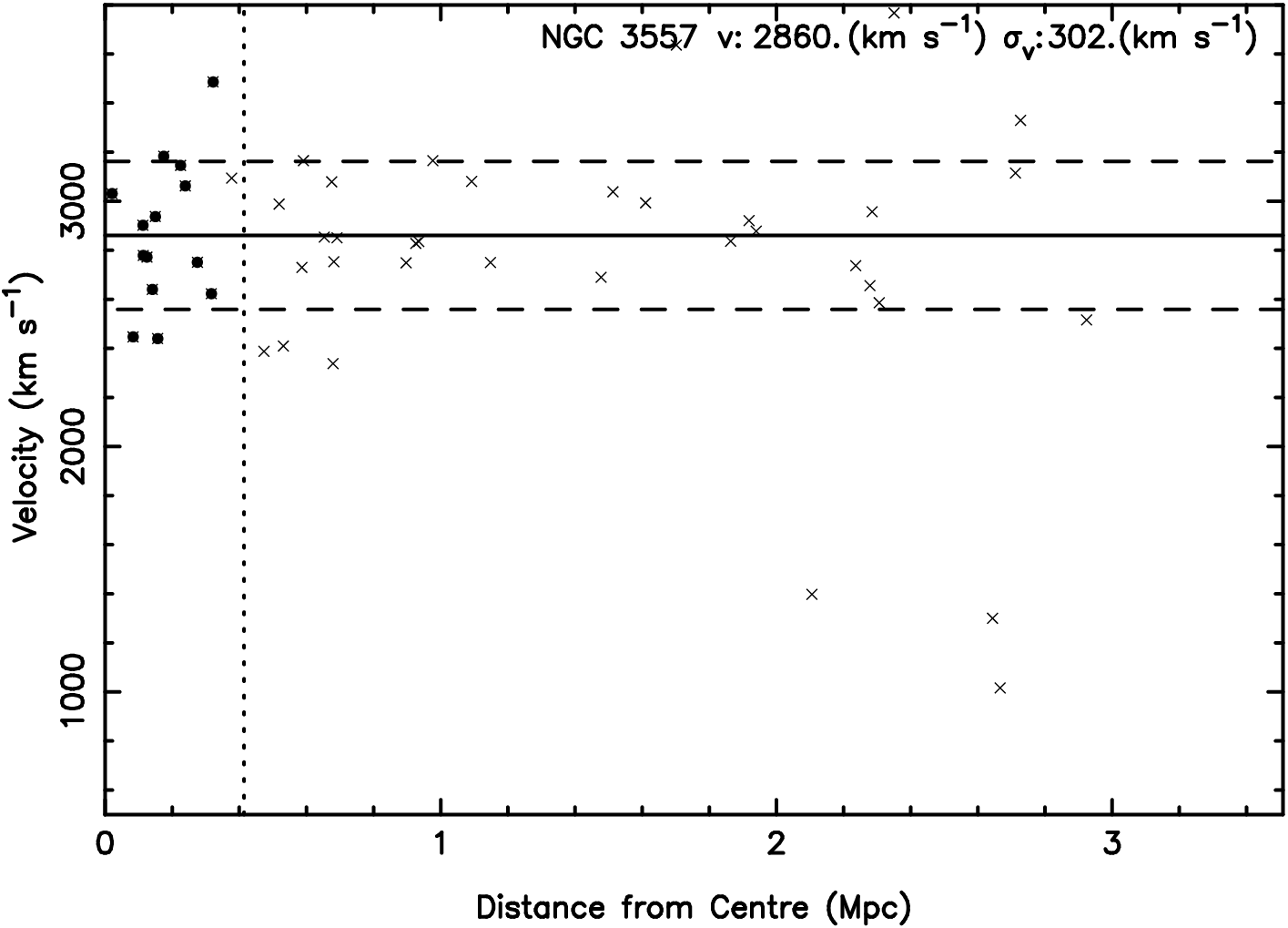}
} 
}
    \resizebox{40pc}{!}{
     \rotatebox{-90}{ 
	\includegraphics{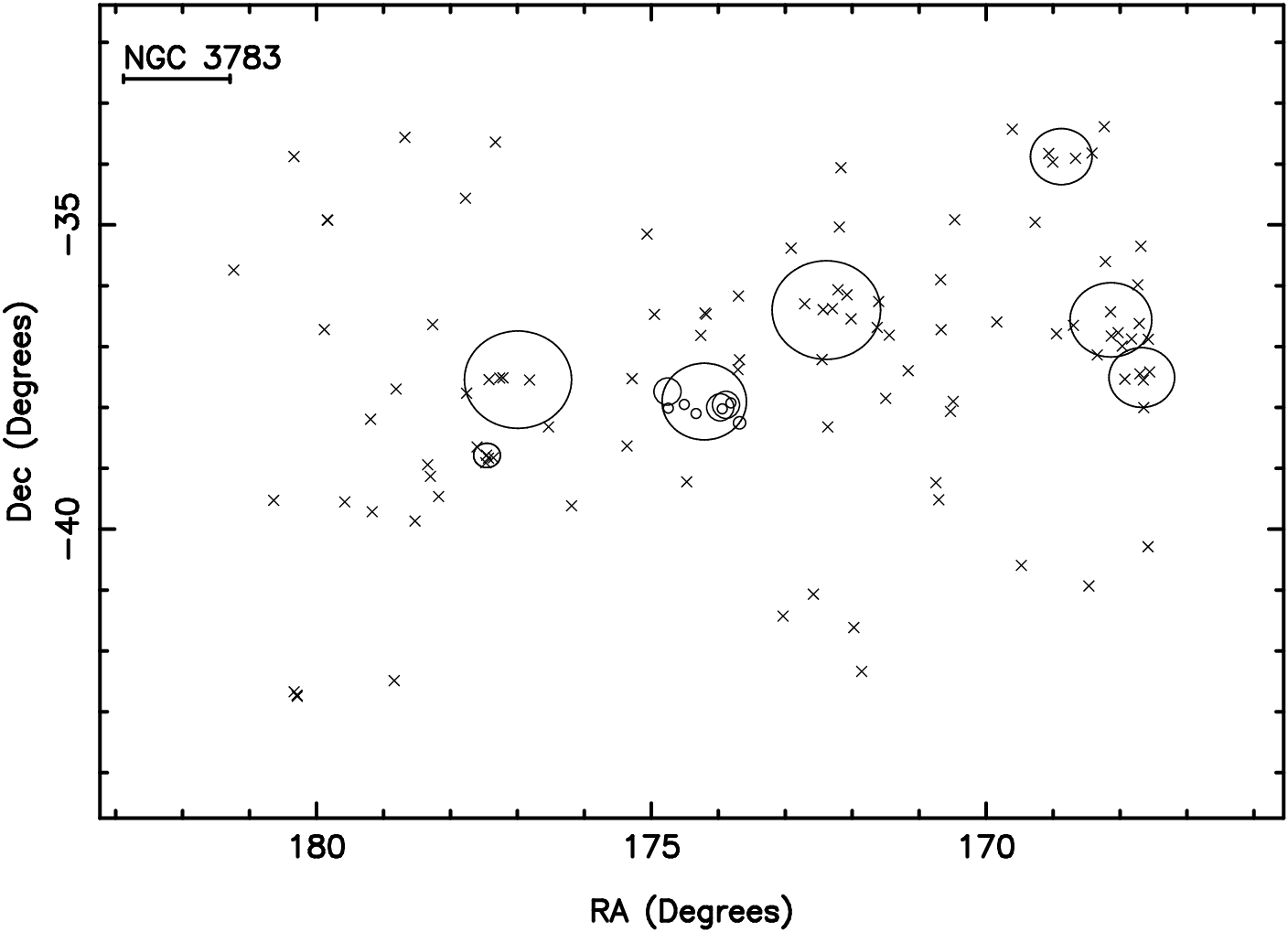} 
    }
    \rotatebox{-90}{ 
	\includegraphics{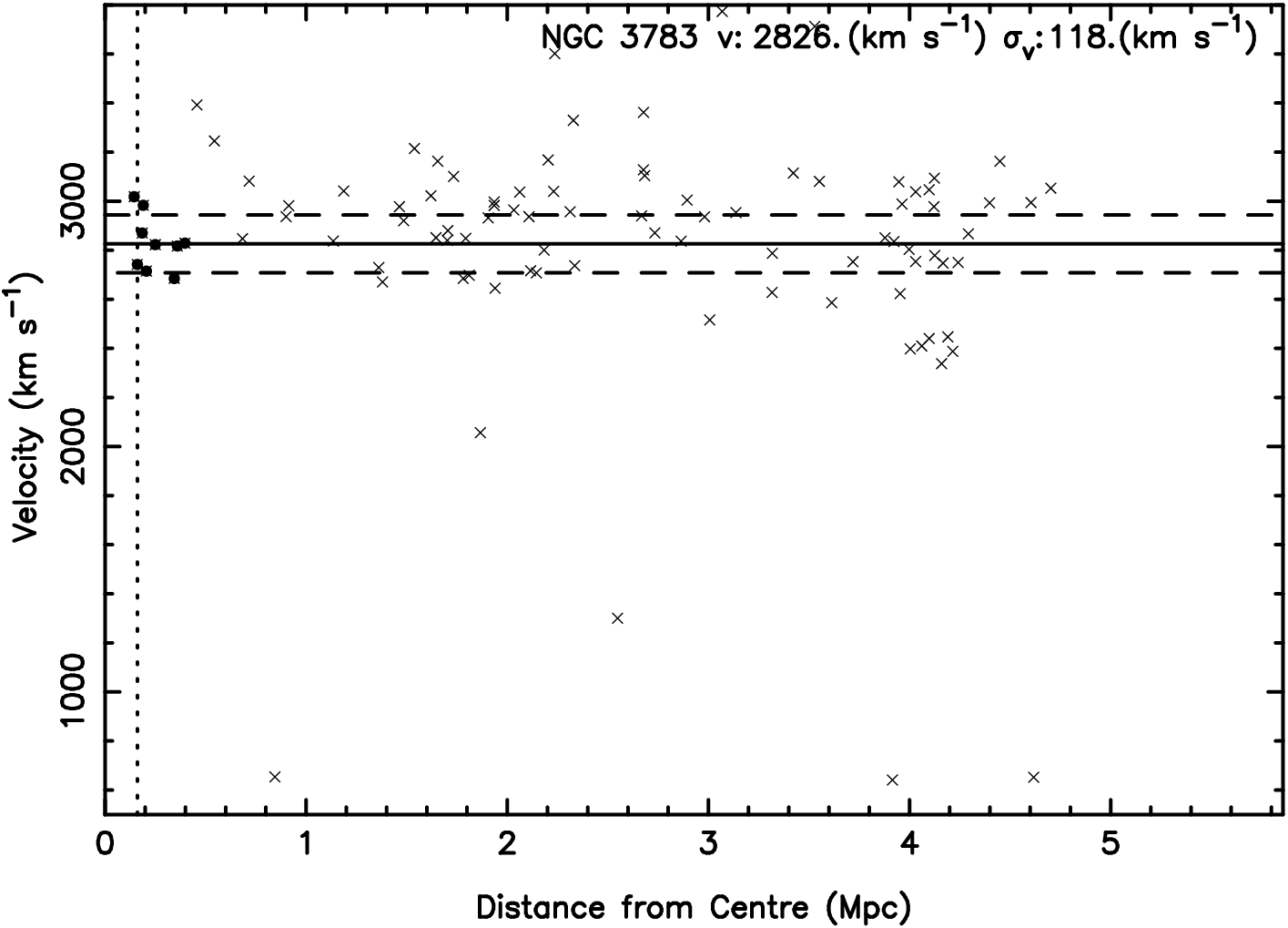} 
    }
}

  \end{center}
\caption {Continued. }
\end{figure*}

\begin{figure*}
\begin{center}
    \resizebox{40pc}{!}{
     \rotatebox{-90}{
	\includegraphics{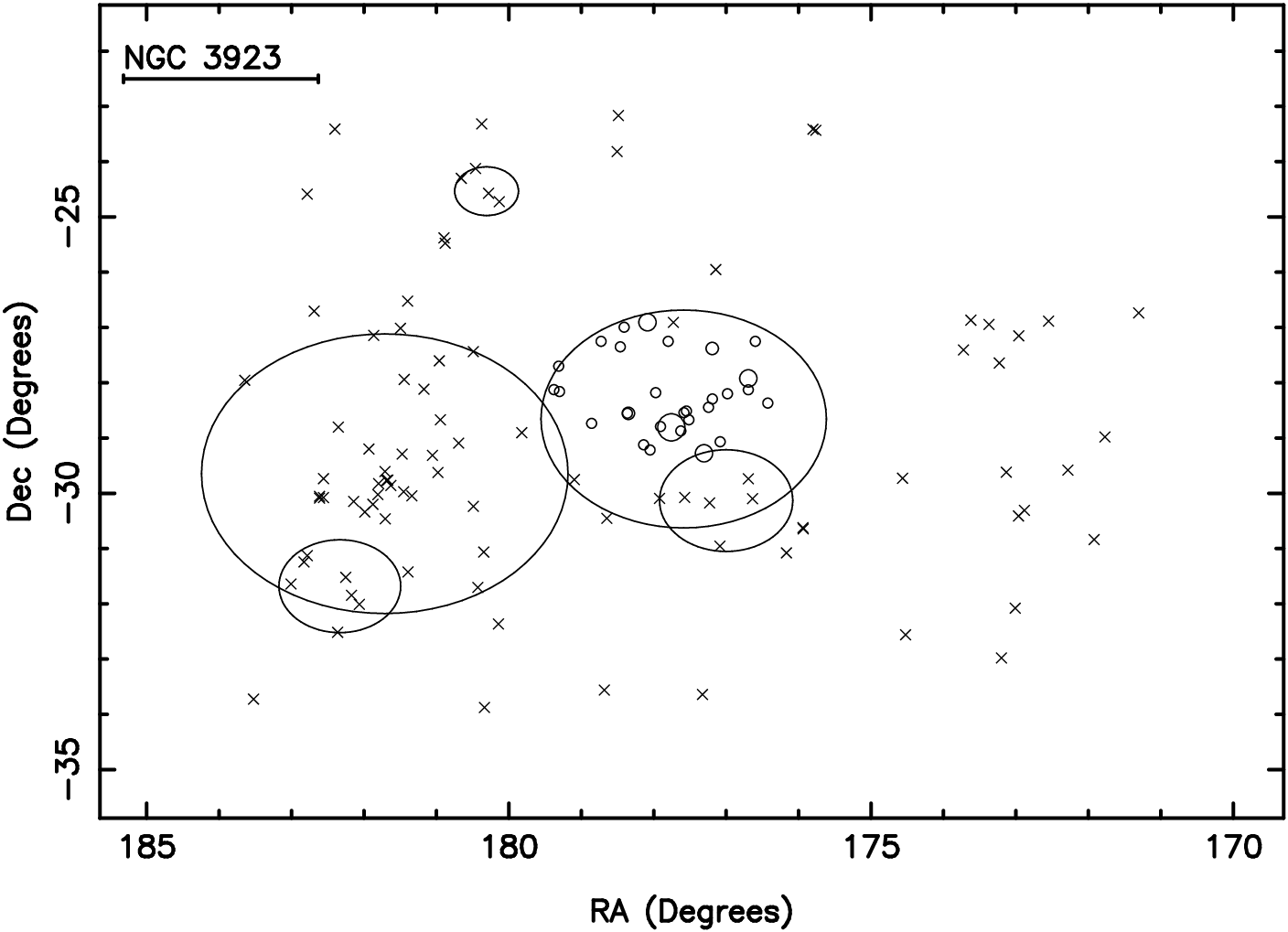}
} 
     \rotatebox{-90}{
	\includegraphics{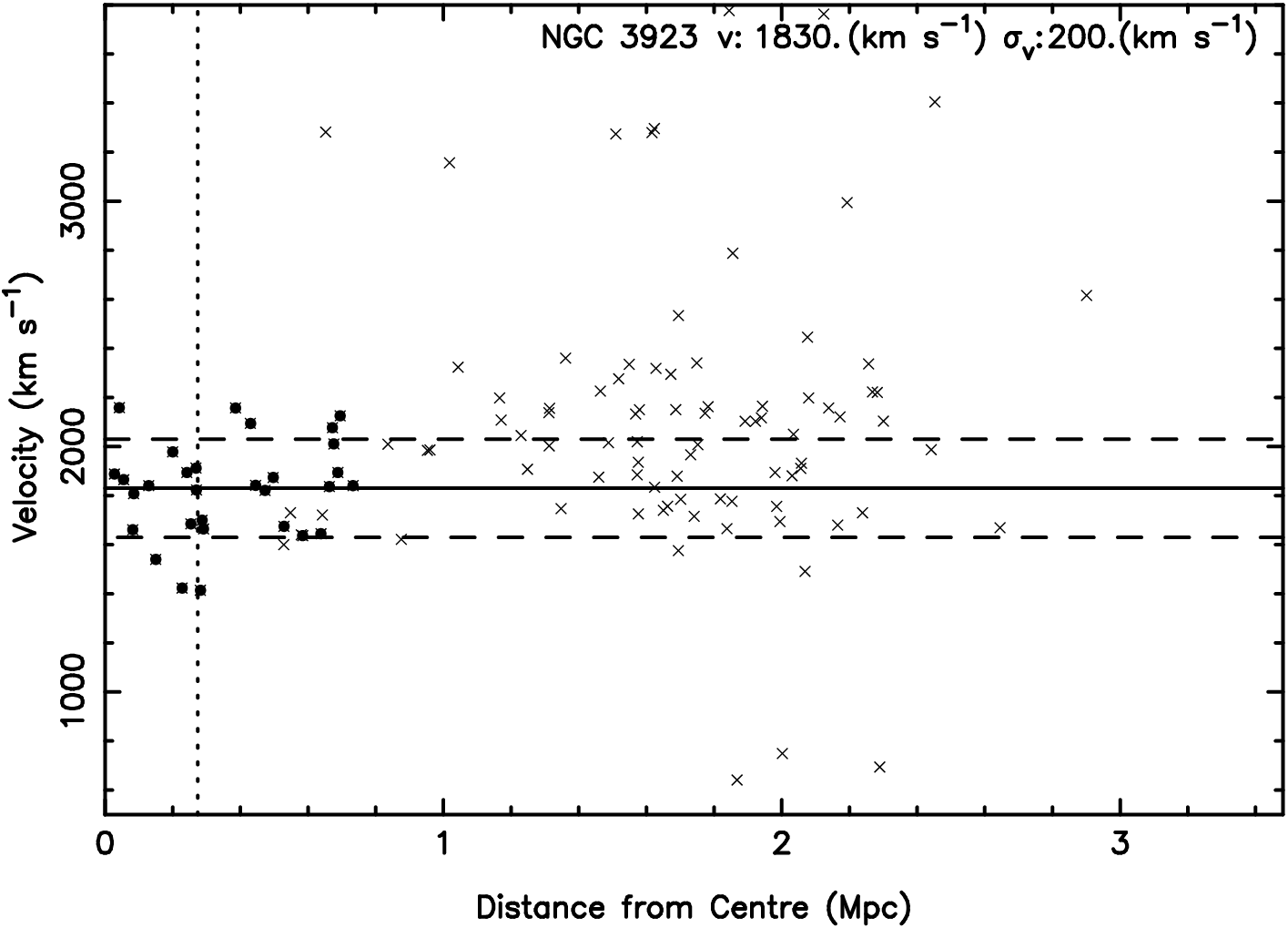}
} }

    \resizebox{40pc}{!}{
     \rotatebox{-90}{ 
	\includegraphics{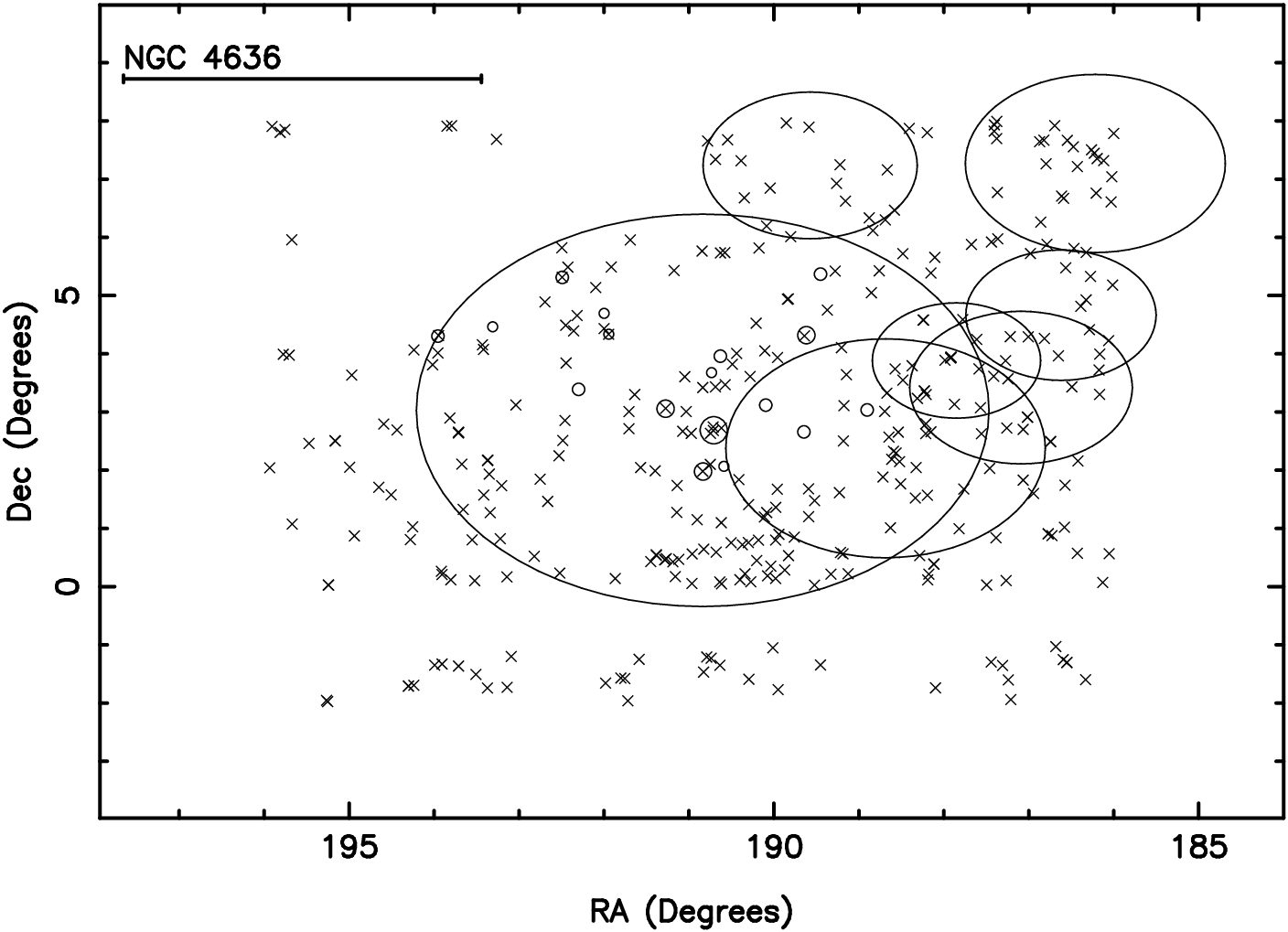} 
    }
     \rotatebox{-90}{ 
	\includegraphics{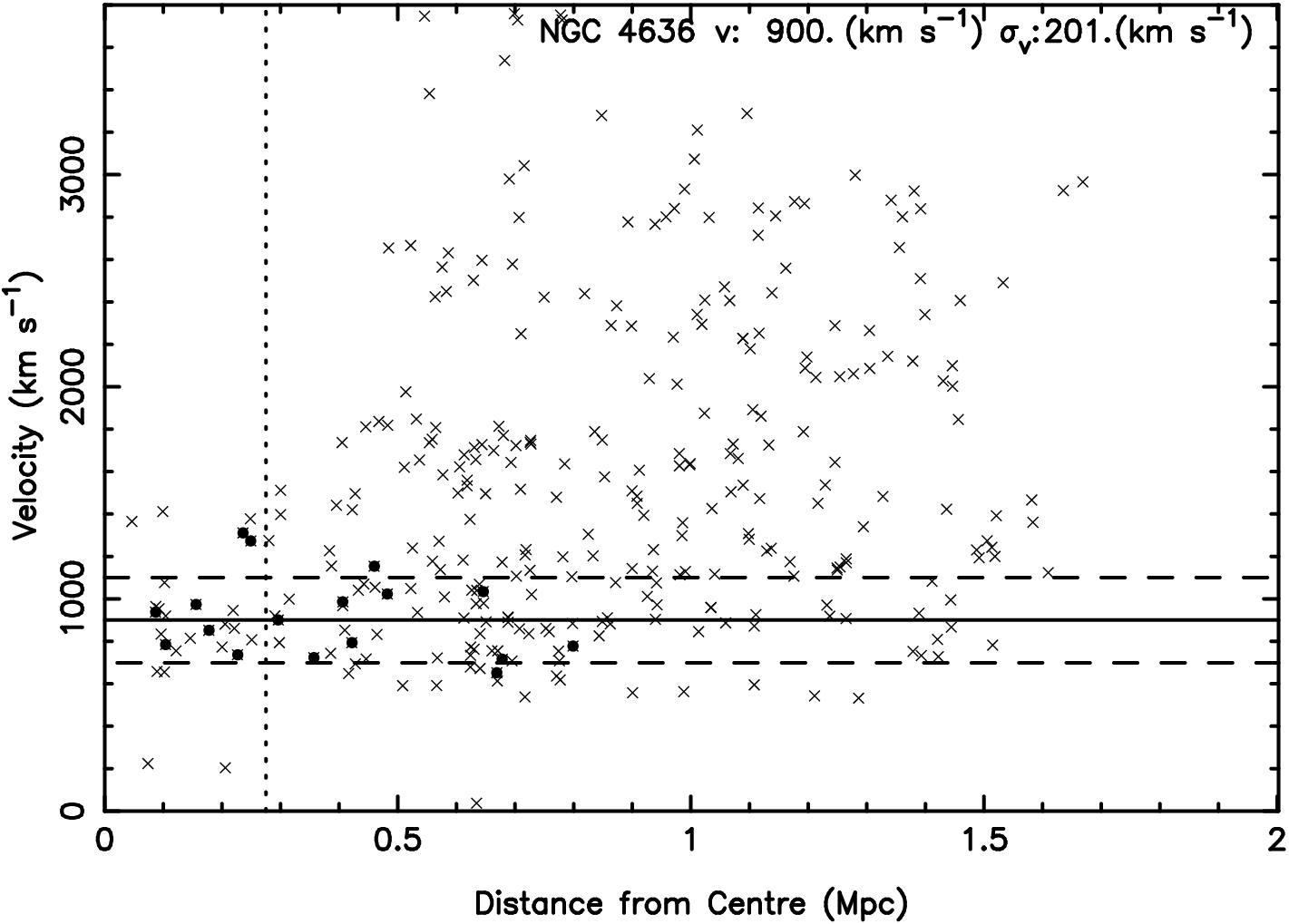} 
    }
}
    \resizebox{40pc}{!}{
     \rotatebox{-90}{
	\includegraphics{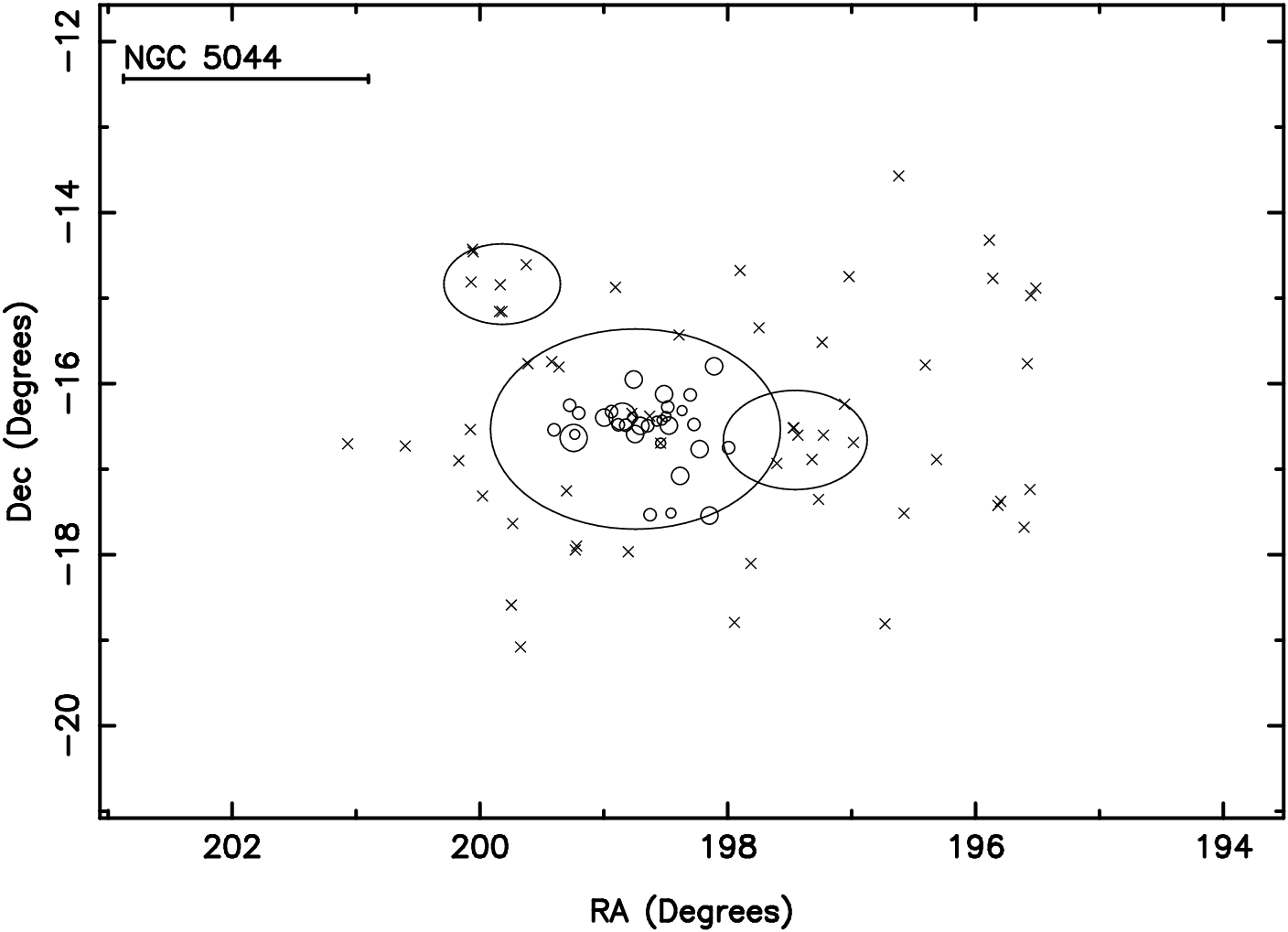} 
} 
     \rotatebox{-90}{
	\includegraphics{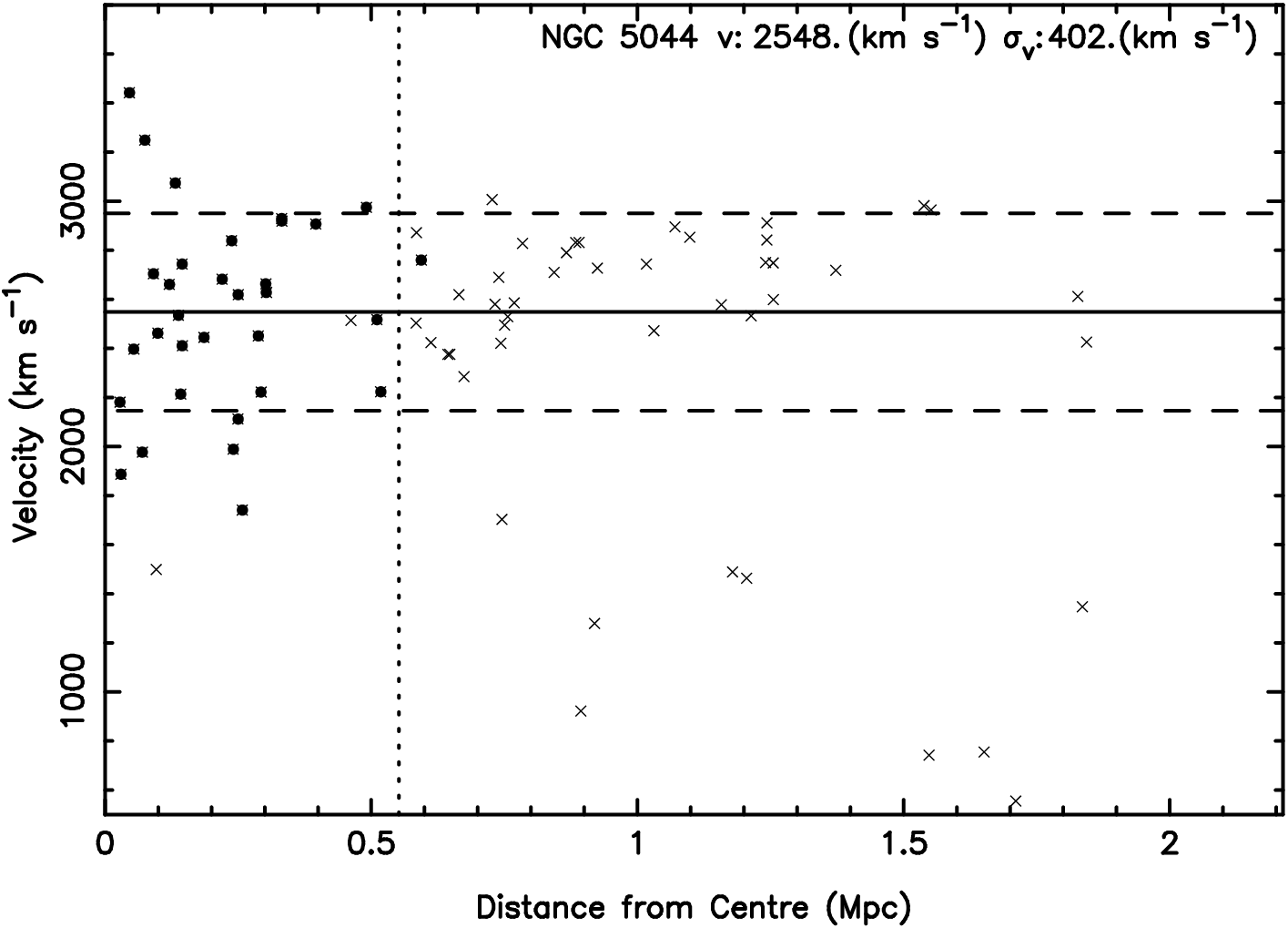} 
} 
}
  \end{center}
\caption {Continued.}
\end{figure*}

\begin{figure*}
\begin{center}
    \resizebox{40pc}{!}{
     \rotatebox{-90}{
	\includegraphics{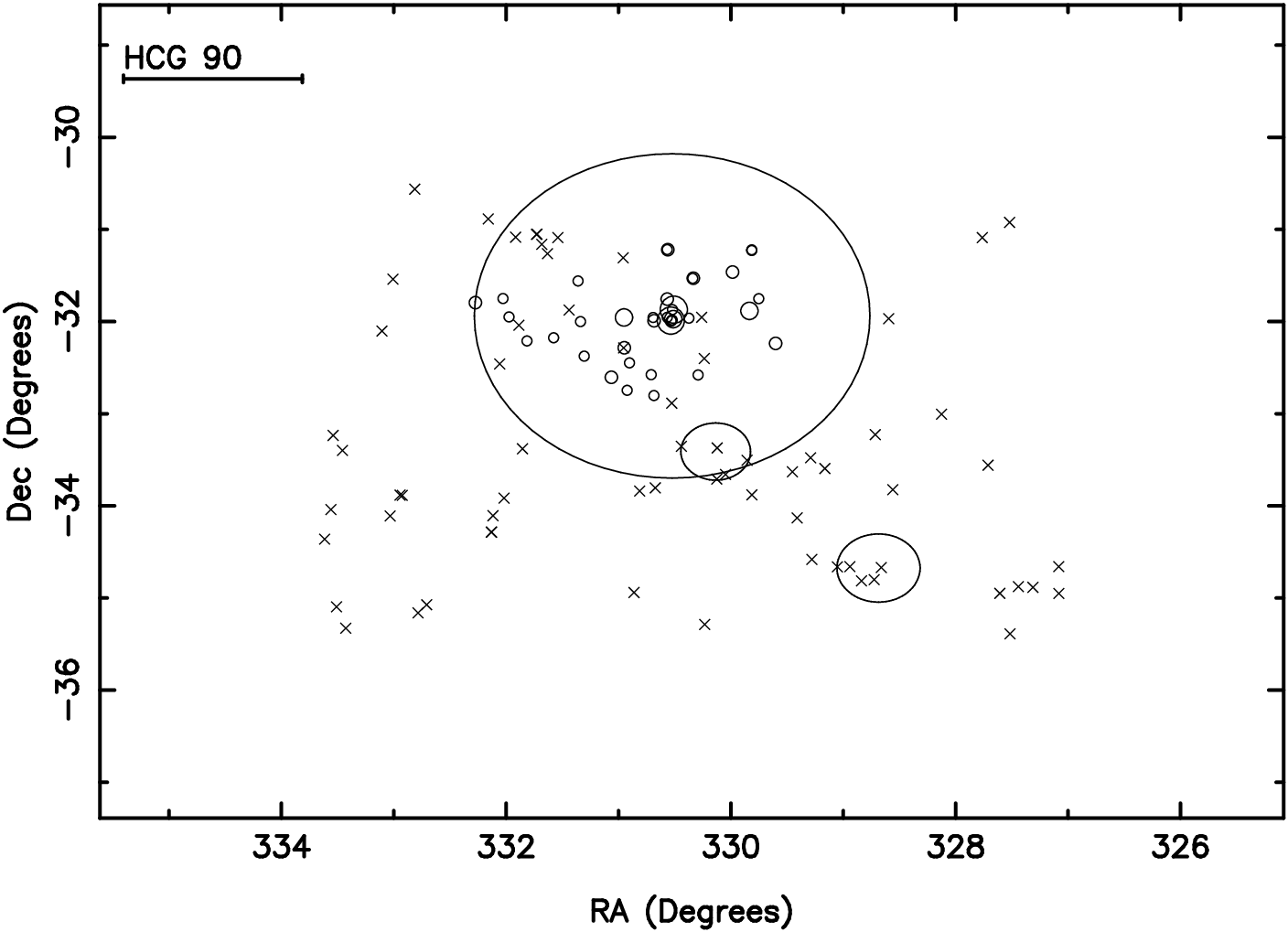} 
    }
     \rotatebox{-90}{
	\includegraphics{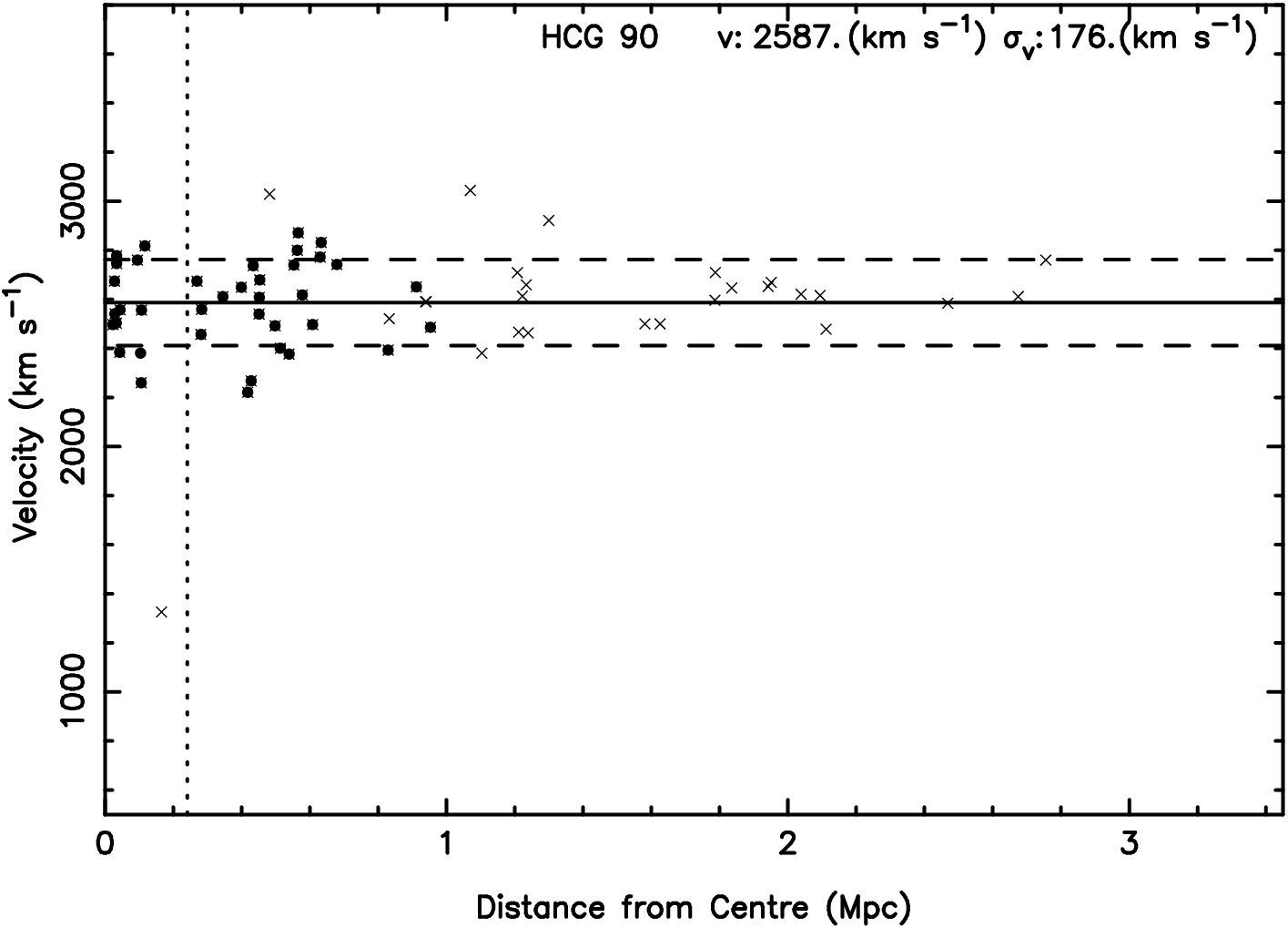} 
    }
}

    \resizebox{40pc}{!}{
     \rotatebox{-90}{
	\includegraphics{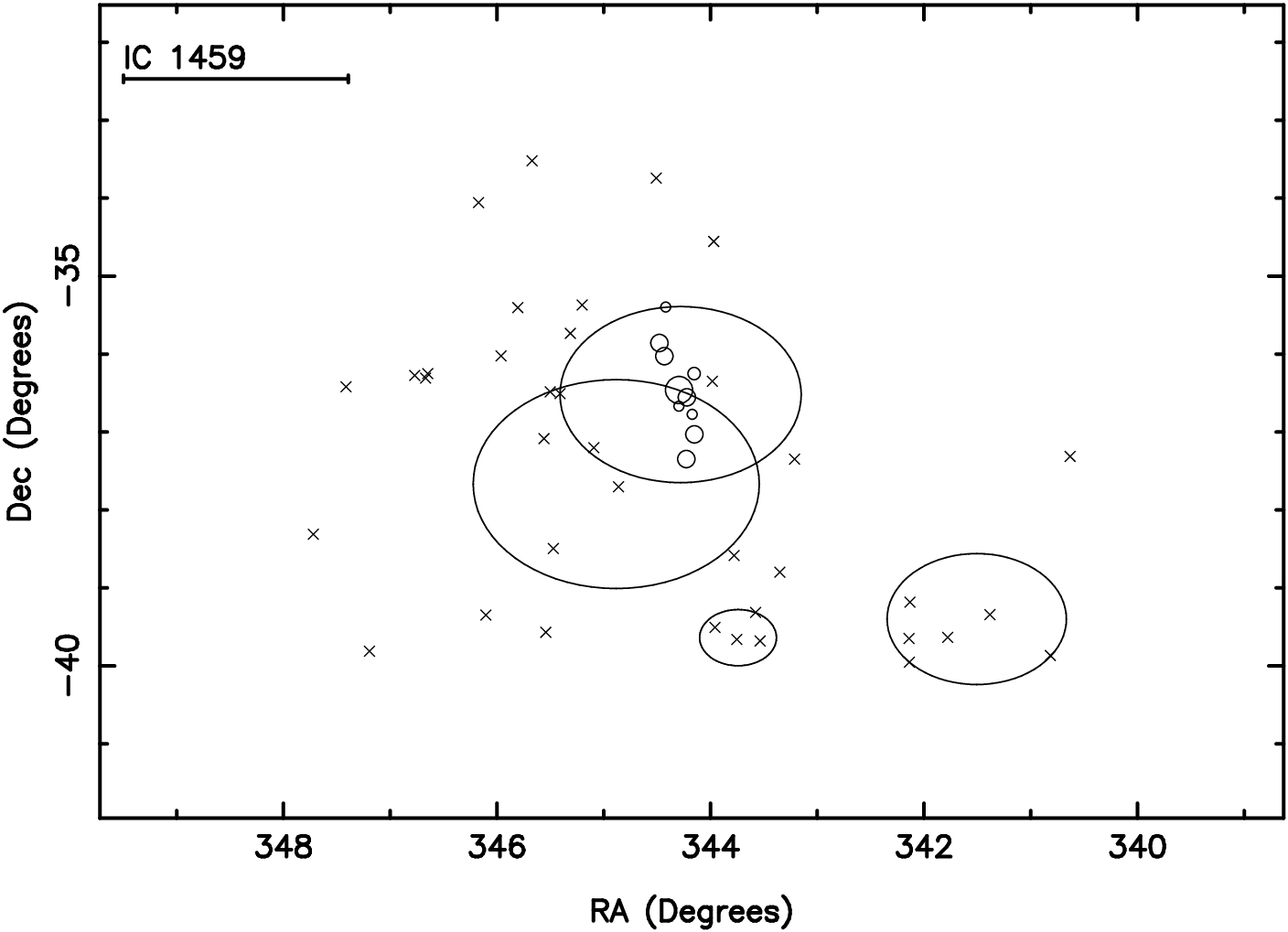}
} 
     \rotatebox{-90}{
	\includegraphics{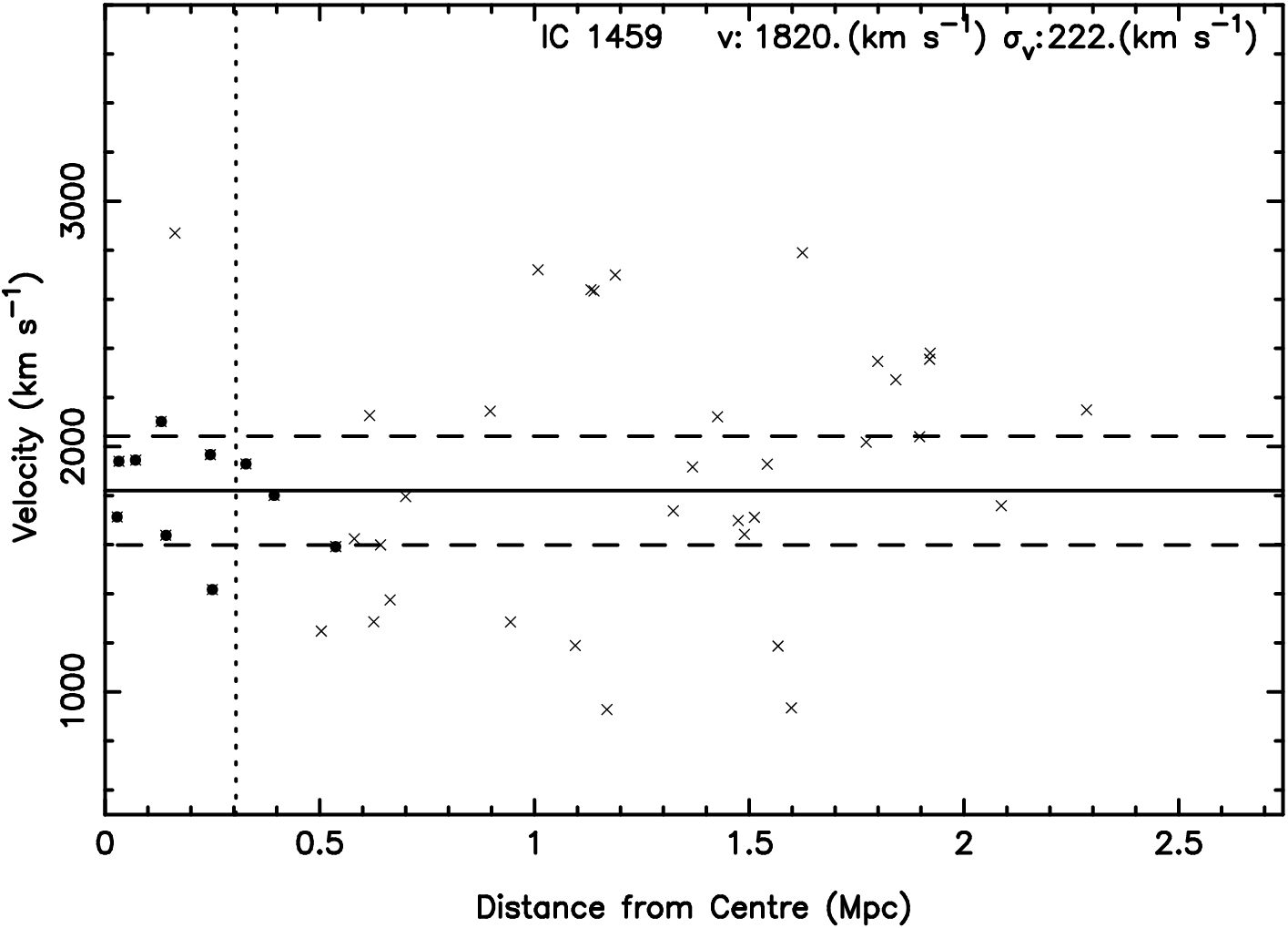}
} 
}

\resizebox{40pc}{!}{
     \rotatebox{-90}{ 
	\includegraphics{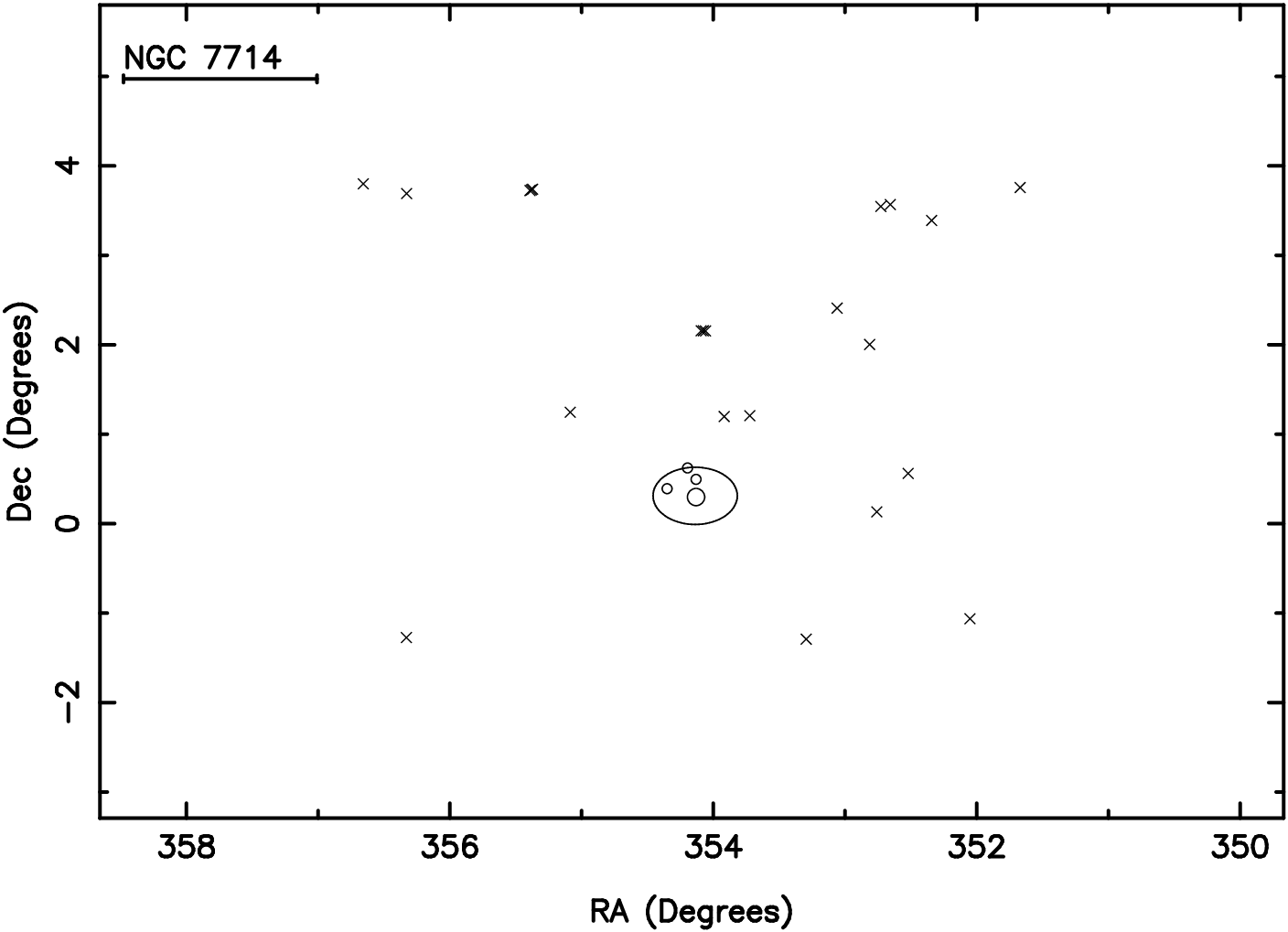} 
    }
     \rotatebox{-90}{ 
	\includegraphics{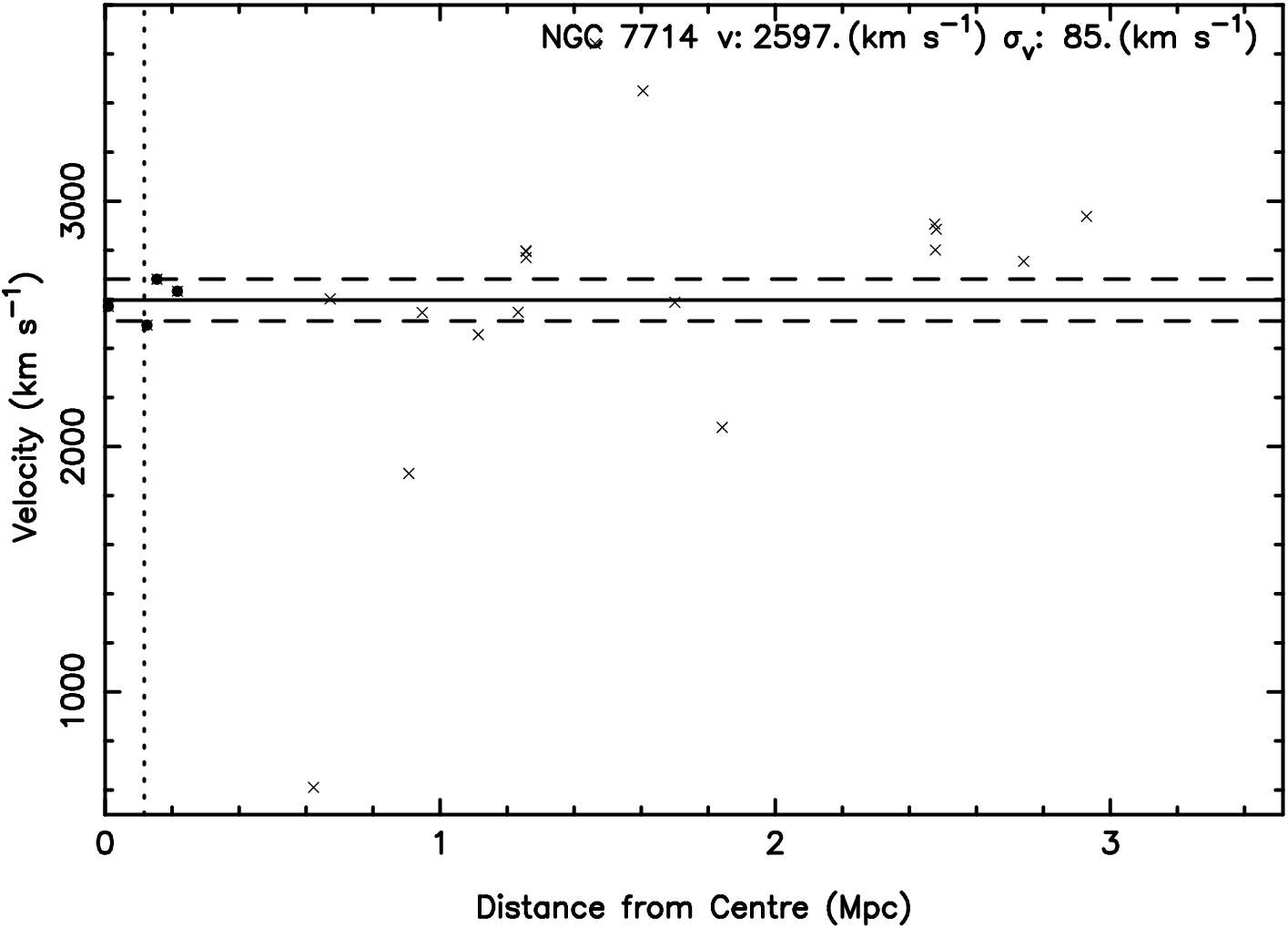} 
    }
}
  \end{center}
\caption {Continued.}
\end{figure*}

\bsp

\label{lastpage}

\end{document}